\documentclass[fleqn,usenatbib]{mnras}

\usepackage[T1]{fontenc}
\defcitealias{2019MNRAS.483.5508F}{1}
\defcitealias{10.1093/mnras/staa1684}{2}
\defcitealias{2021MNRAS.502L..90F}{3}
\defcitealias{2023A&A...673A.114H}{HR2023}
\defcitealias{2025RMxAA..61....3D}{DM2025}
\defcitealias{hunt2025}{Hunt et al. (2025)}

\usepackage{graphicx}	
\usepackage{amsmath}	
\usepackage{amssymb}	
\usepackage{rotating}
\usepackage{textalpha}
\usepackage{pdflscape}
\usepackage{soul}
\usepackage{subcaption}

\title[31 New star clusters discovered]{
New star clusters discovered towards the Galactic anticentre
direction using  {\textit{Gaia}} DR3 data}

\author[F. A. Ferreira et al.]{
F. A. Ferreira$^{1}$\thanks{E-mail: filipe1906@ufmg.br},
M. S. Angelo$^{2}$,
J. F. C. Santos Jr.$^{1}$,
W. J. B. Corradi$^{1,3}$,\newauthor
\ and F. F. S. Maia$^{4}$
\\
$^{1}$Universidade Federal de Minas Gerais, Departamento de F\'isica, Av. Ant\^onio Carlos 6627, 31270-901, Brazil \\
$^{2}$Centro Federal de Educa\c c\~ao Tecnol\'ogica de Minas Gerais, Av. Monsenhor Luiz de Gonzaga, 103, 37250-000, Brazil\\
$^{3}$Laborat\'orio Nacional de Astrof{\'{\i}}sica LNA, R. Estados Unidos, 154, 37504-364, Itajub\'a, MG, Brazil\\
$^{4}$Universidade Federal do Rio de Janeiro, Instituto de F\'isica, 21941-972, Brazil 
}

\date{Accepted XXX. Received YYY; in original form ZZZ}

\pubyear{2025}

\begin{document}
\label{firstpage}
\pagerange{\pageref{firstpage}--\pageref{lastpage}}
\maketitle

\begin{abstract}
We report the discovery of 31 new open clusters (OCs) identified in \textit{Gaia}~DR3 data through a systematic search over 220 adjacent $1^\circ\times1^\circ$ fields towards the Galactic anticentre, in the direction of the Perseus arm gap. Eight of them display low-density structures, possibly indicating open cluster remnants properties. The objects were identified and characterized through a combined analysis of photometric, kinematic, and spatial distributions, a methodology successfully applied in our previous works. Their structural properties, mean proper motions, ages, distances and reddening were derived and their centres cross-matched with the available catalogues. The clusters are low-concentrated systems and are mostly located within $3<d<5$ kpc, exhibiting reddening up to $E(B-V)\approx1.5$, and ages from $\sim$20 Myr to 1 Gyr. The new OCs represent a significant increase in the anticentre cluster census: $31\%$ for $3<d<4$ kpc and $12\%$ for $d>4$ kpc. They do not belong to the Perseus arm, but may be associated with the Outer Norma arm. The Gulf of Camelopardalis region appears as an interruption in the Perseus arm, possibly reflecting low star-formation activity, dust obscuration, or that the Milky Way is a flocculent, rather than a grand-design spiral galaxy.

\end{abstract}




\section{Introduction}
\label{sect:intro}

OCs are gravitationally bound systems composed of coeval stars that form a simple stellar population. They present a wide range of ages ($\log[t({\rm yr})] \lesssim 10.0$) and typically exhibit near-solar metallicities ($-0.5 \lesssim [Fe/H] \lesssim 0.5$, \citealt{kharchenko2013}; \citealp{2016A&A...585A.150N}; \citealt{10.1093/mnras/stab770}). They are predominantly distributed throughout the Galactic disc and are therefore key tracers for studying its formation history and present-day structure. Young OCs are associated with star-forming regions, as they are too young to migrate far from their birthplaces. For this reason they are excellent tracers of the Galactic structure and reveal how stars form in embedded environments as well as the recent Galactic disc history \citep{Lada:2003}. The older OCs establish a lower age limit to the Galactic disc and are fingerprints of its chemical and dynamical evolution (\citealp[e.g.][]{friel1995,2007A&A...476..217C}; \citealp{2016A&A...585A.150N}; \citealp{2020A&A...640A...1C} ).




However, such objects, especially the young OCs, are predominantly located at low Galactic latitudes, close to the Galactic plane, and therefore exhibit small scale heights \citep{kharchenko2013,cjv18}. As a result, the study and characterisation of OCs relies on the challenging task of accurately identifying their members, which are embedded within the field population. This difficulty becomes even more pronounced in the high stellar density environments, typical of low Galactic latitudes, where the contrast between cluster and field populations is significantly reduced.

However, more recently, new data that are crucial to addressing these issues have become available through the latest releases of the \textit{Gaia} catalogue (DR2: \citealp{Lindegren:2018,Evans:2018}; EDR3: \citealp{2021A&A...649A...1G}, DR3: \citealp{2023A&A...674A...1G}). The \textit{Gaia} catalogue provides high-precision proper motions in right ascension and declination, parallaxes and photometric data in three passbands ($G,G_{BP},G_{RP}$) for almost 2 billion stars. With \textit{Gaia} data, there has been a substantial increase of works providing OCs astrophysical parameters and precise memberlists (\citealp[]{cjv18,2019A&A...624A...8A,2021MNRAS.500.4338A,10.1093/mnras/stab770}; \citealp[]{2023A&A...673A.114H} [hereafter \citetalias{2023A&A...673A.114H}]; \citealp[]{2024A&A...689A..18A,2024AJ....167...12C,2025MNRAS.539.2513A}).

The high precision of \textit{Gaia} data has also opened a new dimension in the search for star clusters. Before the \textit{Gaia} era, the efforts to detect new objects were primarily based on identifying spatial stellar overdensities with respect to the background \cite[e.g.][]{Froebrich:2007,Kronberger:2006}. Now, the distinction between OCs and field stars has become possible by analysing the distribution of cluster stars in the astrometric space.

\begin{figure*}
    \centering

\includegraphics[width=0.75\linewidth]{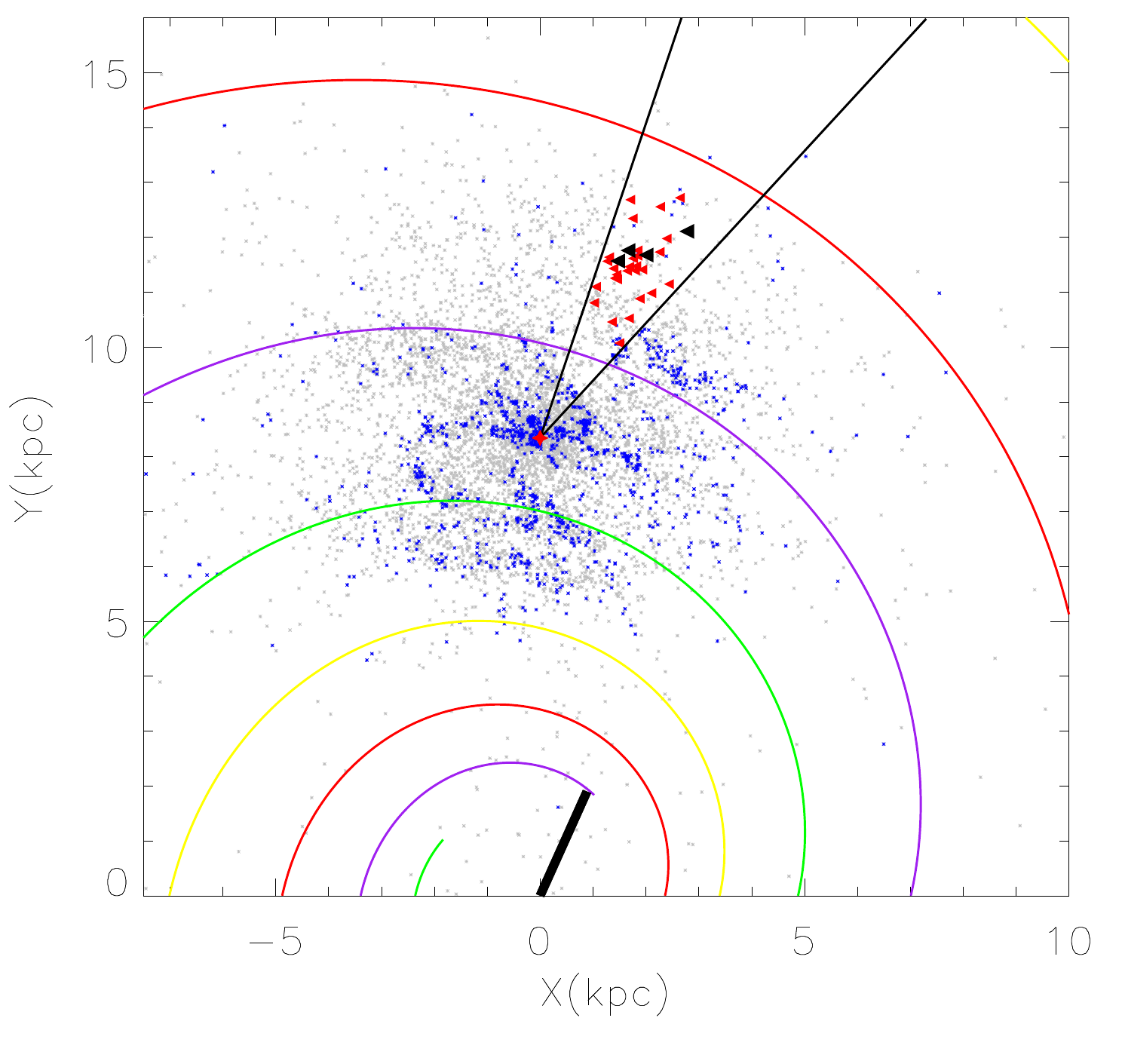}   
\includegraphics[width=0.75\linewidth]{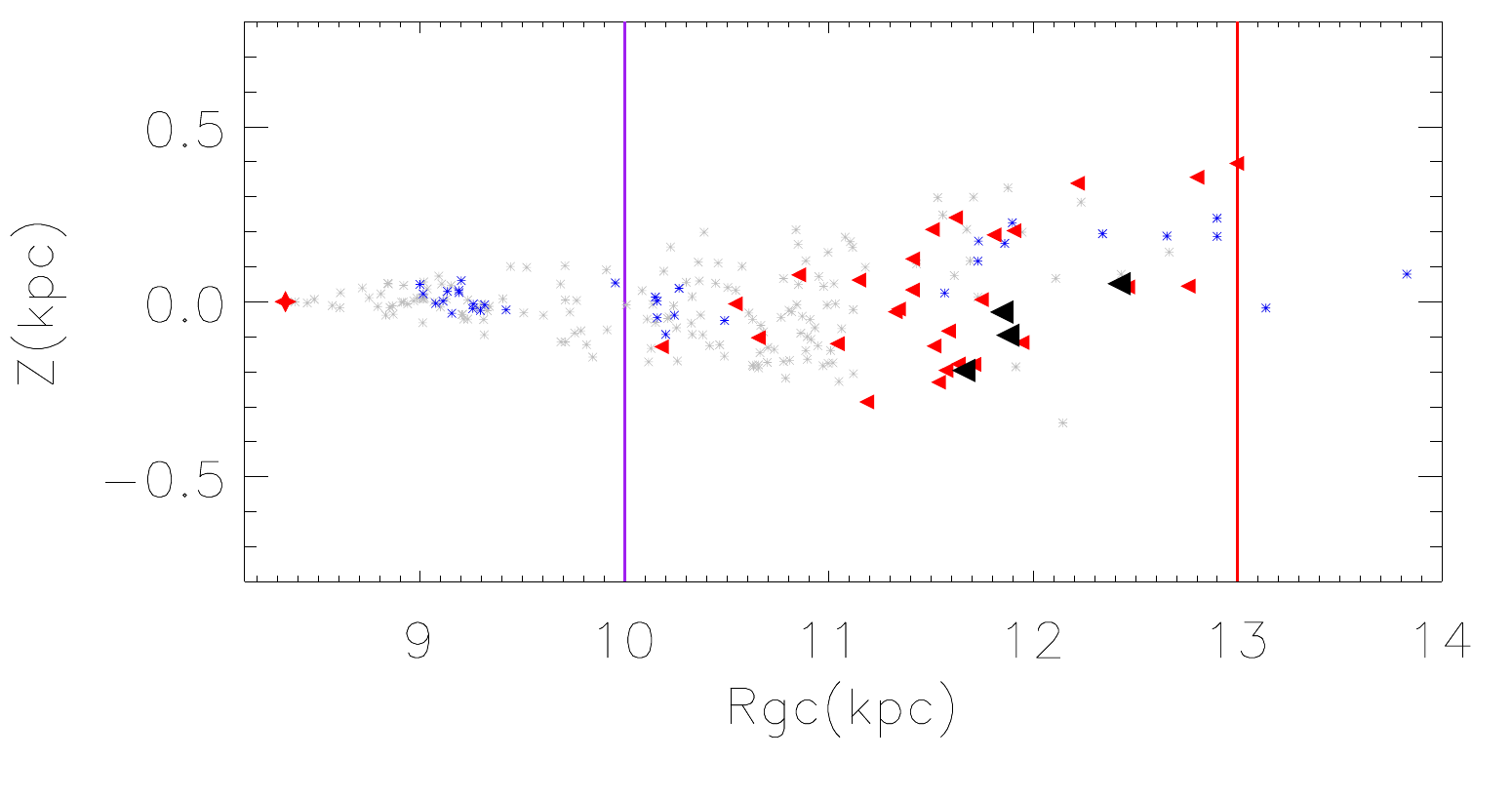}    
     \caption{Top: schematic drawing of the spiral arms and the local population of known OCs according to. The blue dots represent the population of younger OCs ($log(t) < 7.5$) and the grey ones represents their older counterparts. The 31 newly discovered OCs in this work are also shown as black triangles (young) and red triangles (old). The solid lines represent the spiral arms positions: Perseus (purple), Outer Norma (red), Scutum (yellow) and Sagittarius (green). The Galactic bar and the Sun's location (red star, at 8.34 kpc; Reid et al. (2014)) are also represented. The black solid lines represent the limit in Galactic longitude used in this work. Bottom: The Galactocentric distances versus the distance from the Galactic plane for younger and older OCs inside the region of interest. The Sun's location is also represented by a red star. The approximate location of the arms are also represented: Perseus in purple and  Outer Norma in red.}
    \label{fig:galaxy_hunt_ocs}
\end{figure*}

The use of \textit{Gaia} astrometric data in combination with different search methodologies has led to a remarkable increase in the number of star clusters discovered recently, especially those projected against crowded fields along the Galactic disc (\citealp[]{cjv18,cjl18}, \citealp[]{2019MNRAS.483.5508F} [hereafter paper \citetalias{2019MNRAS.483.5508F}]; \citealp[]{2019AJ....158..122K,sla19,2020AJ....160..279K,cjl20}; \citealp[]{10.1093/mnras/staa1684} [hereafter paper \citetalias{10.1093/mnras/staa1684}]; \citealp[]{2021RAA....21..117C}; \citealp[]{2021MNRAS.502L..90F} [hereafter paper \citetalias{2021MNRAS.502L..90F}]; \citealp[]{2022arXiv220908504H}; \citetalias{2023A&A...673A.114H}; \citealp{Chi_2024}; \citealp{10.1093/pasj/psaf082}). Most of the new OCs identified in the \textit{Gaia} era were detected through automated searchers, particularly those employing well-established clustering algorithms such as DBSCAN \citep{cjl19,cjl20,2022A&A...661A.118C} and HDBSCAN (\citetalias{2023A&A...673A.114H}; \citealp[]{2025RMxAA..61....3D} [hereafter \citetalias{2025RMxAA..61....3D}]). On the other hand, a significant number of objects has also been detected by visual inspection of the data, with focus on smaller data subsets and verification of the presence of cluster-like patterns directly in the astrometric parameter space (paper \citetalias{2019MNRAS.483.5508F}; paper \citetalias{10.1093/mnras/staa1684}; paper \citetalias{2021MNRAS.502L..90F}; \citealp{sla19,2021RAA....21..117C}).

Before the \textit{Gaia} era, at least $\sim$2000 OCs had measured parameters reported in the literature \citep{dias2002,kharchenko2013}. After the first year of the second \textit{Gaia} data release, \cite{2019AJ....157...12B} showed that the census of Galactic star clusters, associations and candidates had surpassed 10000 objects (although many of these corresponded to asterisms, young embedded OCs, or globular clusters). Bica et al.'s catalogue, which included only a small number of clusters discovered using \textit{Gaia} DR2 data, listed $\sim 2900$ objects flagged as 'OC'. Subsequently, \cite{2021A&A...652A.102H} compiled a catalogue of 3794 OCs that incorporated several hundreds of newly identified \textit{Gaia} DR2 clusters in addition to pre-\textit{Gaia} era OCs. More recently, with the improved precision in parallaxes and, particularly, in proper motions provided by \textit{Gaia} EDR3 and DR3, thousands of additional OCs have been discovered, and the current census of confirmed OCs in the Milky Way is expected to exceed 6000 objects (\citealp{2022arXiv220908504H}; \citetalias{2023A&A...673A.114H};\citealp{2023MNRAS.526.4107P,2024AJ....167...12C}). This number still represents only a small fraction of the total expected OC population in Galaxy, which may reach  $\sim 1.3 \times 10^{5}$ objects \citep{Bonatto2006,2024A&A...686A..42H}. In a statistical sense, the more OCs are discovered and properly characterized, the better we can constrain the structure and evolution of the Milky Way. In particular, young OCs are crucial tracers of the spiral arms, making their identification and study especially valuable.

The morphology of the Milky Way is still a matter of debate. Since we are deeply embedded within the Galactic disc, multiple structural features overlap along any line of sight, making it a challenging task to trace its global structure. The spiral structure, mainly detected with radio, infrared and optical observations, is commonly described either as a two-arm structure (Norma and Perseus; e.g. \citealp{2000A&A...358L..13D}; \citealp{2009PASP..121..213C}; \citealp{2023ApJ...947...54X}) or as a four-arm structure (Norma, Scutum, Sagittarius and Perseus; e.g. \citealp{1976A&A....49...57G}; \citealp{2003A&A...397..133R}; \citealp{2017PASP..129i4102K}; \citealp{10.1093/mnras/stad3350}). To better constrain this morphology, objects such as young star clusters, masers associated with high-mass star forming regions (HMSFRs), classical Cepheids, H\,II regions and even red clump stars \citep{2014ApJ...783..130R,2021A&A...652A.102H,2022ApJ...931...72L,2025A&A...696A..67S} are among the best and most commonly used tracers capable of revealing the nature of the spiral arms.


The continuity of the Milky Way spiral arms is also matter of debate. The spatial distribution of spiral arms tracers usually outlines segments of the arms, as regions where they are strongly concentrated. However, some arms seem to be interrupted in certain regions, exhibiting apparent gaps in the distribution of such tracers \citep{2022PASJ...74..209S,2024IAUS..380...97S,2023AJ....166..170J}. Such interruptions on the arms structures suggest different scenarios: (i) low star formation rate, (ii) the presence of dense dust clouds or (iii) even that the Milky Way is not a grand design spiral galaxy with prominent and well-defined continuous spiral arms with possibly multi-arm, patchy and flocculent spirals \citep{2016SciA....2E0878X,2024NewAR..9901696C}. 

An important gap occurs in the Perseus arm, in its outer portion, towards the Galactic anticentre. This region, which has been referred to as the Gulf of Camelopardalis \citep{cjl19, 2023AJ....166..170J, 2024MNRAS.532.3480D}, is limited by Galactic longitude $140^{\circ}$ $\lesssim l \lesssim$ $160^{\circ}$, extends over $\sim 1$ kpc, and lies in the II quadrant of the Galactic disc. Its mapping was only recently made possible by \textit{Gaia} DR2 data, as a lack of OB stars \footnote{https://www.cosmos.esa.int/web/gaia/iow$\_$20180614} noted in the region. This gap has been observed {using different tracers}. In particular, the lack of known young OCs in this region has been reported in different studies (\citealp{cjv18}; \citealp{2019A&A...624A.126C}; \citealp{cjl19}; \citealp{2021A&A...652A.162C}; \citealp{2024NewAR..9901696C}; \citealp{10.1093/mnras/stad3350}; \citealp{2023AJ....166..170J}; \citealp{2008ApJ...672..930V}). Figure \ref{fig:galaxy_hunt_ocs} shows a schematic drawing of the spiral arms from \cite{Vallee_2020} overplotted with known OCs from \citetalias{2023A&A...673A.114H}. The lack of young OCs ($log[t](yr)<7.5$), represented by the blue dots, is noted towards the anticentre. The most up-to-date sets of standard candles also highlight the scarcity of other known tracers such as HMSFR masers and H\,II regions (see Figs.~6, 7 and 8 from \citealp{2025A&A...696A..67S}), classical Cepheids \footnote{https://gaia-unlimited.org/map-of-milky-way-cepheid-variables/} (see Fig.~8 from \citealp{2025A&A...698A.230D}) and red supergiants stars (see Fig.~11 from \citealp{2025A&A...698A.282M}) in this region. 

There are also alternative explanations for the apparent lack of objects. According to \cite{2021A&A...651A.104P}, this gap can be naturally explained as simply being the inter-arm region between the Perseus arm, as traced by \cite{2006Sci...312.1773L}, and the Local arm. On the other hand, most of the models tend to trace the Perseus arm through this region, and the apparent scarcity of objects at Galactocentric distances greater than 10 kpc in this direction does not seem to occur due to interstellar extinction, but rather reflects a true physical under-density of objects in this portion of Perseus arm.

In a series of papers, we have discovered new OCs by visually inspecting the astrometric signatures of the clusters in the \textit{Gaia} data (papers \citetalias{2019MNRAS.483.5508F}, \citetalias{10.1093/mnras/staa1684} and \citetalias{2021MNRAS.502L..90F}). This method has proven to be effective to find low concentrated OCs, especially those projected against dense stellar fields. These works were based on \textit{Gaia} DR2 data. At this time, automated algorithms were not capable to recover such clusters. The improved astrometric precision of \textit{Gaia} DR3 data with respect to \textit{Gaia} DR2, allowed those automated searchers to recover such low contrast objects, and our discoveries have since been independently recovered by other authors. For example, \citetalias{2023A&A...673A.114H} successfully recovered $25/25$ ($100\%$) and $32/34$ ($94.1\%$) of our discovered OCs in papers \citetalias{10.1093/mnras/staa1684} and \citetalias{2021MNRAS.502L..90F}, respectively. Besides the three OCs discovered on paper \citetalias{2019MNRAS.483.5508F}, which were not recovered by their method, these objects have been fully re-detected in other works using \textit{Gaia} data \citep{lp19,cjl20,2022ApJS..259...19L}.

According to \citetalias{hunt2025}, the detectability of OCs projected towards the anticentre is highly influenced by mass, distance, extinction and age. They showed that massive young OCs are highly detectable (see Figs. 6 and 9 in \citetalias{hunt2025}) and that the absence of such objects at a Galactocentric radius of $R_{GC}\sim 13$\,kpc is more likely due to a limit for massive cluster formation rather than a cut-off radius of in star formation in the Milky Way. In this way, the \textit{Gaia}-DR3-based cluster census remains highly incomplete for distant low-mass clusters of all ages. Our methodology has proven capable of finding such low contrast and sparse clusters, typically, reddened, distant and low mass systems (papers \citetalias{2019MNRAS.483.5508F},\citetalias{10.1093/mnras/staa1684} and \citetalias{2021MNRAS.502L..90F}). In this work, we applied our method to perform a systematic search for OCs towards the Perseus gap using \textit{Gaia} DR3. The main goal of this search, focused on low Galactic latitudes, is to identify primarily young clusters that are likely associated with the Perseus arm in this region. If there exist remaining OCs there, the youngest ones would provide valuable tracers for mapping the Perseus arm in this direction.

This paper is structured as follows. In Section \ref{sect:search} the data and the method used to search for new OCs is presented.  Section \ref{sect:validation} we validated the newly discovered OCs. In Section \ref{sect:analysis} the analysis procedures are discussed, including membership assessment and determination of structural and astrophysical parameters. In Section  \ref{sect:discu}, the properties of our OCs sample are explored in the Milky Way context. Concluding remarks are given in Section \ref{sect:concl}.

\begin{figure*}
\centering
\includegraphics[width=0.98\linewidth]{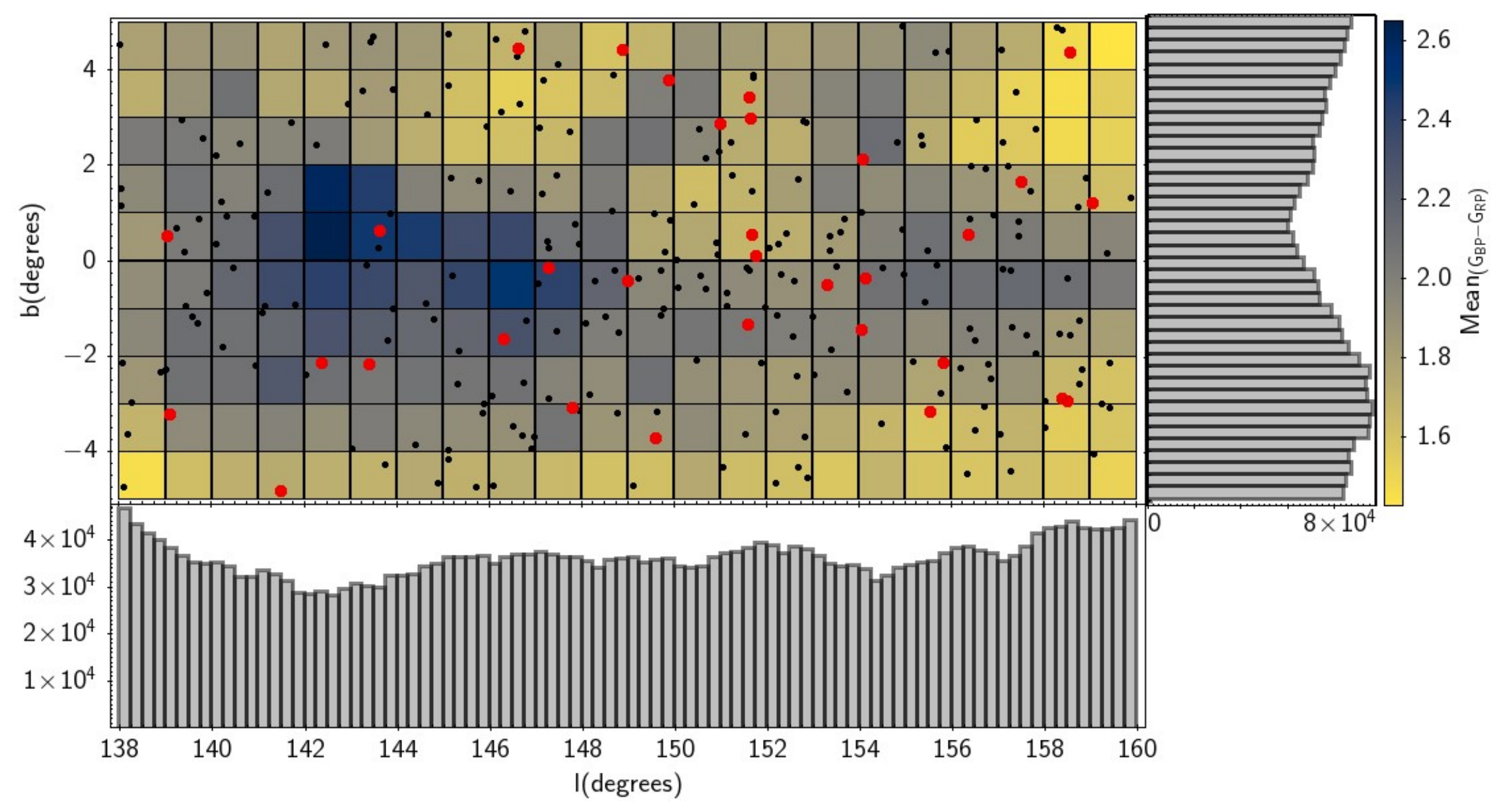}
\caption{Spatial coverage of the Galactic fields surveyed in this work. The size of the black squares indicates the surveyed regions.  The colours indicate the mean colour index ($G_{BP}-G_{RP}$) of the samples inside each square. The histograms show the local number of stars in bins of Galactic longitude and latitute. The red filled circles indicate the position of the 31 newly discovered OCs in this work. The black filled circles indicate the position of the 225 known OCs detected by our methodology.}
\label{fig:tiles_menores}
\end{figure*}

\section{Method: Searching for star clusters}
\label{sect:search}

\subsection{Data}

In this work we have made use of the high-precision astrometric and photometric data from the \textit{Gaia DR3} catalogue. The \textit{Gaia} DR3 parallaxes present a zero-point offset and a correction over the photometric flux excess factor parameter is also required. Therefore, to ensure the correct use of the data, we applied the following procedures:

\begin{itemize}

 \item we have used the Gaia@AIP\footnote{https://gaia.aip.de/} online services to extract \textit{Gaia} DR3 data with the corrected photometric flux excess factors \citep{2021A&A...649A...3R}. The query was restricted to the Galactic coordinates $-5^{\circ}$ $\leq b \leq$ $5^{\circ}$ and $138^{\circ}$ $\leq l \leq$ $160^{\circ}$, covering the Perseus gap region. We opt to acquire data through a small range of latitudes to assure that potential new OCs are not located far from the Galactic plane at the expected position of the Perseus arm. The query closely follows the example provided in the \textit{Gaia} EDR3 documentation (Appendix B, \citeauthor{2021A&A...649A...1G} \citeyear{2021A&A...649A...1G});
 
 \item to ensure astrometric and photometric completeness, a filter to remove stars fainter than the nominal \textit{Gaia} magnitude limit ($G < 19$) have been adopted \citep{2021A&A...649A...2L};
 
\item the parallax zero-point correction was applied following the prescription provided by \cite{2021A&A...649A...2L}, where the offset depends on the source sky position, magnitude, and colour.
 
\end{itemize}

In order to remove spurious astrometric and photometric solutions, quality filters have been applied to our database, by keeping sources consistent with the following conditions:

\begin{equation}
|C^{*}|<5 \sigma_{C^{*}}
\end{equation}
\begin{equation}
 RUWE<1.4,
\end{equation}

\noindent where $C^{*}$ is the corrected BP/RP flux excess factor and $\sigma_{C^{*}}$ is given by equation 18 of \cite{2021A&A...649A...3R}. The RUWE (renormalised unit weight error) selection follows \cite{2021A&A...649A...2L}. 

\subsection{Tiling and searching for OC candidates}
\label{sect:tile_detect}

In the present work, we searched for new clusters candidates following similar procedures to those described in paper \citetalias{2021MNRAS.502L..90F}. The surveyed area was sectioned into 220 tiles of 1\,$\times$\,1\,deg$ ^2$ area in Galactic coordinates. For each tile, we determined the mean stars colour index $G_{BP}-G_{RP}$ (Fig. \ref{fig:tiles_menores}). Based on these mean colour values, we built two smaller subsamples: one containing stars bluer than the mean colour tile and another containing stars redder than this value. The star distributions of both subsamples were analysed on vector point diagrams (VPDs) and skycharts, where we searched for overdensities by inspecting the stars distribution on both diagrams. 

\begin{figure*}
\includegraphics[page=1,width=0.24\linewidth]{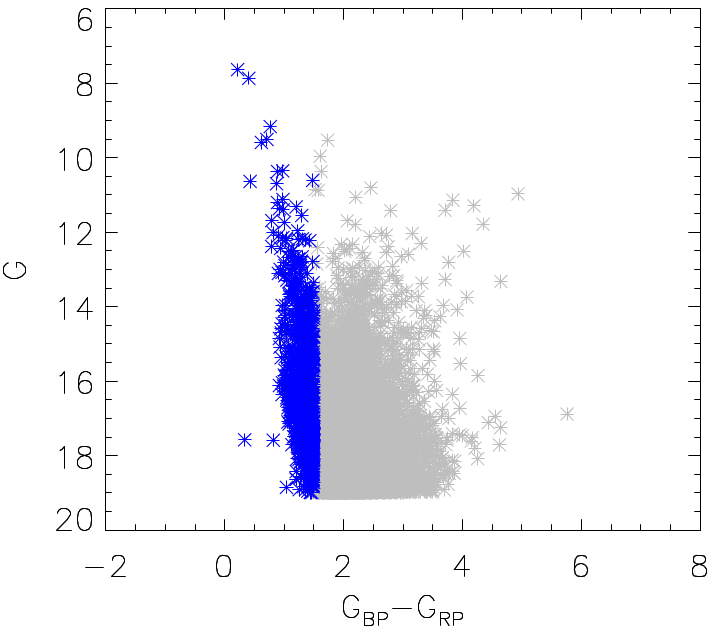}
\includegraphics[page=1,width=0.25\linewidth]{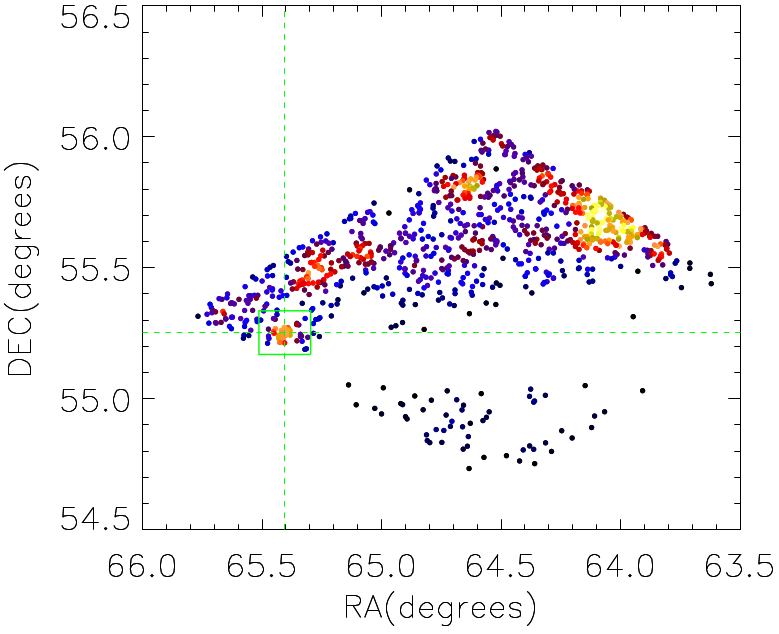}
\includegraphics[page=1,width=0.24\linewidth]{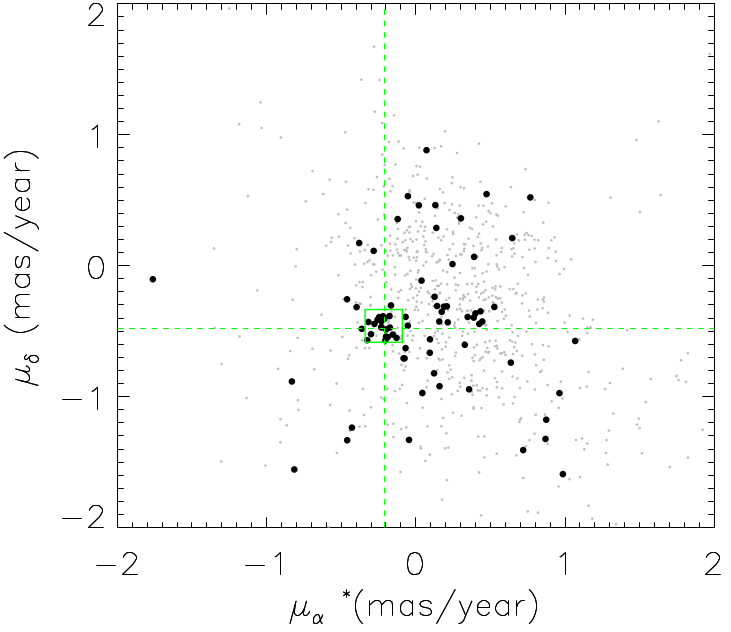}
\includegraphics[width=0.20\linewidth]{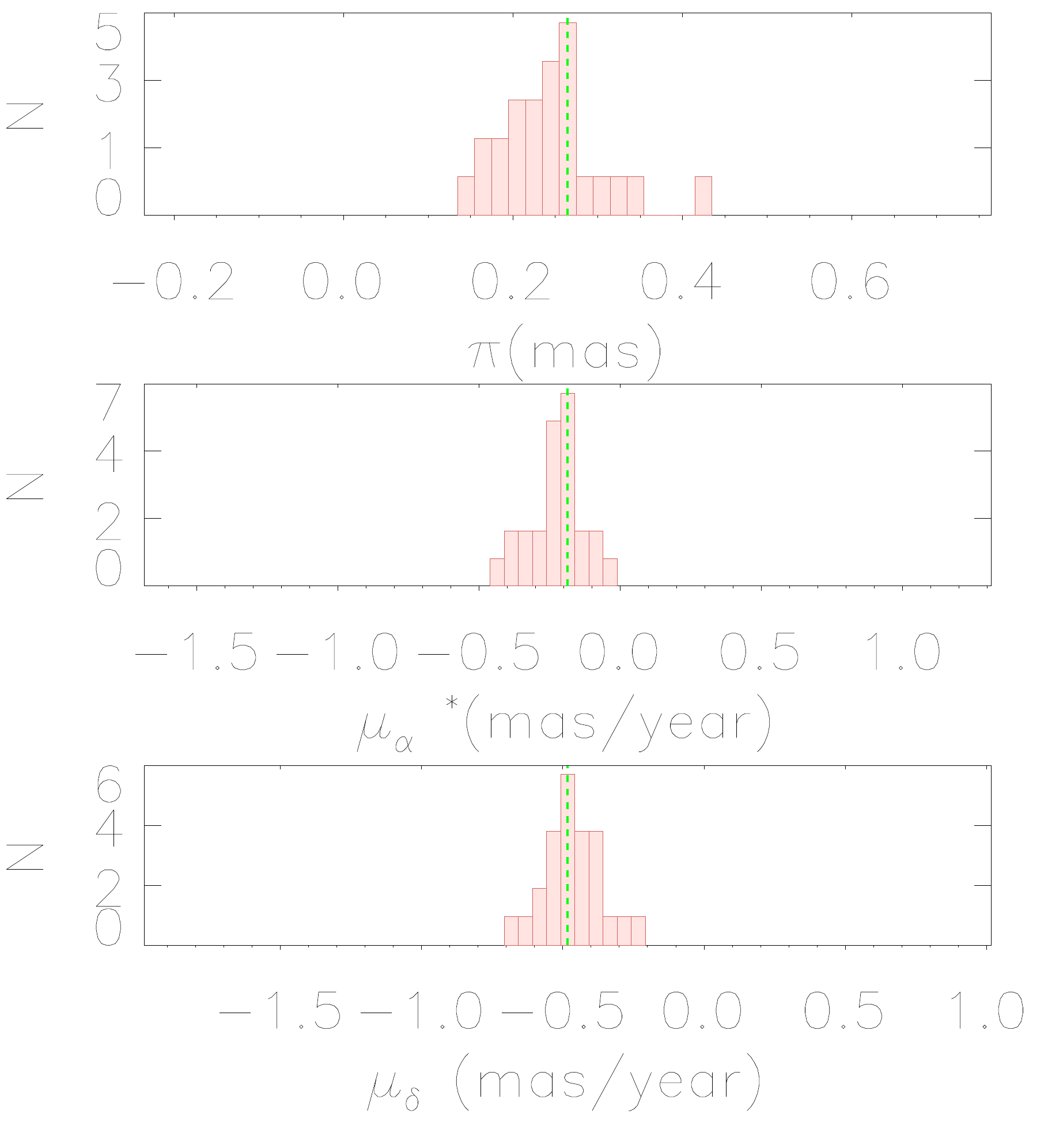}
\includegraphics[page=1,width=0.24\linewidth]{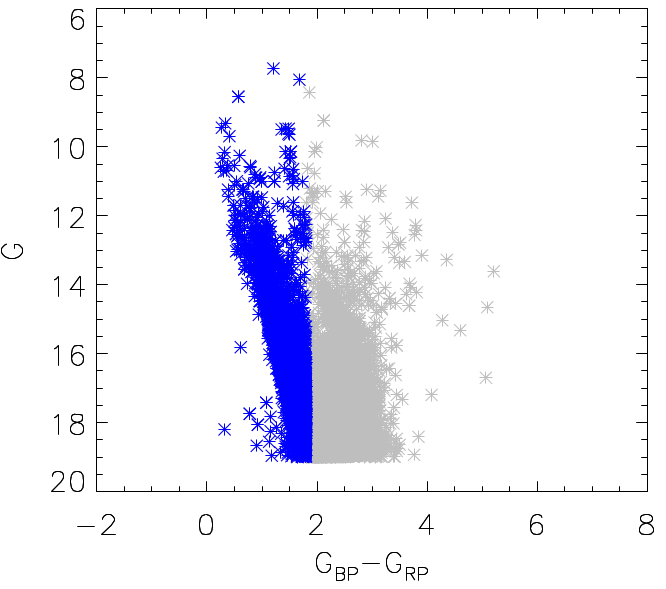}
\includegraphics[page=1,width=0.24\linewidth]{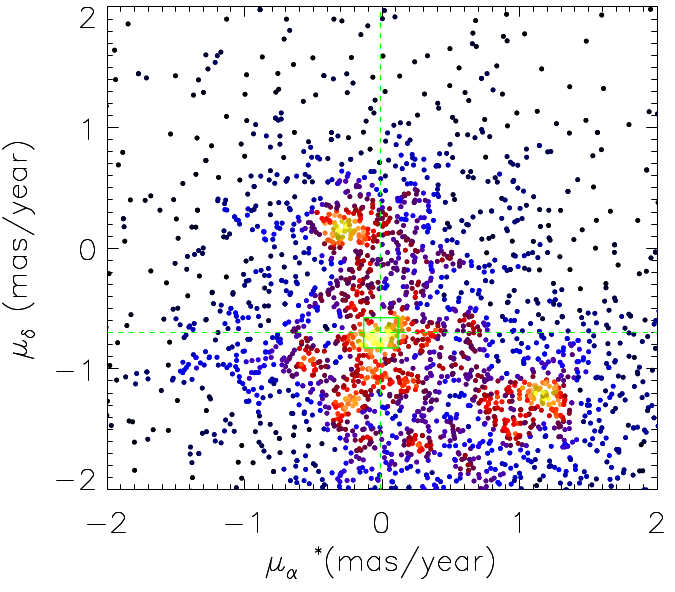}
\includegraphics[page=1,width=0.25\linewidth]{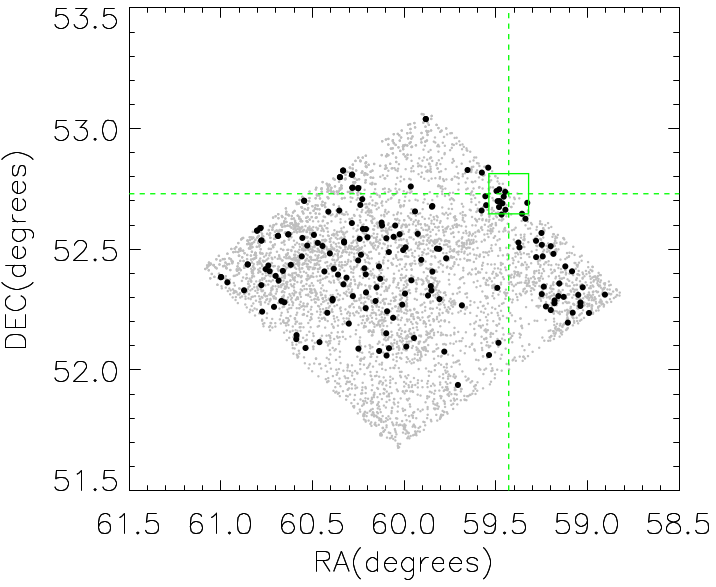}
\includegraphics[width=0.20\linewidth]{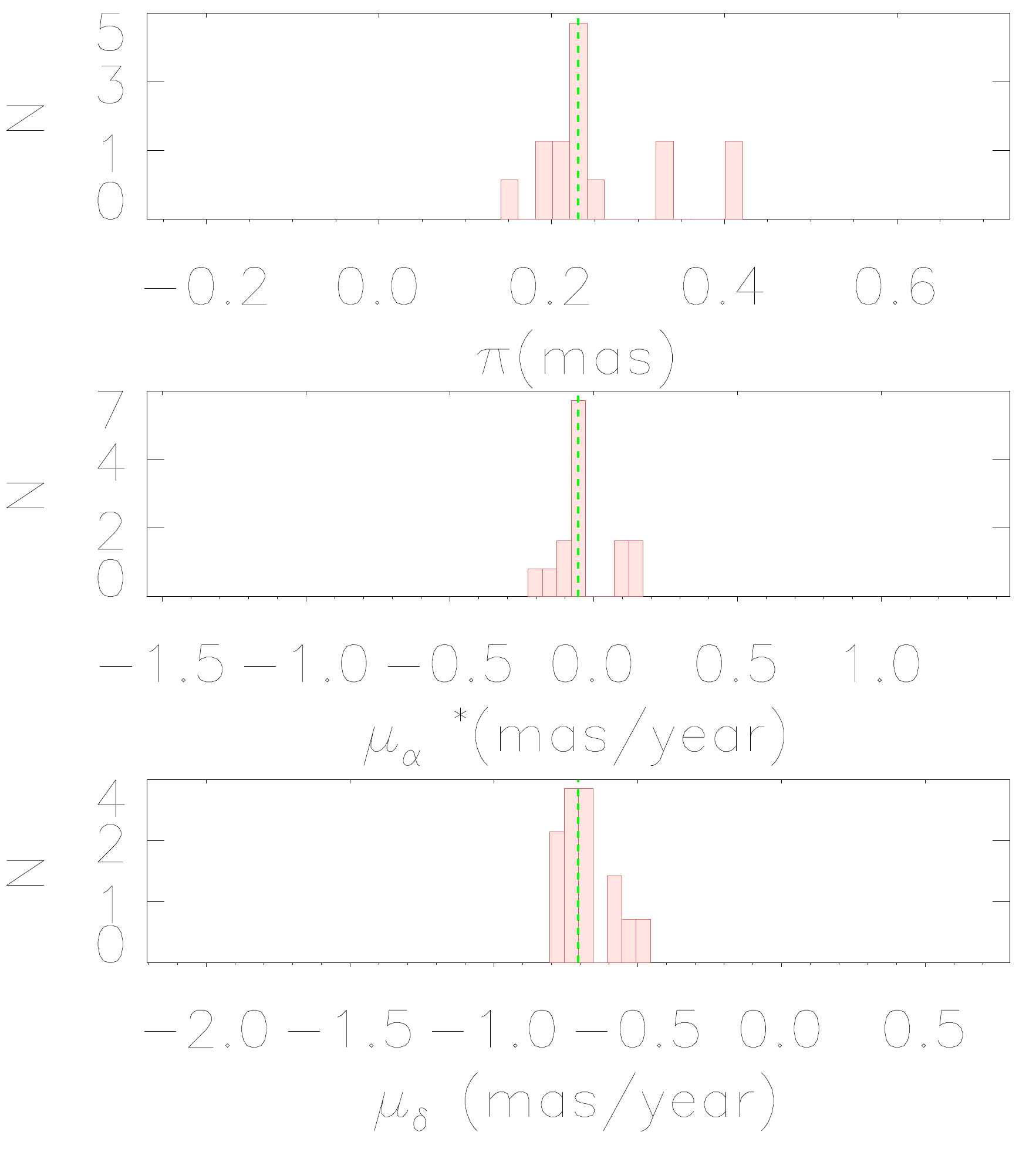}
\includegraphics[page=1,width=0.24\linewidth]{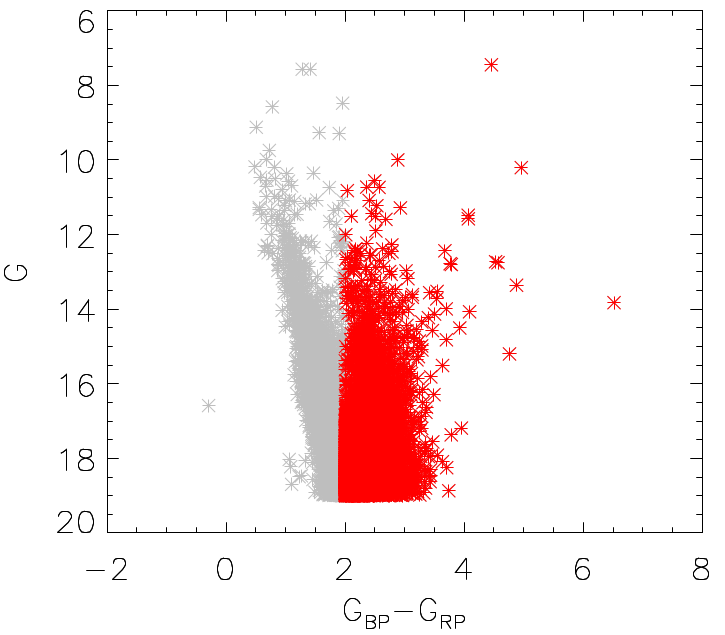}
\includegraphics[page=1,width=0.24\linewidth]{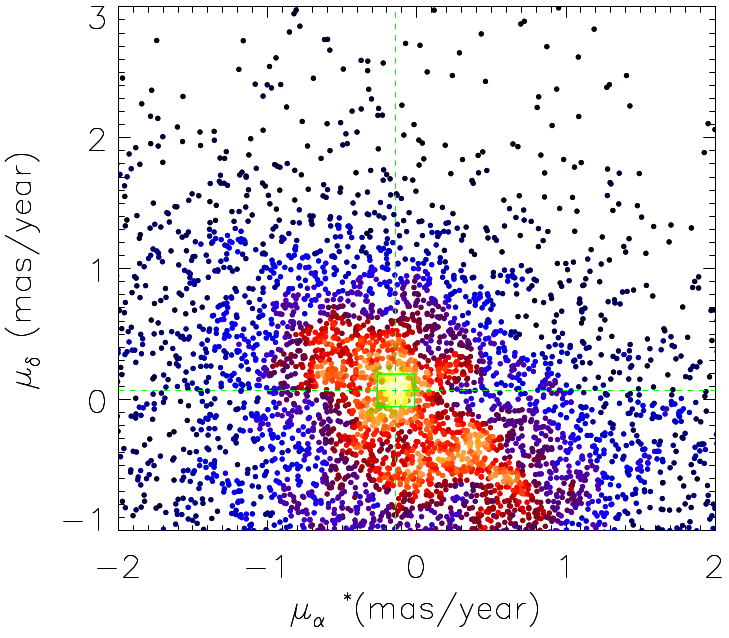}
\includegraphics[page=1,width=0.25\linewidth]{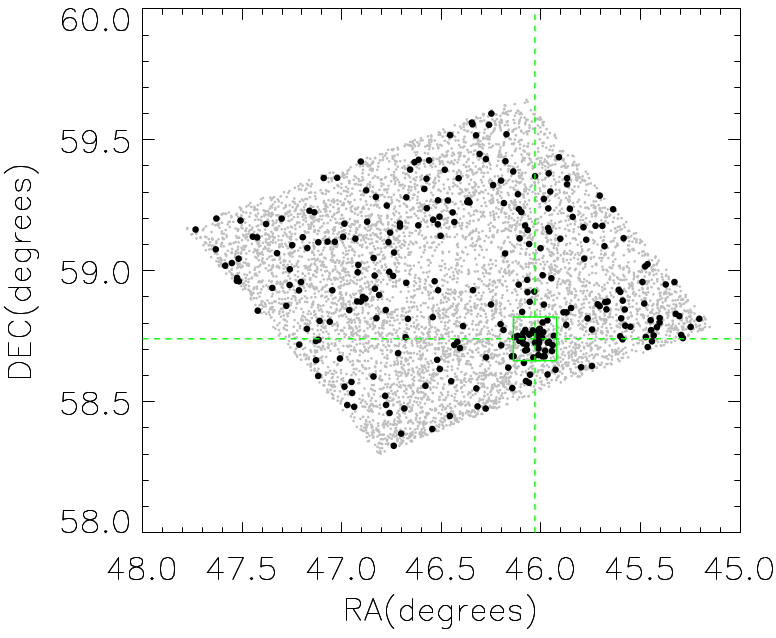}
\includegraphics[width=0.20\linewidth]{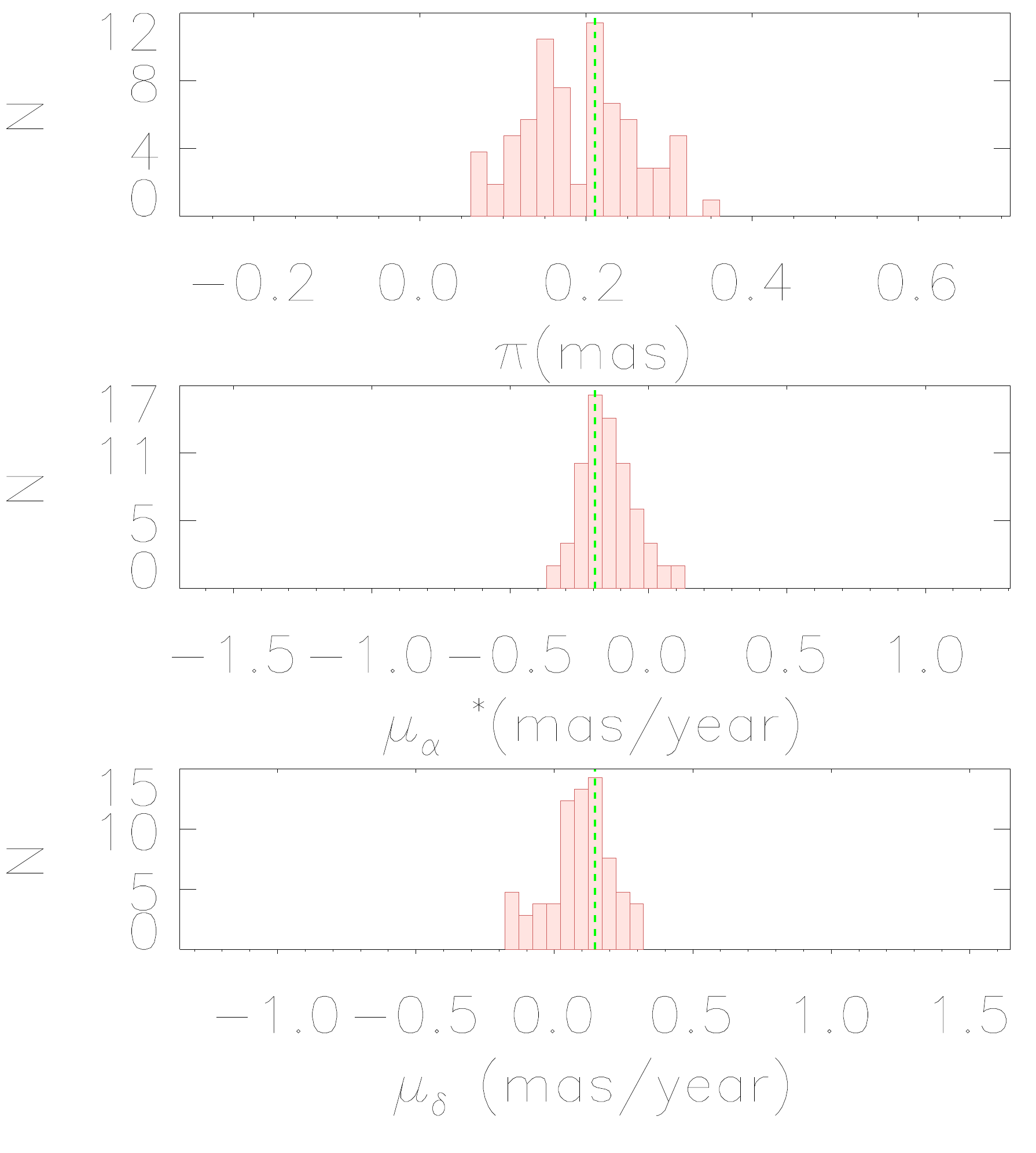}
\caption{Sequence of steps for OC detection applied to UFMG100 (top), UFMG113 (middle), and Berkeley 66 (bottom). First column: CMD of the entire tile (grey), overplotted with the colour–filtered samples (blue and red). Second column: Density maps of the sky chart (top) and VPDs (middle and bottom) for the colour–filtered subsample. Third column: VPD (top) and sky charts (middle and bottom) for stars filtered by colour and by the selection boxes from the previous step, shown as black filled circles. Fourth column: Histograms of parallaxes and proper motions in right ascension and declination for the final sample of stars filtered by colour, proper motion, and sky position. Modal values are shown by green dashed lines.}
\label{fig:pmra_detect}
\end{figure*}

The first line of panels in Fig \ref{fig:pmra_detect} illustrated the detection of OC candidates UFMG100. On the first panel, the colour-filtered sample is represented by the blue sample in the colour-magnitude diagram (CMD). This subsample was then analysed in skycharts, and a clustered structure was spatially trimmed by applying a box-shaped mask of size 10 arcmin, calculating the mean positions of right ascension and declination (middle panel). A VPD was then constructed after applying the colour and spatial filters and an overdensity found was extracted by applying a box-shaped mask of size 0.5\,mas\,yr$^{-1}$ (right panel).

The second line of panels in Fig \ref{fig:pmra_detect} shows the detection of the OC candidate UFMG113. The same procedure described for the previous example was applied: the sample was first restricted to stars bluer than the mean tile colour, highlighted in blue in the CMD (left panel). A VPD was then constructed with this subsample, and a proper-motion overdensity was isolated by applying a 0.5 mas,yr$^{-1}$ box-shaped mask (middle panel). Finally, the sample filtered by colour and proper motion was analysed on the skycharts, and a clump of clustered structure was spatially trimmed using a box-shaped mask of size 10 arcmin (right panel). 

The third line of panels in Fig \ref{fig:pmra_detect} shows how the known OC Berkeley 66 was detected. Differently from the previous examples, this cluster is highly reddened and therefore it is revealed by selecting stars redder than the mean tile colour, highlighted in the CMD (red sample, first panel). The procedure then follows the same steps as for UFMG100 and UFMG113.

 For all cases, in which a candidate is detected, we use the final sample filtered by colour, proper motion and skycharts to compute the mean star positions $\alpha$ and $\delta$ and the modal values of $\mu_{\alpha}^{*}$, $\mu_{\delta}$ and $\varpi$ (fourth column of panels, Fig \ref{fig:pmra_detect}).  These quantities are interactively refined during the subsequent analysis of confirmed candidates. For example, the clusters centres and radii are updated after construction of the radial density profiles (RDPs), and proper motion and parallaxes are updated after assessing membership (see Sects.~\ref{sect:4.2} and \ref{sect:membership}).

In total, we surveyed a projected area of $\sim$220 deg$^{2}$. All detected signatures are initially considered OC candidates, including repeated detections of the same object in adjacent tiles or spatial substructures of already-identified objects. To remove such duplicates, we performed an internal cross-match within the candidate list, resulting in a final set of 256 objects.

\begin{figure*}
\includegraphics[page=1,width=0.33\linewidth]{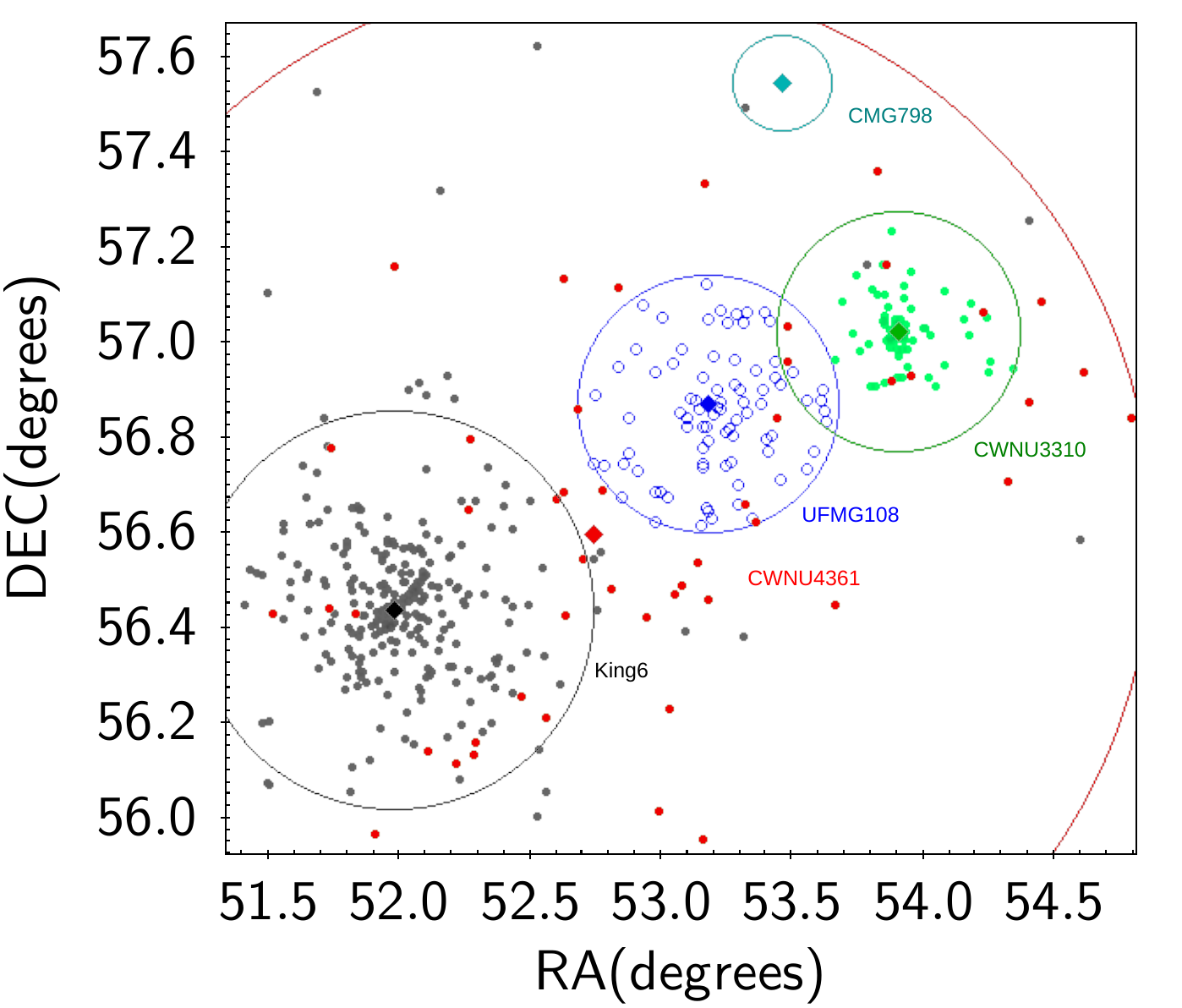}
\includegraphics[page=1,width=0.33\linewidth]{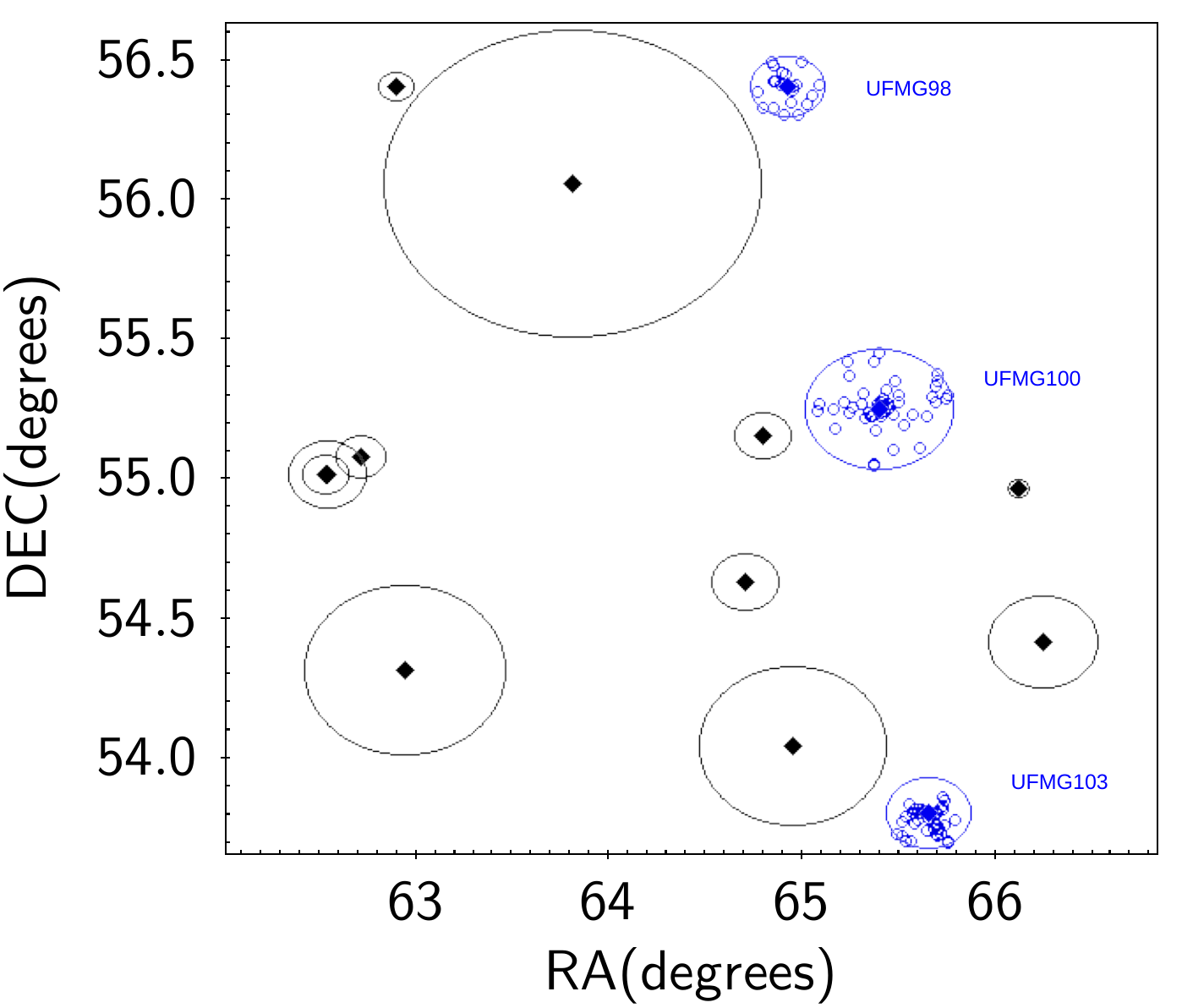}
\includegraphics[page=1,width=0.33\linewidth]{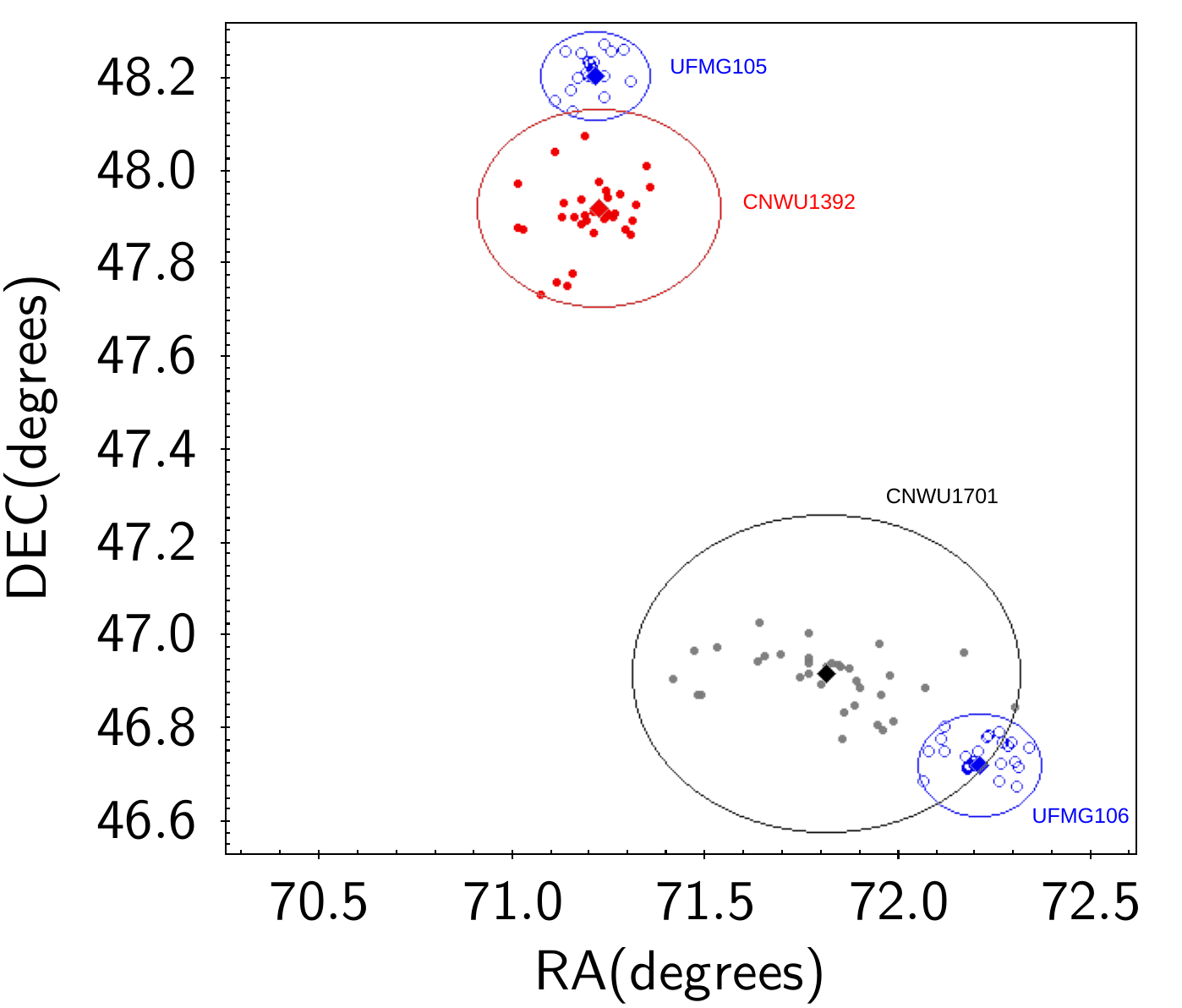}

\caption{Spatial distribution of the discovered OCs and nearby known OCs. I this figure, the limiting radii are represented by open large circles, the central coordinates by diamonds and OCs members by filled small circles (known OCs) and open blue small circles (new OCs). Left: the discovered OC UFMG108 with its neighbors CMG798 (cyan), CWNU3310 (green), CWNU4361 (red) and King6 (black). Middle: the discovered OCs UFMG98, UFMG100 and UFMG103 together with distant literature OCs (black). Right: the discovered OC UFMG105 and UFMG106 with their respective neighbors CNWU1392 (red) and  CNWU1701(black).}
\label{fig:cluster_lit_compare}
\end{figure*}

\section{Literature catalogues comparison and new candidates}
\label{sect:validation}

Efforts to establish a comprehensive and homogeneous star cluster census have been made in different works, including \citealp{dias2002,kharchenko2013,2019AJ....157...12B,2023MNRAS.526.4107P}; \citetalias{2023A&A...673A.114H}. To the best of our knowledge, the largest database of these objects so far was published by \cite{2023MNRAS.526.4107P} and \citetalias{2023A&A...673A.114H}.

In order to build an up-to-date reference catalogue, we have made efforts to compile the available information on OC censuses and memberlists. We included the databases used in our previous works (papers \citetalias{2019MNRAS.483.5508F}, \citetalias{10.1093/mnras/staa1684} and \citetalias{2021MNRAS.502L..90F}) and added the following available databases that had not been incorporated into this reference catalogue:
 
 \cite{2020AJ....160..279K}, \cite{2021RAA....21..117C}, \cite{2022ApJS..259...19L},  \cite{2022A&A...661A.118C},   \cite{2022A&A...660A...4H},    \cite{2022ApJS..260....8H}, \cite{2022arXiv220908504H}, \cite{2022ApJS..262....7H},  \cite{2023ApJS..265...20C}, \cite{2023MNRAS.526.4107P}, \citetalias{2023A&A...673A.114H}, \cite{2023RAA....23f5008C}, \cite{2023ApJS..266...36C}, \cite{2023ApJS..267...34H}, \cite{2024RAA....24e5014L}, \cite{Chi_2024},\citetalias{2025RMxAA..61....3D},\cite{10.1093/pasj/psaf082}  and \cite{Dias_2026}. 

We compared the 256 detected objects with the information available in the reference catalogue. The comparison procedure involved some steps, based on the information collected along the detection process of the candidates (Sect.~\ref{sect:search}): mean positions, mean astrometric parameters and partial memberlists inferred by filters based on proper motion, skycharts and colour. These informations were then compared with those of known OCs, such as mean astrometric parameters, radii, astrophysical parameters and memberlists.

Firstly, we crossmatched the central coordinates with those on the literature, adopting a search radius of 1 degree. For matched objects, where astrometry information is available, we compare proper motions and parallax differences and the centres separation. Objects for which these differences were within 3 $\sigma$ and the centres separation smaller than 5 arcmin are assumed as known OCs. Otherwise, we adopt this object as a candidate for further investigation.

\begin{figure}
\centering
\includegraphics[width=0.45\linewidth]{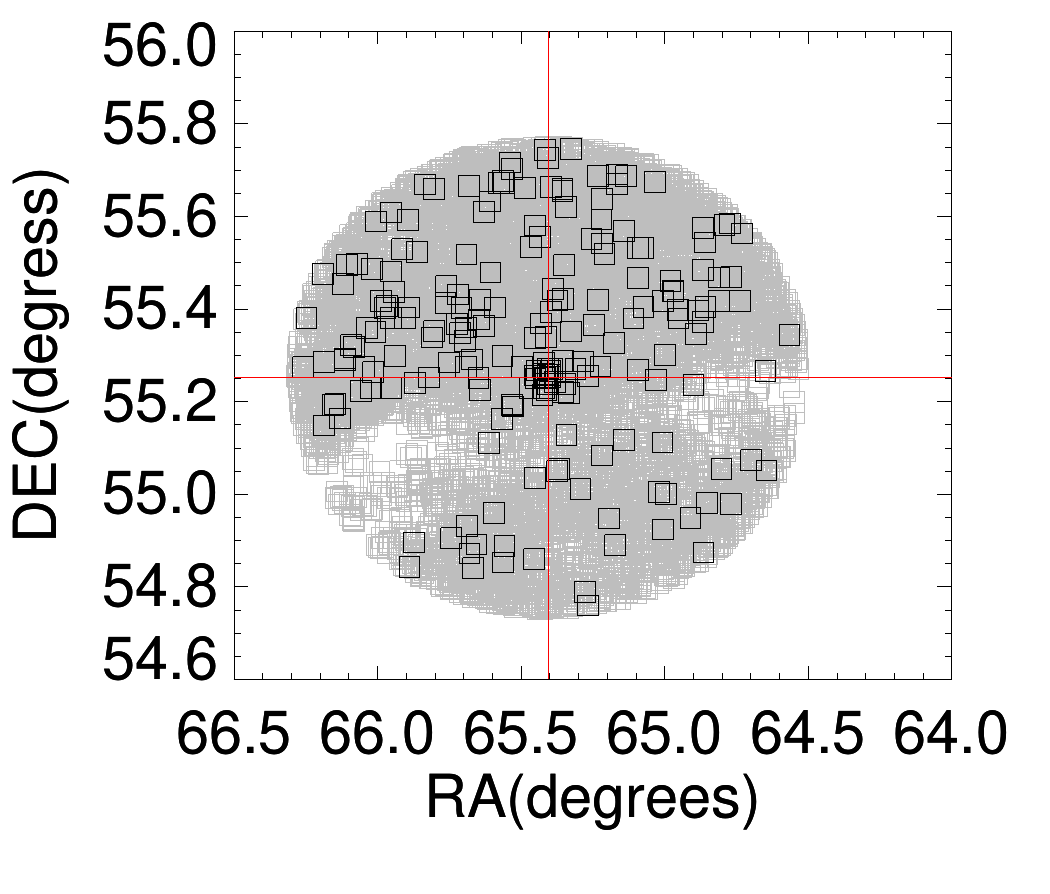}  \hspace{0.0cm}  
\includegraphics[width=0.45\linewidth]{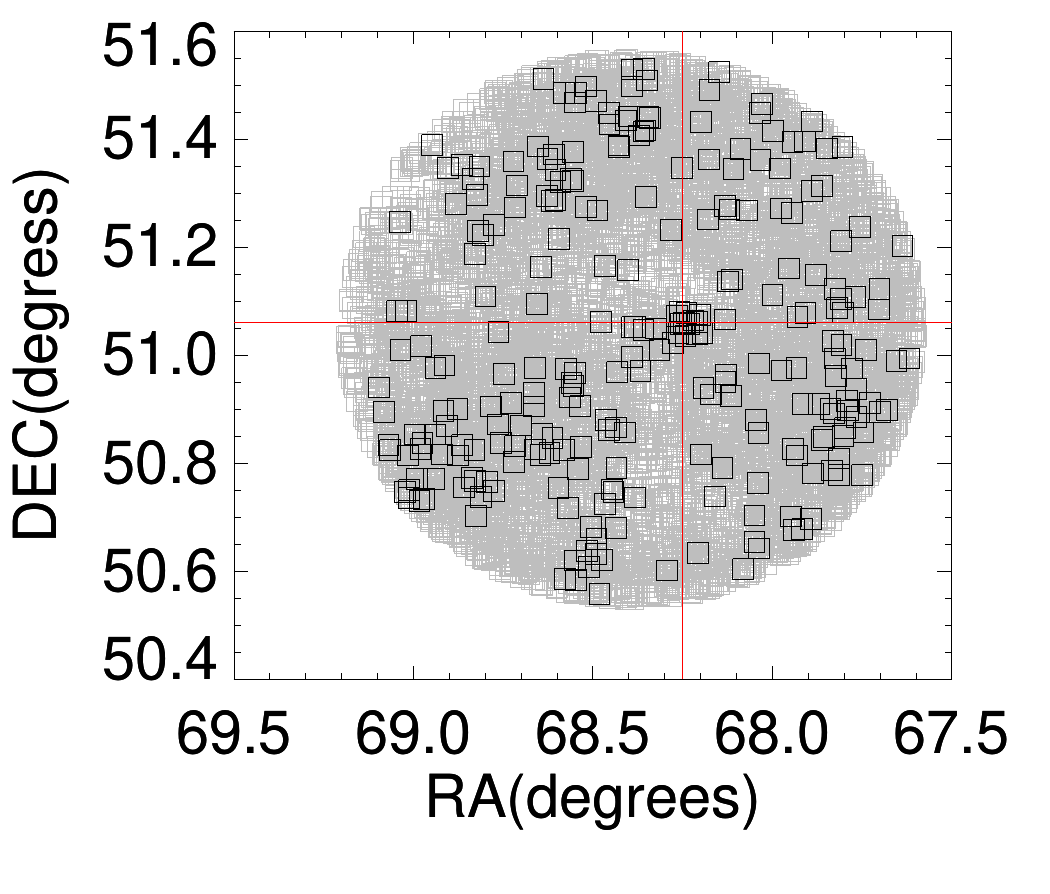}  \vspace{0.0cm}  
\includegraphics[width=0.45\linewidth]{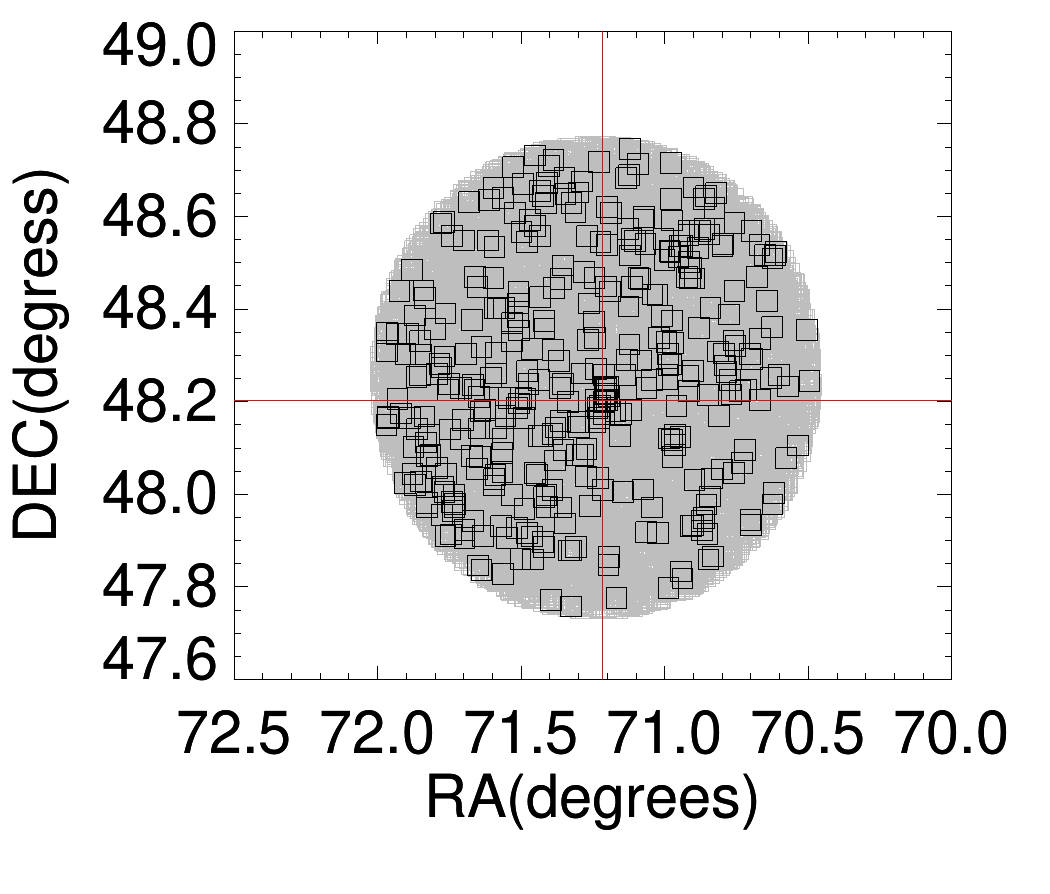}                  
\includegraphics[width=0.45\linewidth]{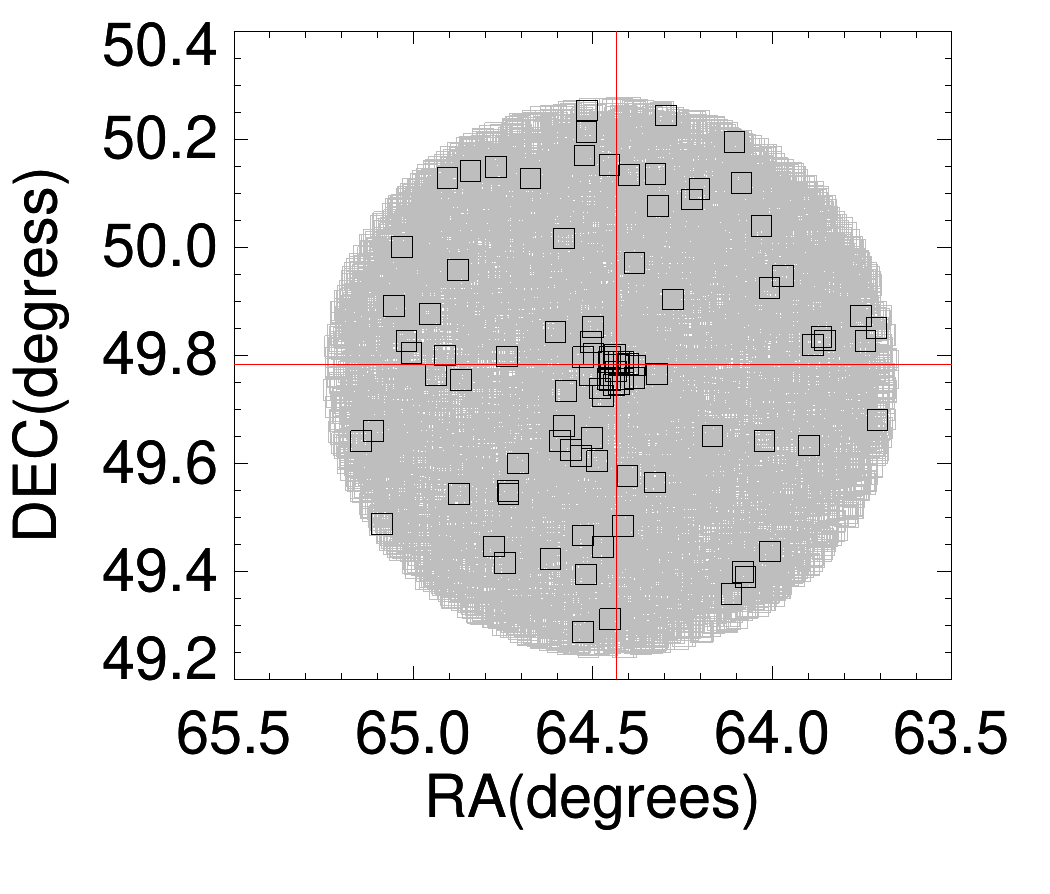}   \hspace{0.0cm}  
\caption{Centre determination for the discovered clusters: UFMG100 (top-left), UFMG104 (top-right), UFMG105 (bottom-left) and UFMG114 (bottom-right). The sample of stars filtered by proper motion e parallax boxes are represented by black opened squares. The entire field population are represented by the grey opened squares. The final centre position is ploted as two red solid lines. }
\label{fig:rdp_centre}
\end{figure}

For the candidates selected to be investigated, we established their astrometric, astrophysical and structural parameters (see Sect \ref{sect:analysis}). Once the clusters sizes were determined, we compared the OCs with their neighbours. Figure \ref{fig:cluster_lit_compare} shows examples of some candidates detected in this work and nearby literature OCs, showing skycharts with their radii and member stars overplotted. 

If the separation of the centres is greater than or equal the sum of our limiting radius and the comparison literature OC radius, we assume it as a new cluster (see middle panel of Fig.~\ref{fig:cluster_lit_compare}). Otherwise, we compare the other parameters, such as mean proper motions and parallax. Besides the astrometric comparison, we also crossmatched our cluster memberlist with the members available in the literature. As illustrated in the first and third panels of Fig.~\ref{fig:cluster_lit_compare}, some objects may show overlapping radii, however, taking into account their memberlists, we note that they correspond to distinct stellar systems.

Initially, we found 32 targets that were not reported in the literature. However, during the elaboration of this paper, \citetalias{2025RMxAA..61....3D} and \cite{Dias_2026} have reported the discovery of sets of 28 and 178 new OCs, respectively. Among these, \citetalias{2025RMxAA..61....3D} reported the cluster named Dias 17 that we identified as UFMG102. Both objects present consistent astrometric properties, exhibiting proper motions and parallaxes differences within 1 $\sigma$. The authors also found similar astrophysical parameters: $d=3137\pm 615$\,pc, $log(t)= 8.225 \pm 0.763$, $A_{V}=2.480 \pm 0.220$ and 31 members, while in this work we found $d=3890\pm 269$\,pc, $log(t)=  8.40 \pm 0.30$, $A_{V}=2.54 \pm 0.15$ and 34 members. However, \citetalias{2025RMxAA..61....3D} did not derive structural parameters for the coincident cluster, making this work the first one to provide them for this target.

\begin{figure*}
\centering
\includegraphics[width=0.32\linewidth]{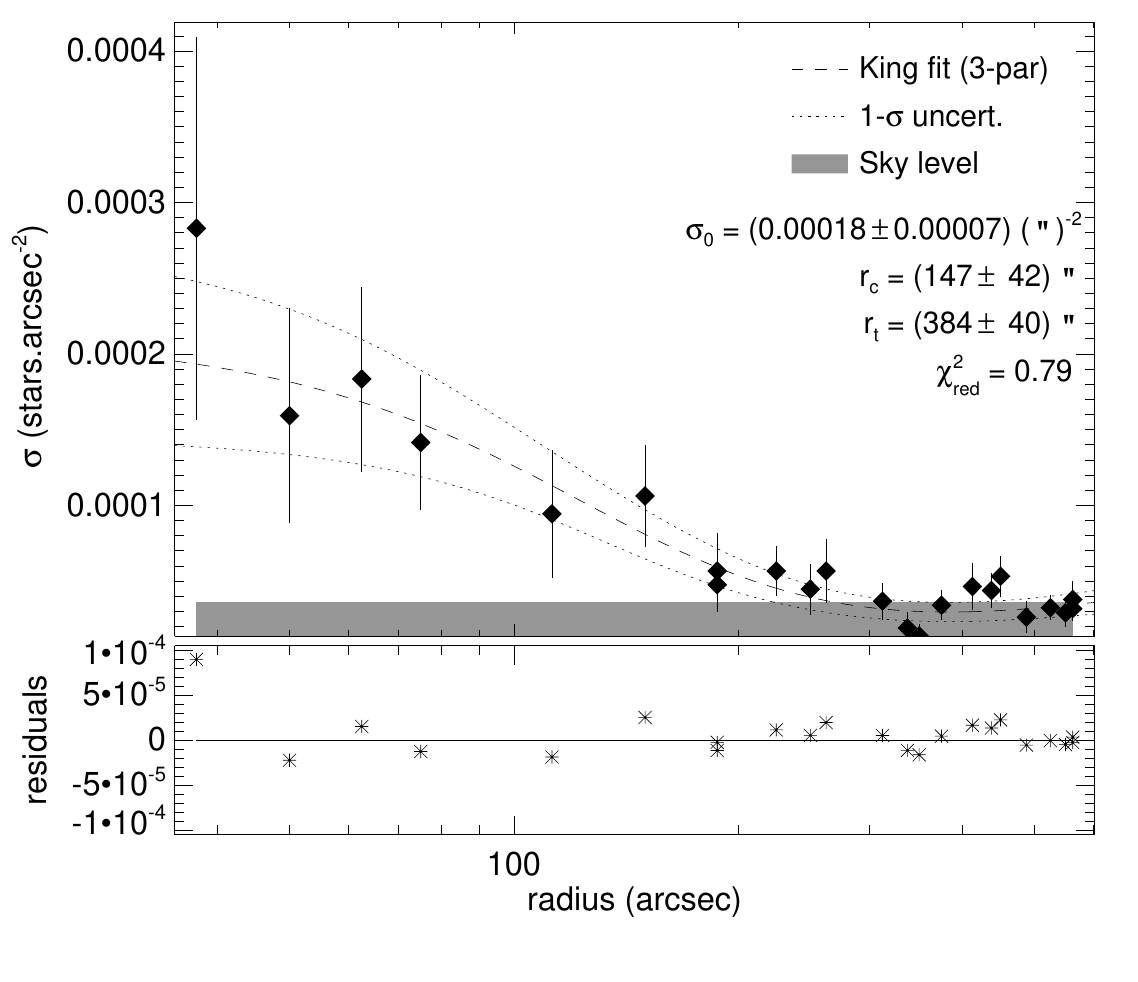} \hspace{0.0cm}  
\includegraphics[width=0.32\linewidth]{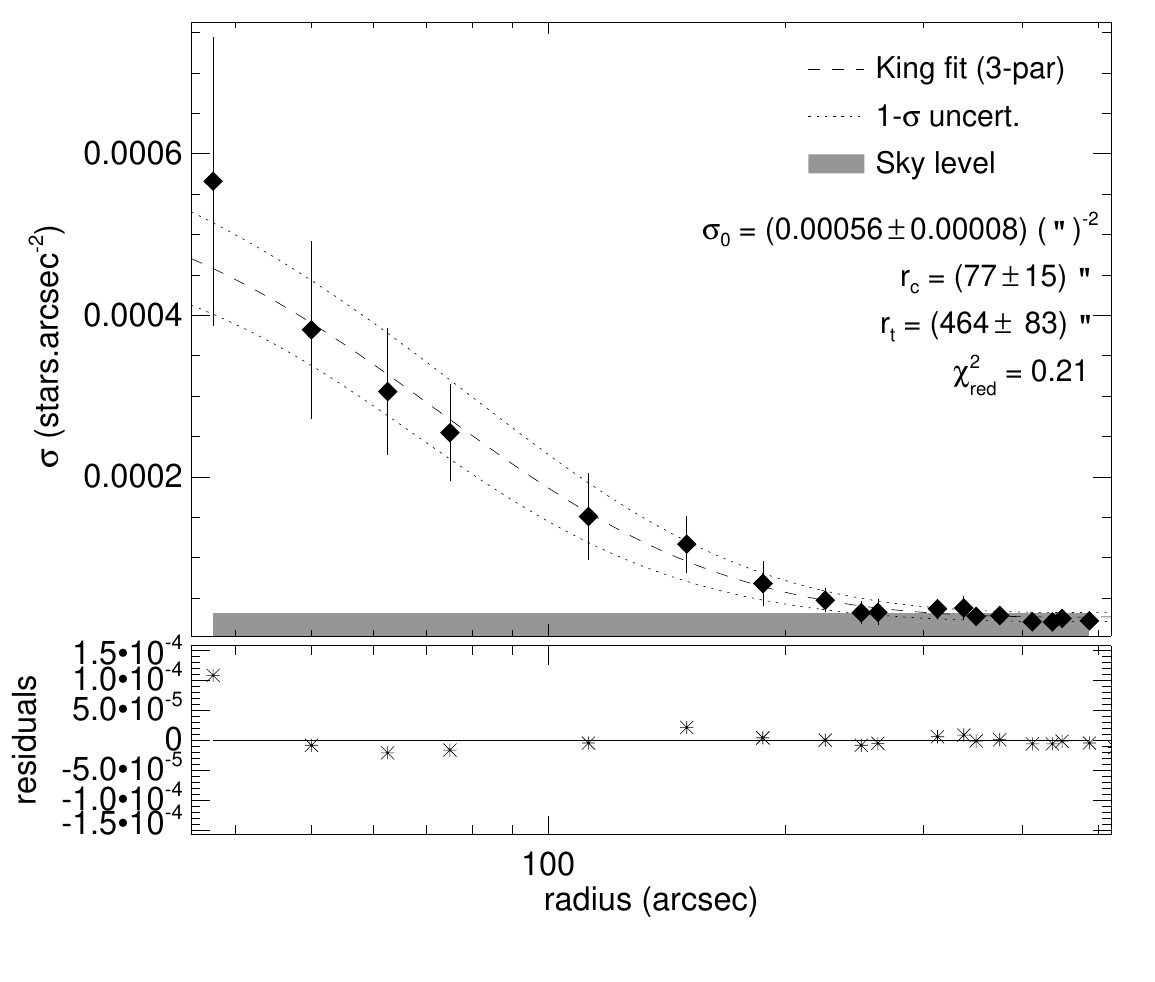}  \hspace{0.0cm}  
\includegraphics[width=0.32\linewidth]{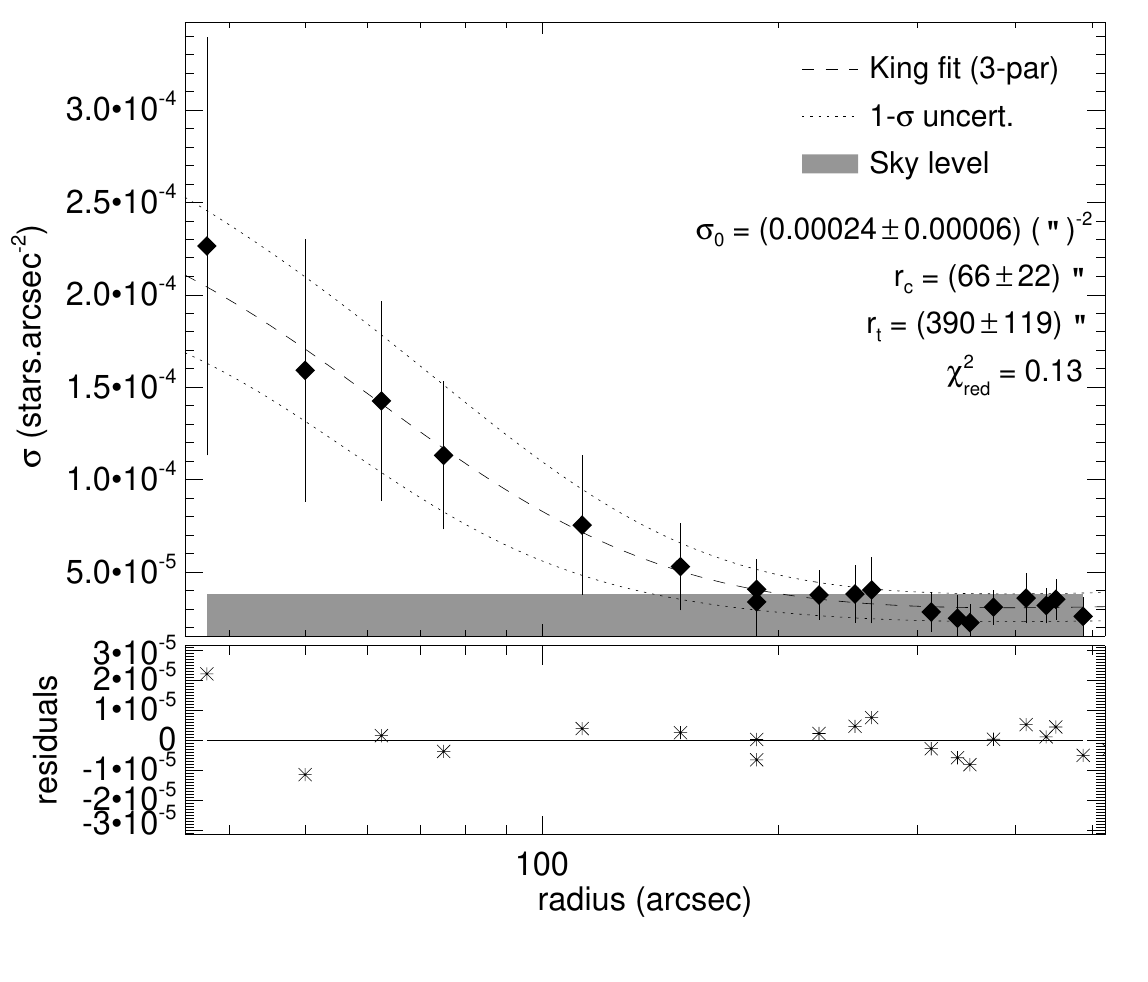}   \hspace{0.0cm}  
\includegraphics[width=0.32\linewidth]{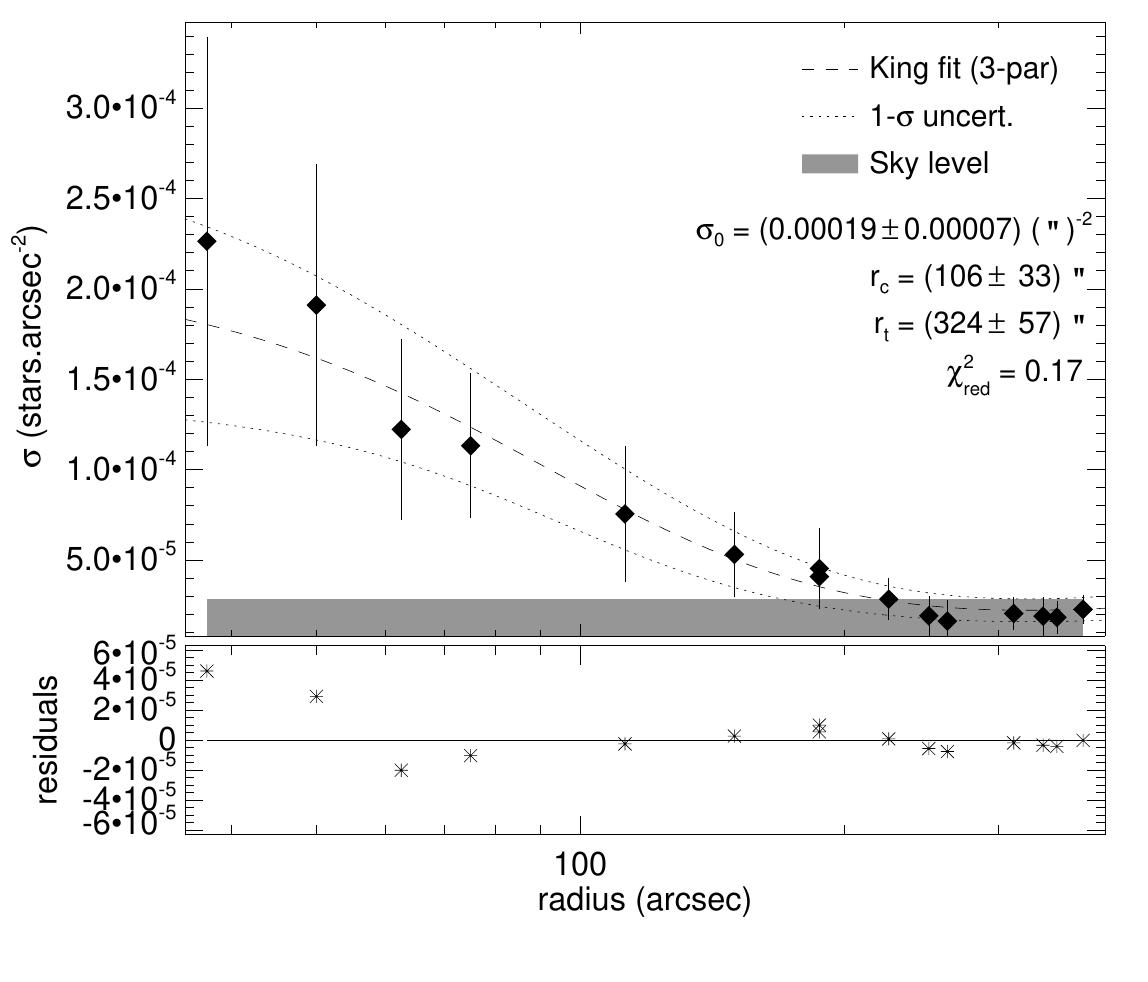}  \hspace{0.0cm}  
\includegraphics[width=0.32\linewidth]{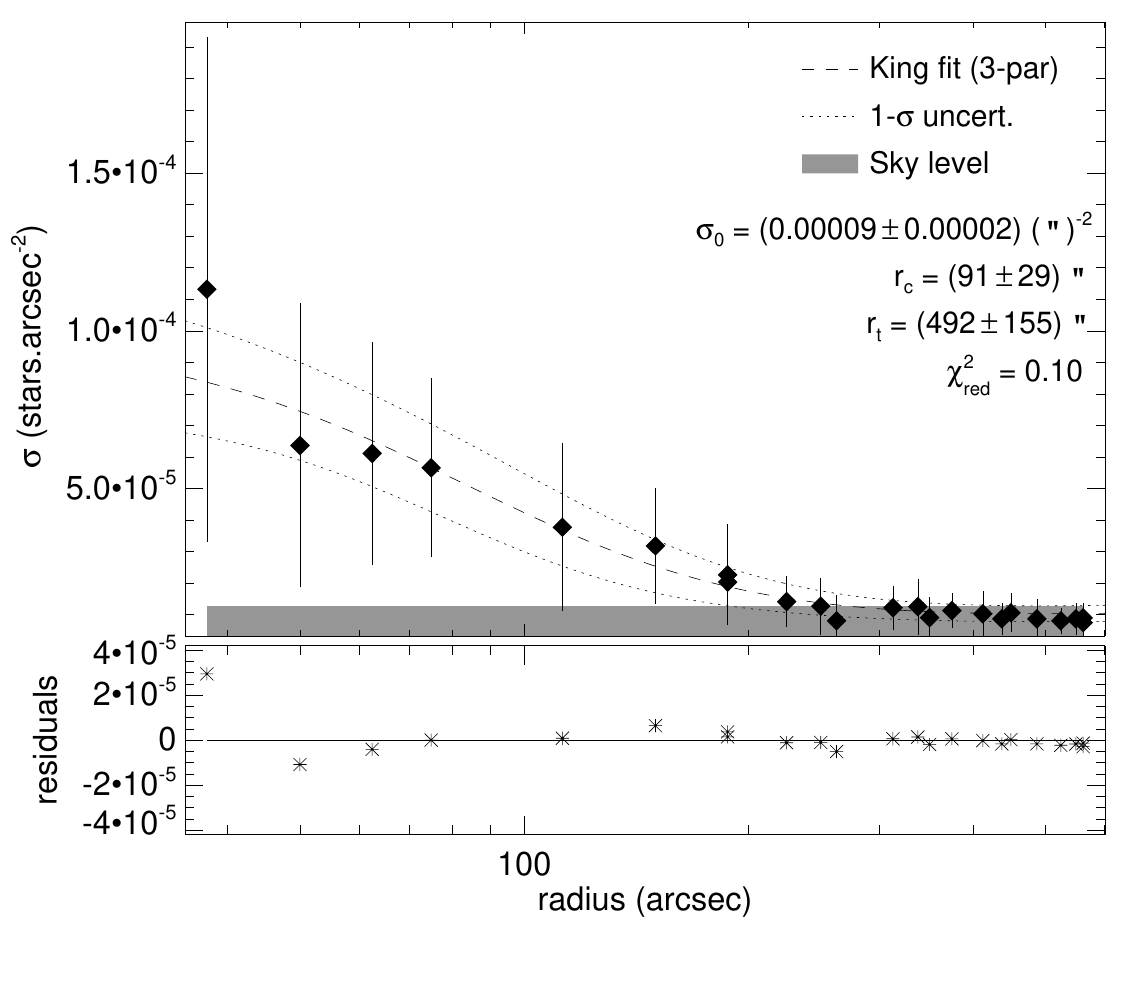}  \vspace{0.3cm}  
\includegraphics[width=0.32\linewidth]{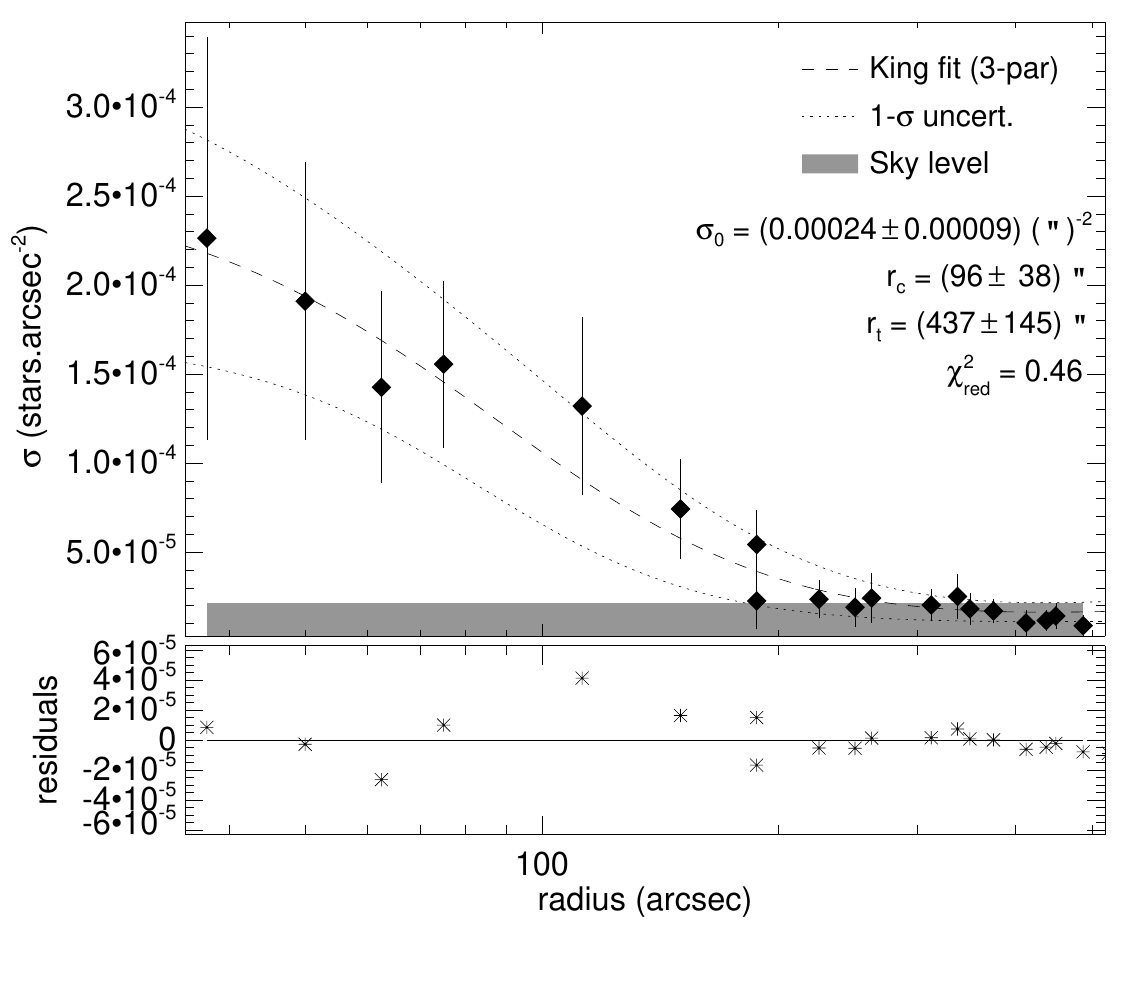}  \hspace{0.0cm}  
\includegraphics[width=0.32\linewidth]{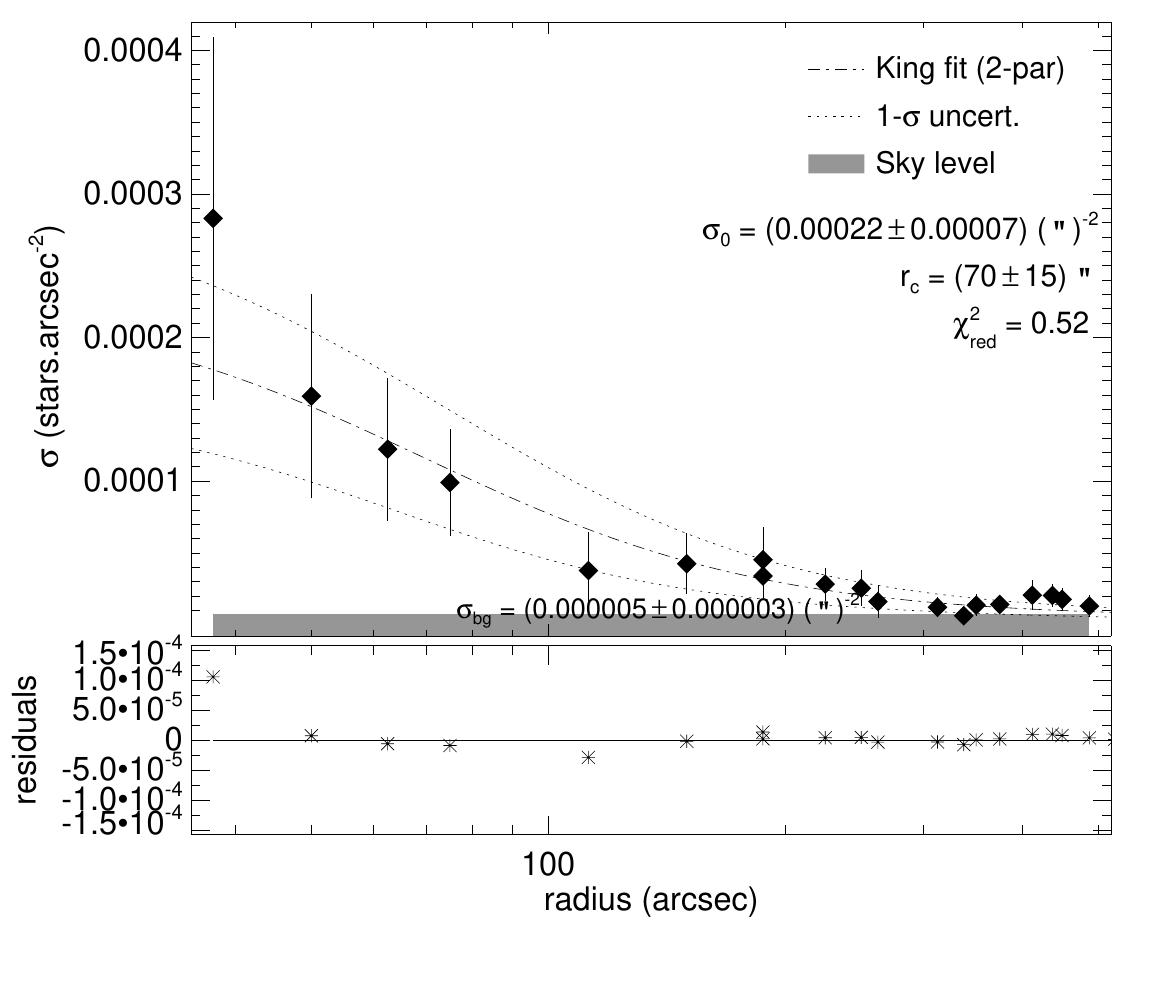}                  
\includegraphics[width=0.32\linewidth]{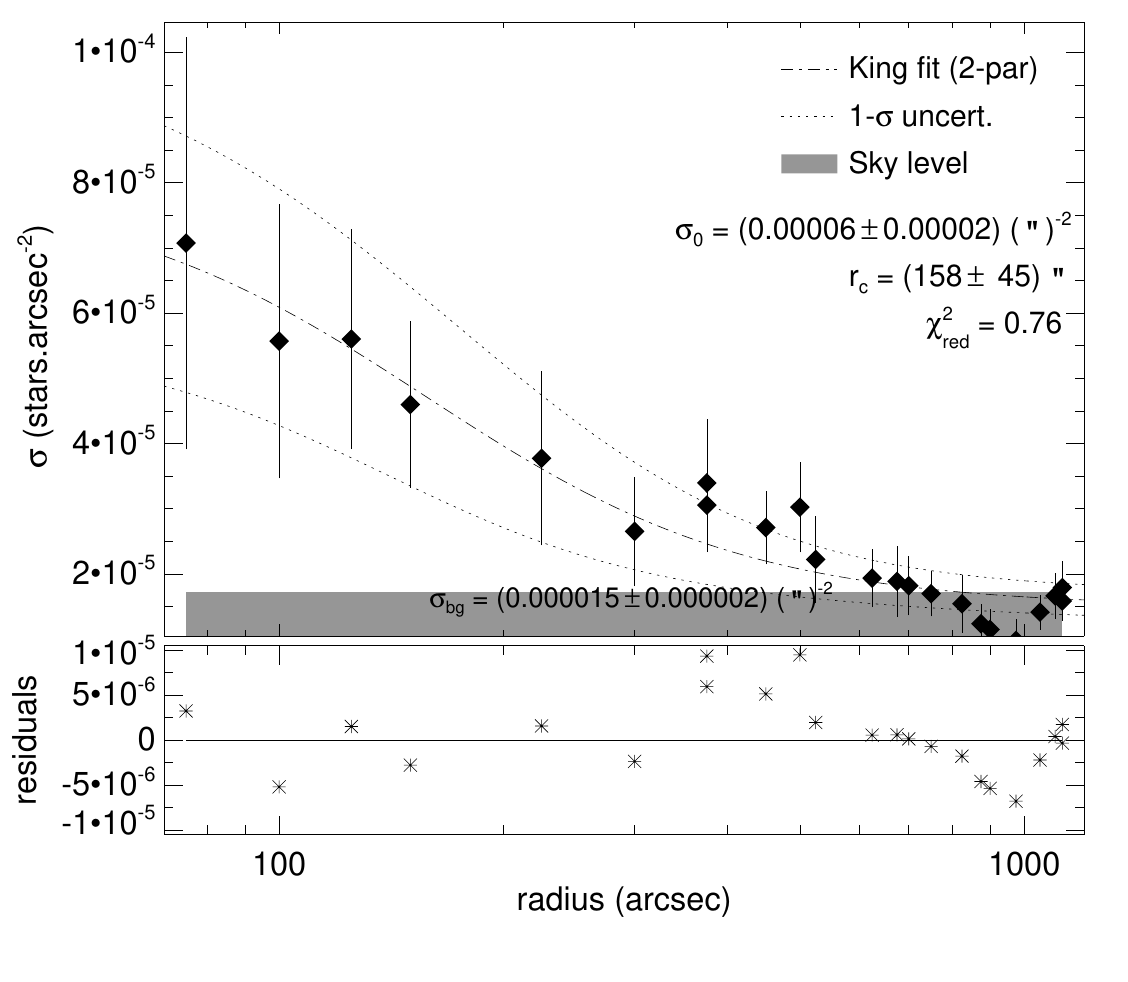}  \vspace{0.3cm}  
\includegraphics[width=0.32\linewidth]{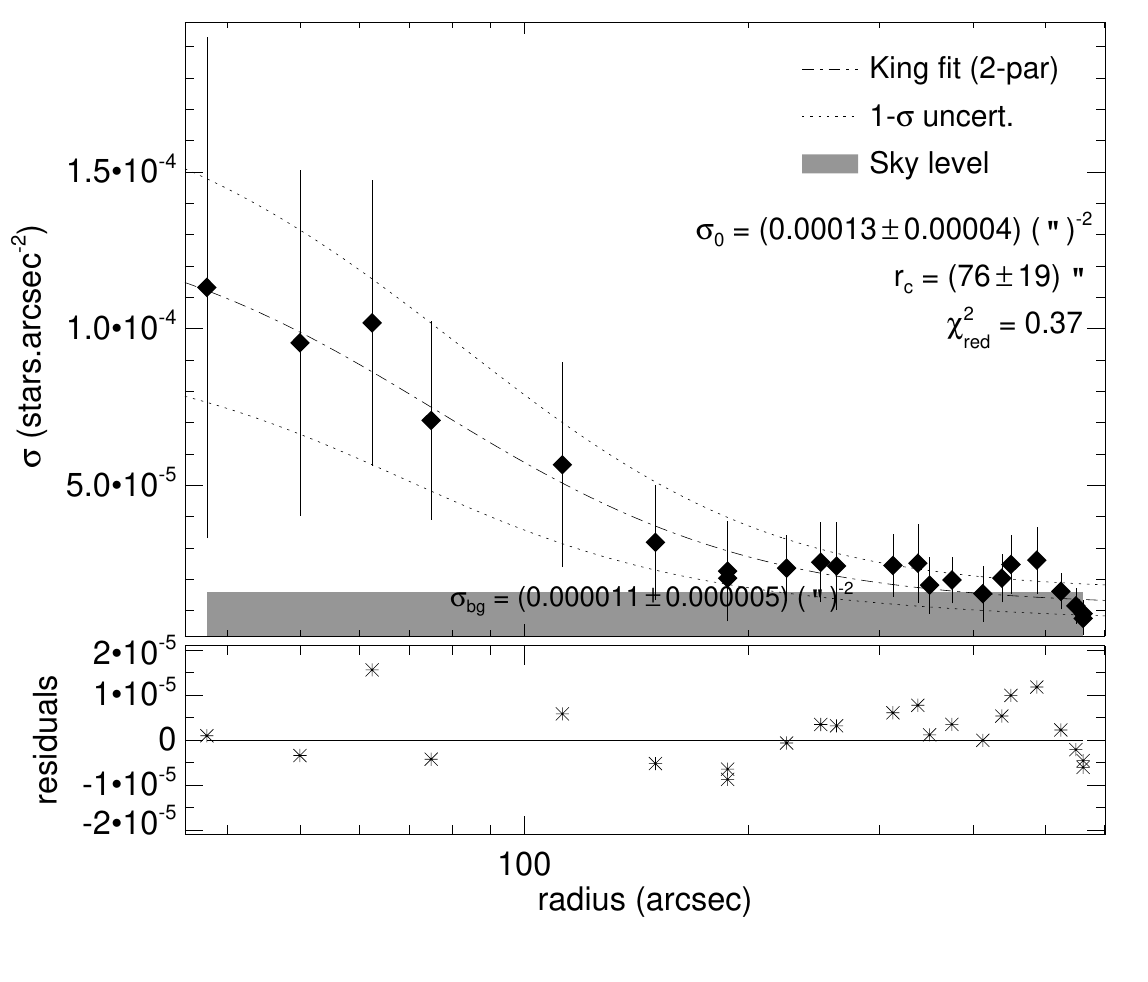}                  

\caption{RDPs and best-fitting King models (dashed line) adopted for some discovered OCs. The 1-$\sigma$ uncertainty envelope is also represented by dotted lines. The first and second lines of panels represent the OCs for which the 3-parameters King models achieved satisfactory convergence. For those OCs on the third line of panels, the 2-parameters King models was adopted. First line of panels: UFMG99, UFMG102 and UFMG106. Second line of panels: UFMG111, UFMG116 and UFMG124. Third line of panels: UFMG117, UFMG125 and UFMG128.  The background level and its fluctuation are indicated by the grey bar and the error bars correspond to poissonian noise. The fitting residuals are also displayed on the bottom of each profile.}
\label{fig:rdp_king}
\end{figure*}

By making this comparison, we assume they are identical objects and we give the discovery credit to \citetalias{2025RMxAA..61....3D}, reducing our discovered cluster sample from 32 to 31 objects. To the best of our knowledge, the remaining sample of 31 OCs are newly discovered objects and are not catalogued in the literature.

\section{ANALYSIS}
\label{sect:analysis}

\subsection{Centre and structural parameters}
\label{sect:4.2}

With the 5-astrometric parameters determined during the OCs detection process ($\alpha, \delta, \mu_{\alpha}^{*},\mu_{\delta}$ and $\varpi$; Sect \ref{sect:search}), we followed a procedure similar to that adopted in papers  \citetalias{2019MNRAS.483.5508F}, \citetalias{10.1093/mnras/staa1684} and \citetalias{2021MNRAS.502L..90F}, using these values as initial guesses for the subsequent analysis. The discovered OCs in this work present similar properties of those found in paper \citetalias{10.1093/mnras/staa1684} and \citetalias{2021MNRAS.502L..90F} for objects located at distances $d\gtrsim 2$ kpc: low contrast due to sparse structure and few members, with tidal radii typically smaller than 15 arcmin. For this reason, for each candidate, we selected stars within 30\arcmin\ from the adopted cluster centre in order to establish a sample of stars capable to encompass both the cluster region and a surrounding comparison field.

For each OC candidate, we established a subsample of stars based on their distribution in the astrometric space, restricting a sample of stars inside a box around the mean values of $\mu_{\alpha}^{*},\mu_{\delta}$ and $\varpi$, computed previously. This filter reduces the amount of field stars, increasing the contrast between the cluster population and the background, and allows the construction of the RDPs. Distant OCs ($d\gtrsim 2$ kpc) characterised in recent works using \textit{Gaia} DR3 astrometry typically present proper motion dispersions of the order of 0.1 mas\,yr$^{-1}$ and parallaxes around $\sim 0.05$ mas \citep{2023A&A...673A.114H,2025MNRAS.539..265F,2025MNRAS.539.2513A}. 

The objects analysed here exhibit mean parallaxes of order of $\varpi \lesssim 0.4 $ mas. Therefore, for such distant clusters, a 0.5\,mas\,yr$^{-1}$ box around the mean proper motion values is expected to encompass more than 2$\sigma$ of the distribution, assuming an average dispersion of 0.1 mas\,yr$^{-1}$, as found by \citep{2025MNRAS.539..265F}. Similarly, we adopted a parallax restriction width of 0.4 mas around the cluster mean values. In summary, the astrometric space used to isolate the cluster population is defined as a cube of dimensions ($L_{\mu_{\alpha}^{*}},L_{\mu_{\delta}}, L_{\varpi}$)=(0.5 mas\,yr$^{-1}$,0.5 mas\,yr$^{-1}$,0.4 mas).

The filtered subsample was then used to build RDPs, by counting stars within concentric rings with same thickness as function of the distance to the cluster centre (previously determined during the detection procedure, Sect \ref{sect:search}). In order to mitigate binning effects on the density distribution, we repeated this procedure using four different ring widths (75, 100, 125 and 150 arcseconds) and subsequently merged the resulting profiles into a single RDP. The stellar density of each ring was calculated by dividing the number of stars by the corresponding ring area. 

We also computed the the local background density using stars located far from the cluster centre, typically within the range $\sim 15$\arcmin\ to $\sim 25$\arcmin. During this process, we refined the cluster centre coordinates by making small adjustments to the initial values and selecting the solution that maximised the central stellar density as the new centre (see Fig.~\ref{fig:rdp_centre}).

In order to derive structural parameters central density ($\sigma_0$), core radius ($r_c$) and tidal radius ($r_t$), we performed weighted fittings of the \cite{king1962} analytical model to the clusters RDPs. We present examples of some fitted profiles  in Fig.~\ref{fig:rdp_king}, where the sky background level ($\sigma_{bkg}$) and its uncertainty are represented by the grey bar. The uncertainty in each RDP data point corresponds to its Poisson error, which was used as the weight in the fitting procedure.

Since our targets are projected against dense stellar fields, background contamination was a recurrent issue. For 8 OCs, the 3-parameter King model did not converge. In these cases, we adopted the 2-parameter version of the model. For these specific objects, in order to establish the clusters radii, we determined their limiting radii ($r_{lim}$), as the distance at which the density level reaches the mean value computed for the sky background. For UFMG122, although it clearly presents a significant spatial overdensity, both 2 and 3-parameters King models failed to converge due to background fluctuations. For this specific case, we report only its limiting radius. The Galactic coordinates $l$ and $b$; structural parameters $r_{c}$ and $r_{t}$ and mean astrometric parameters $\mu_{\alpha}^{*}$,  $\mu_{\delta}$ and $\varpi$ are listed in the Table \ref{Tab:clusters_prop1}. This table is also available electronically through Vizier\footnote{http://cdsarc.u-strasbg.fr/vizier/cat/J/MNRAS/{\bf vol/page}}.

\begin{table*}
\caption{Positions, structural parameters and mean astrometric parameters of the new clusters. }
\label{Tab:clusters_prop1}
\begin{tabular}{|l|r|r|r|r|r|r|r|r|r|r|r|r|}
\hline
  \multicolumn{1}{|c|}{name} &
  \multicolumn{1}{c|}{$l$} &
  \multicolumn{1}{c|}{$b$} &
  \multicolumn{1}{c|}{$r_{t}$} &
  \multicolumn{1}{c|}{$\delta_{r_{t}}$} &
  \multicolumn{1}{c|}{$r_{c}$} &
  \multicolumn{1}{c|}{$\delta_{r_{c}}$} &
  \multicolumn{1}{c|}{$\mu_{\alpha}^{*}$} &
  \multicolumn{1}{c|}{$\delta_{\mu_{\alpha}^{*}}$} &
  \multicolumn{1}{c|}{$\mu_{\delta}$} &
  \multicolumn{1}{c|}{$\delta_{\mu_{\delta}}$} &
  \multicolumn{1}{c|}{$\varpi$} &
  \multicolumn{1}{c|}{$\delta{\varpi}$} \\
\hline
  \multicolumn{1}{|c|}{} &
  \multicolumn{1}{c|}{degrees} &
  \multicolumn{1}{c|}{degrees} &
  \multicolumn{1}{c|}{pc} &
  \multicolumn{1}{c|}{pc} &
  \multicolumn{1}{c|}{pc} &
  \multicolumn{1}{c|}{pc} &
  \multicolumn{1}{c|}{mas/yr} &
  \multicolumn{1}{c|}{mas/yr} &
  \multicolumn{1}{c|}{mas/yr} &
  \multicolumn{1}{c|}{mas/yr} &
  \multicolumn{1}{c|}{mas} &
  \multicolumn{1}{c|}{mas} \\
\hline
  UFMG97 & 146.641 & 4.446 & 6.2 & 0.3 & 2.6 & 0.4 & -0.023 & 0.102 & 0.451 & 0.095 & 0.186 & 0.036\\
  UFMG98 & 148.895 & 4.419 & 9.6 & 3.4 & 2.0 & 0.8 & 0.234 & 0.16 & -0.065 & 0.114 & 0.239 & 0.044\\
  UFMG99 & 158.580 & 4.361 & 8.7 & 0.9 & 3.3 & 0.9 & 0.492 & 0.191 & -1.093 & 0.318 & 0.217 & 0.030\\
  UFMG100 & 149.898 & 3.795 & 13.7 & 4.3 & 0.7 & 0.2 & -0.115 & 0.243 & -0.460 & 0.147 & 0.280 & 0.049\\
  UFMG101 & 151.643 & 3.412 & 6.8 & 0.8 & 1.8 & 0.4 & 0.310 & 0.125 & -0.337 & 0.112 & 0.236 & 0.064\\
  UFMG102 (Dias17) & 151.665 & 2.989 & 8.8 & 1.6 & 1.5 & 0.3 & 0.073 & 0.210 & -0.999 & 0.167 & 0.271 & 0.042\\
  UFMG103 & 151.028 & 2.876 & 8.4 & 1.0 & 3.9 & 1.1 & 0.446 & 0.107 & -0.446 & 0.106 & 0.235 & 0.056\\
  UFMG104 & 154.106 & 2.118 & 4.4 & 0.4 & 1.0 & 0.2 & 0.365 & 0.175 & -0.976 & 0.136 & 0.305 & 0.031\\
  UFMG105 & 157.518 & 1.648 & 4.4 & 0.7 & 0.9 & 0.2 & 0.348 & 0.123 & -1.078 & 0.123 & 0.374 & 0.058\\
  UFMG106 & 159.083 & 1.199 & 5.6 & 1.7 & 1.0 & 0.3 & 0.289 & 0.195 & -1.186 & 0.190 & 0.359 & 0.051\\
  UFMG107 & 139.029 & 0.518 & 13.0 & 1.8 & 2.5 & 0.3 & -0.680 & 0.088 & -0.326 & 0.204 & 0.325 & 0.059\\
  UFMG108 & 143.644 & 0.630 & 22.2 & 2.9 & 7.6 & 1.3 & -0.443 & 0.124 & -0.197 & 0.161 & 0.310 & 0.083\\
  UFMG109 & 151.781 & 0.092 & 5.3 & 0.7 & 0.9 & 0.2 & 0.818 & 0.102 & 0.063 & 0.112 & 0.215 & 0.059\\
  UFMG110 & 151.707 & 0.532 & 6.7 & 0.8 & 1.1 & 0.3 & 0.846 & 0.038 & 0.139 & 0.182 & 0.224 & 0.064\\
  UFMG111 & 156.372 & 0.554 & 6.9 & 1.2 & 2.2 & 0.7 & 0.044 & 0.153 & -1.578 & 0.097 & 0.259 & 0.058\\
  UFMG112$^{a}$ & 147.309 & -0.140 & 3.7 & 0.9 & 0.7 & 0.2 & -0.315 & 0.107 & -0.495 & 0.077 & 0.275 & 0.085\\
  UFMG113 & 149.023 & -0.438 & 14.6 & 2.6 & 8.8 & 4.2 & -0.009 & 0.152 & -0.701 & 0.142 & 0.273 & 0.061\\
  UFMG114 & 153.320 & -0.512 & 7.0 & 0.4 & 2.0 & 0.2 & -0.478 & 0.106 & -0.134 & 0.082 & 0.346 & 0.054\\
  UFMG115$^{a}$ & 154.151 & -0.372 & 8.6 & 2.1 & 2.6 & 0.6 & -0.404 & 0.180 & -0.947 & 0.175 & 0.303 & 0.069\\
  UFMG116 & 146.328 & -1.639 & 9.7 & 3.1 & 1.8 & 0.6 & -0.351 & 0.227 & -0.624 & 0.144 & 0.274 & 0.122\\
  UFMG117$^{a}$ & 151.623 & -1.353 & 6.1 & 0.7 & 1.2 & 0.3 & -0.160 & 0.130 & -0.810 & 0.108 & 0.385 & 0.095\\
  UFMG118 & 154.075 & -1.444 & 9.6 & 2.6 & 1.7 & 0.4 & -0.483 & 0.121 & -0.602 & 0.089 & 0.330 & 0.073\\
  UFMG119 & 155.843 & -2.146 & 4.8 & 1.2 & 0.9 & 0.3 & -0.266 & 0.253 & -0.689 & 0.141 & 0.319 & 0.063\\
  UFMG120 & 142.384 & -2.143 & 7.5 & 1.3 & 2.7 & 0.6 & 0.380 & 0.091 & -0.935 & 0.084 & 0.361 & 0.105\\
  UFMG121 & 143.425 & -2.178 & 5.3 & 1.4 & 2.2 & 1.0 & 0.526 & 0.210 & -1.086 & 0.282 & 0.350 & 0.114\\
  UFMG122$^{a}$ & 158.425 & -2.896 & 12.0 & 0.8 & - & - & 0.253 & 0.148 & -1.321 & 0.157 & 0.373 & 0.077\\
  UFMG123$^{a}$ & 158.517 & -2.944 & 12.6 & 0.7 & 3.5 & 0.7 & 0.789 & 0.111 & -1.699 & 0.152 & 0.317 & 0.060\\
  UFMG124 & 139.109 & -3.222 & 4.9 & 1.6 & 1.1 & 0.4 & 0.665 & 0.069 & -1.446 & 0.105 & 0.392 & 0.034\\
  UFMG125$^{a}$ & 147.818 & -3.098 & 11.7 & 1.3 & 2.8 & 0.8 & -0.237 & 0.077 & -0.517 & 0.057 & 0.266 & 0.065\\
  UFMG126$^{a}$ & 149.605 & -3.720 & 10.3 & 0.8 & 2.2 & 0.8 & 0.603 & 0.055 & -1.274 & 0.133 & 0.359 & 0.054\\
  UFMG127 & 155.565 & -3.175 & 11.0 & 2.7 & 3.1 & 0.8 & 1.033 & 0.124 & -1.418 & 0.168 & 0.348 & 0.070\\
  UFMG128$^{a}$ & 141.491 & -4.849 & 9.9 & 0.7 & 1.3 & 0.3 & -0.512 & 0.201 & -0.308 & 0.156 & 0.285 & 0.095\\
\hline\end{tabular}
\vspace{2mm}
\parbox{0.95\textwidth}{\footnotesize
$^{a}$ OCs for which $r_{t}$ was replaced by their $r_{lim}$.}
\end{table*}

\subsection{Assessing membership and isochrone fittings}
\label{sect:membership}

In order to derive membership likelihoods to the clusters stars, we applied a routine described in \cite{2019A&A...624A...8A}, which statistically evaluates the overdensity of stars in the cluster region relative to a nearby  comparison field in the 3D astrometric space ($\mu_{\alpha}^{*}$, $\mu_{\delta}$, $\varpi$). The sky region limited from the cluster's centre to its $r_{t}$ (or, in few cases, its limiting radius, see Tab ~\ref{Tab:clusters_prop1}) was defined as the cluster region. The comparison field was taken as a ring-like region with an inner radius  $r_{t}$+3 arcmin.
 
 Since a large number of foreground and background stars are projected towards the cluster region, the astrometric space typically contains a broad distribution of stars with values far from the cluster locus. For this reason, in order to improve the performance of the decontamination algorithm and reduce the presence of such outliers, we defined the astrometric space by applying a box-like filter on it, centred on the peak values of $\mu_{\alpha}^{*}$, $\mu_{\delta}$ and $\varpi$, computed in the Sect \ref{sect:tile_detect}. We restricted the astrometric space, for both cluster and field regions, within a box with width of 1.5\,mas\,yr$^{-1}$ for $\mu_{\alpha}^{*}$ and $\mu_{\delta}$ and 0.5 mas for $\varpi$. These limits are significantly broader than the clusters members expected dispersion and also encompass enough field stars for comparison. This step helps to mitigate possible density fluctuations of field stars with non-zero memberships likelihoods outside of the cluster astrometric space.
 
Both cluster and field 3D astrometric space were then divided into cells with widths of 0.25\,mas\,yr$^{-1}$, 0.25\,mas\,yr$^{-1}$ and 0.15\,mas in $\mu_{\alpha}^{*}$, $\mu_{\delta}$ and $\varpi$, respectively. These dimensions are large enough to accommodate a statistically significant number of stars in each cell, while still being small enough to detect local variations in the stellar density. To attribute the final membership likelihoods to the cluster stars, we adopted the same procedure used in papers \citetalias{10.1093/mnras/staa1684} and \citetalias{2021MNRAS.502L..90F}.

\begin{figure*}
\centering
\includegraphics[width=0.49\linewidth]{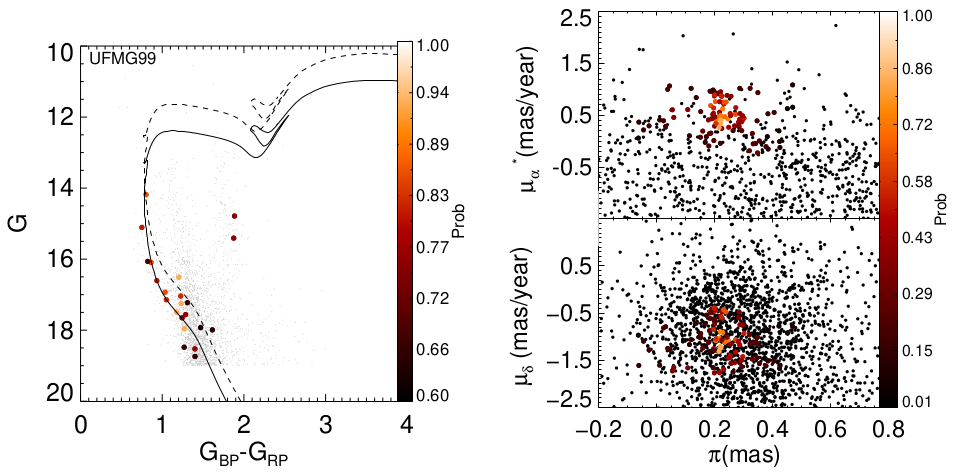 } \hspace{0.0cm}  
\includegraphics[width=0.49\linewidth]{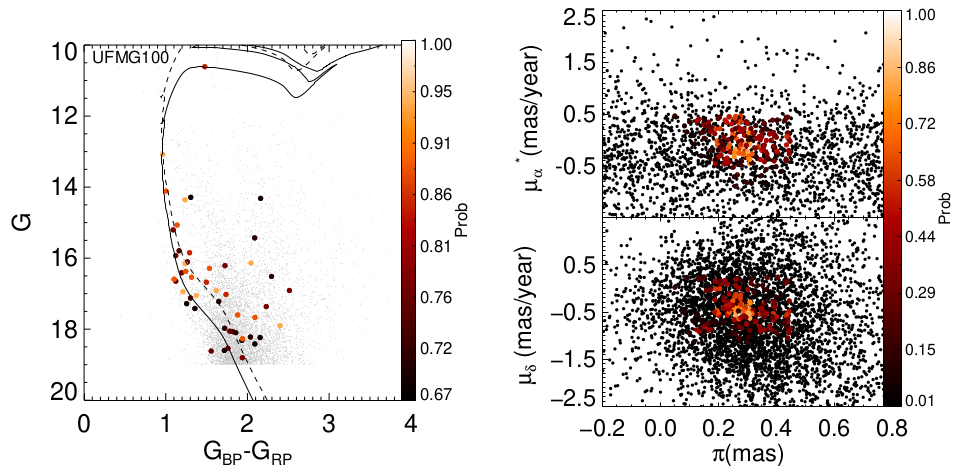 } \hspace{0.0cm}  
\includegraphics[width=0.49\linewidth]{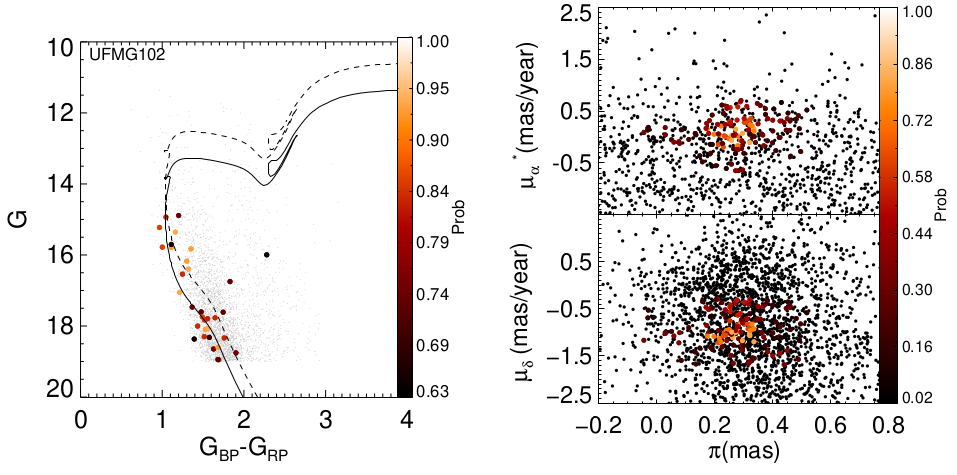 } \hspace{0.0cm}  
\includegraphics[width=0.49\linewidth]{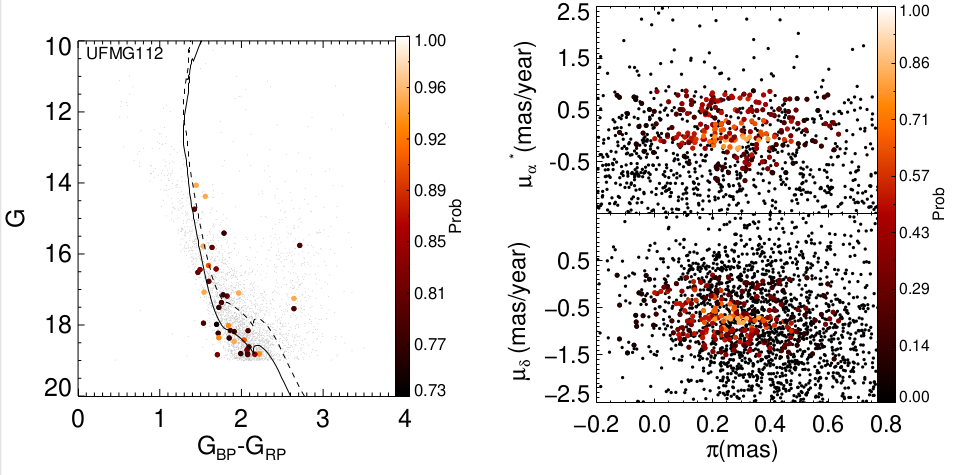 } \hspace{0.0cm}  
\includegraphics[width=0.49\linewidth]{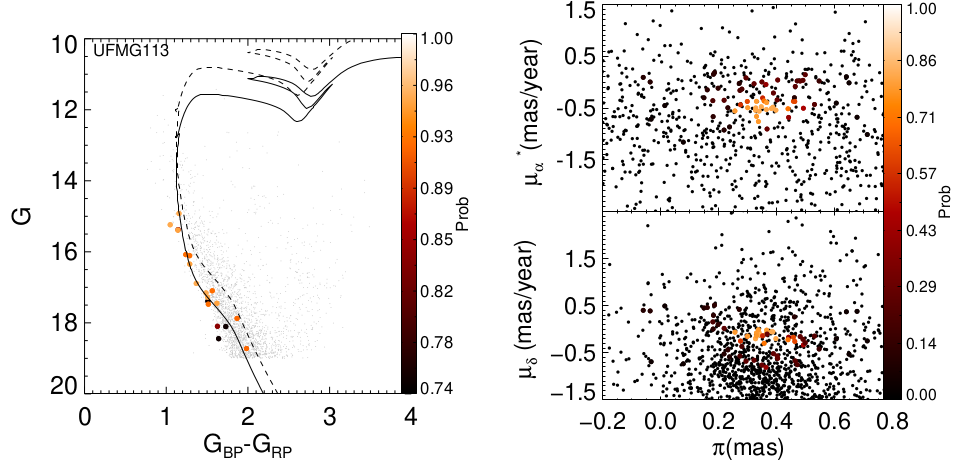 } \hspace{0.0cm}  
\includegraphics[width=0.49\linewidth]{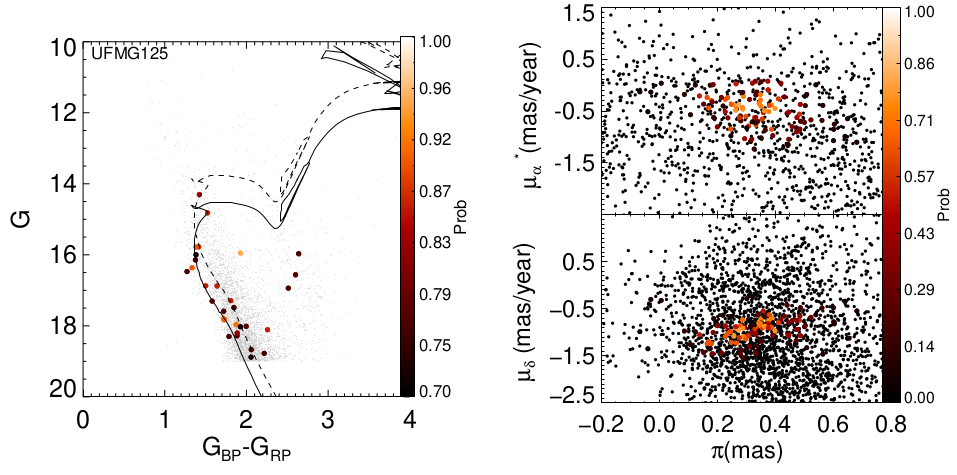 } \hspace{0.0cm}  
\includegraphics[width=0.49\linewidth]{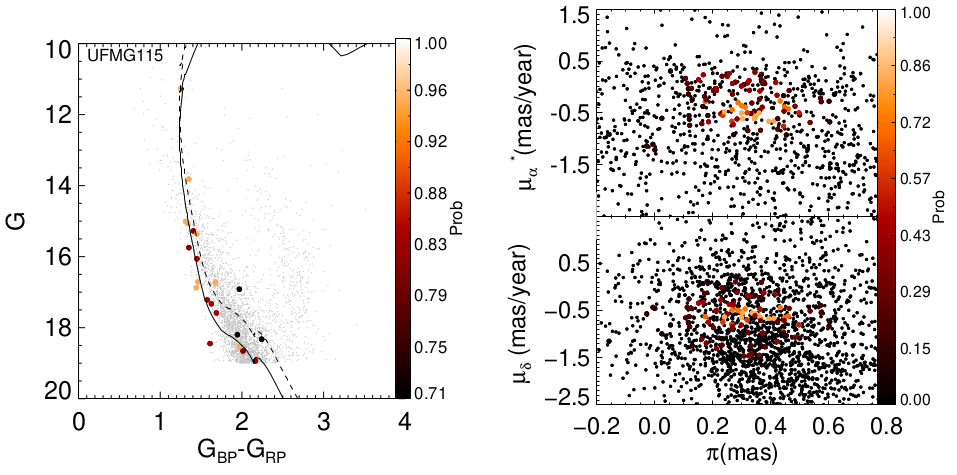 } \hspace{0.0cm}  
\includegraphics[width=0.49\linewidth]{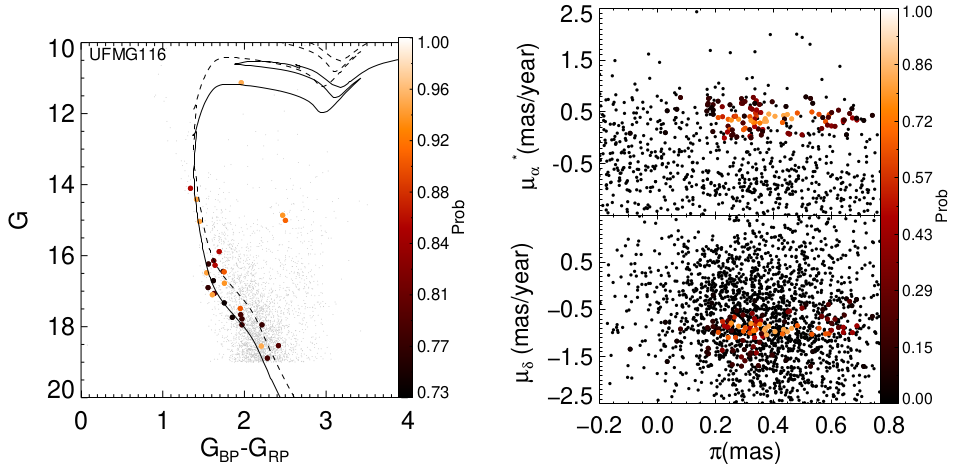 } \hspace{0.0cm}  
\includegraphics[width=0.49\linewidth]{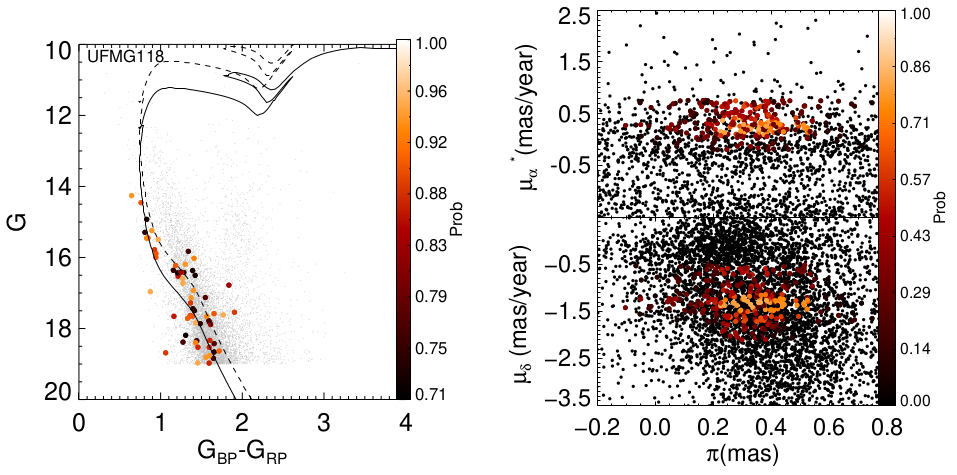 } \hspace{0.0cm}  
\includegraphics[width=0.49\linewidth]{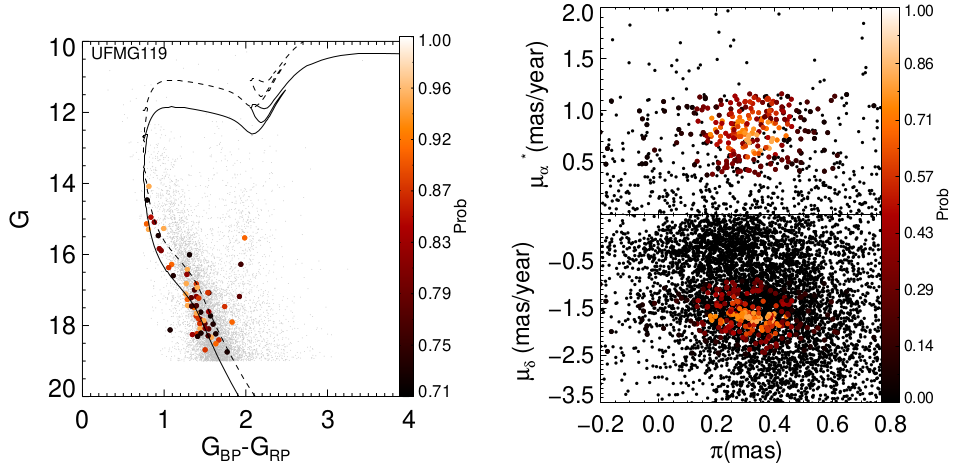 } \hspace{0.0cm}  

\caption{Decontaminated CMDs and proper motion \textit{versus} parallax diagrams for 10 example OCs. On both diagrams, clusters stars are represented by coloured filled circles, which membership probabilities are represented on the colour bar. On the CMDs (left diagrams) a probability cut of $60\%$ is applied, where PARSEC-COLIBRI isochrone  fitting (solid line) over the cleaned CMDs are performed. The corresponding binary sequence (dashed lines) is also established by deducing 0.75 mag from the G magnitude values. The field population is represented by the grey sample. On the right of the CMDs, diagrams of proper motion in right ascension \textit{versus} parallax (top) and proper motion in declination \textit{versus} parallax (bottom) are presented.}
\label{fig:isocrones_vpds}
\end{figure*}

To establish the final cluster members sample, we adopted a membership likelihood cut by keeping only stars with membership probabilities higher than 60$\%$. Fig. \ref{fig:isocrones_vpds} shows examples of decontaminated CMDs and proper motion \textit{versus} parallax diagrams after the decontamination procedure. The high-probability member candidates clearly delineate the cluster locus in the CMDs and form a compact clump in the 3D astrometric space ($\mu_{\alpha}^{*}$, $\mu_{\delta}$ and $\varpi$), revealing the cluster nature. From the memberlists, we also calculated the number of OC members $N$, $r_{50}$ (radius containing $50 \%$ of the members) and the mean values of proper motions in right ascension, declination ($\mu_{\alpha}^{*}$ and $\mu_{\delta}$), parallaxes ($\varpi$) and their dispersions (see Tab. \ref{Tab:clusters_prop2}).

We performed isochrone fittings on the decontaminated samples (i.e., stars with membership likelihoods $> 60\%$) to derive ages, distances and reddening, employing solar-metallicity PARSEC isochrones \citep{Marigo:2017}. We converted the \textit{Gaia} colour excess $E(G_{BP}-G_{RP})$ to $E(B-V)$ by adopting a reddening law according to \cite{Cardelli:1989} and \cite{Odonnell:1994}. 

A set of several isochrones spanning a range of ages was visually compared to the observed cluster sequences in the CMDs, ensuring a consistent fit across key evolutionary features such as the main sequence, the turnoff, and the red giant branch. Uncertainties were estimated by varying the reddening, distance modulus $(m-M)$ and the isochrone age simultaneously searching for the maximum acceptable departure from the best-fitting solution. Fig.~\ref{fig:isocrones_vpds} shows examples of the procedure applied to some of the clusters.

The clusters centres ($\alpha$ and $\delta$), their astrophysical parameters ($\log[t({\rm yr})]$, $d$, $E(B-V)$) , the number of members and $r_{50}$ are listed in the Table \ref{Tab:clusters_prop2}. A more complete version of this table, including additional cluster parameters and the full memberlists, is also available as electronic tables through Vizier\footnote{http://cdsarc.u-strasbg.fr/vizier/cat/J/MNRAS/{\bf vol/page}}.

\begin{table*}
\caption{Positions, radius ($R_{50}$), astrophysical parameters and number of members of the investigated clusters. For each OC, a flag is attributed: OC or OCc (open cluster candidate).}
\label{Tab:clusters_prop2}
\begin{tabular}{|l|r|r|r|r|r|r|r|r|r|r|r|}
\hline
  \multicolumn{1}{|c|}{name} &
  \multicolumn{1}{c|}{$\alpha_{J2000}$} &
  \multicolumn{1}{c|}{$\delta_{J2000}$} &
  \multicolumn{1}{c|}{$R_{50}$} &
  \multicolumn{1}{c|}{$D$} &
  \multicolumn{1}{c|}{$\delta{D}$} &
  \multicolumn{1}{c|}{$\log[t({\rm yr})]$} &
  \multicolumn{1}{c|}{$\delta_{\log[t({\rm yr})]}$} &
  \multicolumn{1}{c|}{$E(B-V)$} &
  \multicolumn{1}{c|}{$\delta_{E(B-V)}$} &
  \multicolumn{1}{c|}{$N_{mem}$} &
  \multicolumn{1}{c|}{$Flag$}  \\
\hline
  \multicolumn{1}{|c|}{} &
  \multicolumn{1}{c|}{degrees} &
  \multicolumn{1}{c|}{degrees} &
  \multicolumn{1}{c|}{degrees} &
  \multicolumn{1}{c|}{pc} &
  \multicolumn{1}{c|}{pc} &
  \multicolumn{1}{c|}{dex} &
  \multicolumn{1}{c|}{dex} &
  \multicolumn{1}{c|}{mag} &
  \multicolumn{1}{c|}{mag} &
  \multicolumn{1}{c|}{}&
  \multicolumn{1}{c|}{} \\
\hline
  UFMG97 & 61.960 & 57.970 & 0.037 & 4365 & 302 & 9.00 & 0.10 & 0.55 & 0.05 & 14&  OC\\
  UFMG98 & 64.930 & 56.400 & 0.070 & 5129 & 473 & 8.40 & 0.15 & 0.70 & 0.05 & 21&  OCc \\
  UFMG99 & 75.415 & 49.095 & 0.069 & 4677 & 431 & 8.15 & 0.10 & 0.68 & 0.03 & 22&  OC\\
  UFMG100 & 65.405 & 55.252 & 0.086 & 3631 & 335 & 7.70 & 0.15 & 0.88 & 0.03 & 50& OC\\
  UFMG101 & 67.030 & 53.738 & 0.056 & 3467 & 160 & 8.70 & 0.10 & 0.73 & 0.05 & 22&  OC\\
  UFMG102 (Dias17) & 66.541 & 53.429 & 0.064 & 3890 & 269 & 8.40 & 0.30 & 0.82 & 0.05 & 34& OC \\
  UFMG103 & 65.655 & 53.804 & 0.051 & 3802 & 351 & 7.70 & 0.15 & 0.93 & 0.03 & 40& OC \\
  UFMG104 & 68.250 & 51.060 & 0.019 & 3311 & 305 & 8.90 & 0.15 & 1.19 & 0.03 & 11& OC \\
  UFMG105 & 71.217 & 48.203 & 0.047 & 2667 & 432 & 7.55 & 0.45 & 0.87 & 0.03 & 20& OC \\
  UFMG106 & 72.210 & 46.720 & 0.065 & 2951 & 341 & 8.40 & 0.20 & 0.78 & 0.04 & 25& OC \\
  UFMG107 & 45.670 & 59.225 & 0.070 & 3715 & 429 & 8.65 & 0.20 & 1.03 & 0.05 & 29& OC \\
  UFMG108 & 53.181 & 56.870 & 0.122 & 4677 & 648 & 7.50 & 0.30 & 1.63 & 0.06 & 36& OCc \\
  UFMG109 & 63.345 & 51.285 & 0.025 & 3715 & 429 & 9.05 & 0.10 & 0.68 & 0.06 & 18& OC \\
  UFMG110 & 63.745 & 51.655 & 0.020 & 4786 & 663 & 8.80 & 0.30 & 0.90 & 0.06 & 15& OC \\
  UFMG111 & 68.845 & 48.334 & 0.049 & 4365 & 605 & 8.90 & 0.30 & 0.95 & 0.06 & 13& OC \\
  UFMG112 & 57.530 & 54.050 & 0.030 & 2512 & 290 & 8.30 & 0.10 & 1.50 & 0.06 & 15& OC \\
  UFMG113 & 59.430 & 52.730 & 0.086 & 3890 & 449 & 7.05 & 0.25 & 1.24 & 0.07 & 41& OC \\
  UFMG114 & 64.435 & 49.784 & 0.031 & 3236 & 373 & 7.95 & 0.25 & 0.96 & 0.07 & 19& OC \\
  UFMG115 & 65.497 & 49.300 & 0.092 & 3236 & 448 & 8.75 & 0.25 & 0.95 & 0.07 & 30& OC \\
  UFMG116 & 54.670 & 53.460 & 0.041 & 4074 & 850 & 8.20 & 0.50 & 1.58 & 0.07 & 11& OC \\
  UFMG117 & 61.610 & 50.330 & 0.044 & 3548 & 409 & 7.95 & 0.30 & 1.37 & 0.07 & 17& OC \\
  UFMG118 & 64.270 & 48.590 & 0.051 & 3802 & 263 & 7.15 & 0.10 & 1.17 & 0.03 & 42& OC \\
  UFMG119 & 65.410 & 46.850 & 0.028 & 3388 & 234 & 8.55 & 0.10 & 0.97 & 0.03 & 17& OC \\
  UFMG120 & 48.570 & 55.240 & 0.062 & 2754 & 254 & 7.80 & 0.15 & 1.17 & 0.03 & 28& OC \\
  UFMG121 & 50.075 & 54.660 & 0.068 & 3162 & 438 & 7.95 & 0.30 & 1.06 & 0.03 & 26& OC \\
  UFMG122 & 67.220 & 44.480 & 0.136 & 3548 & 492 & 8.00 & 0.30 & 0.68 & 0.03 & 64& OCc \\
  UFMG123 & 67.260 & 44.380 & 0.146 & 3467 & 320 & 8.20 & 0.20 & 0.65 & 0.03 & 66& OCc \\
  UFMG124 & 42.580 & 55.870 & 0.042 & 2291 & 211 & 8.10 & 0.20 & 0.67 & 0.03 & 13& OC \\
  UFMG125 & 55.200 & 51.400 & 0.093 & 3631 & 335 & 8.05 & 0.20 & 1.20 & 0.03 & 23& OCc \\
  UFMG126 & 56.830 & 49.820 & 0.083 & 3548 & 409 & 8.95 & 0.15 & 0.89 & 0.04 & 15& OCc \\
  UFMG127 & 64.073 & 46.310 & 0.107 & 3548 & 409 & 7.50 & 0.30 & 0.69 & 0.04 & 33& OCc \\
  UFMG128 & 44.940 & 53.335 & 0.083 & 3388 & 470 & 8.25 & 0.50 & 0.71 & 0.05 & 24& OCc \\
\hline\end{tabular}
\end{table*}

\section{Discussion}
\label{sect:discu}

In this section, we present evidence that the newly found objects are real physical systems and place their derived properties in the Milky Way context.

\subsection{The detectability of the newly found object}

As suggested by \citetalias{hunt2025}, the \textit{Gaia}-DR3-based cluster census is highly incomplete for distant, low-mass clusters of all ages and that mass, extinction, and distance are the main parameters influencing cluster detectability. The mean colour values of the samples within each tile, established in Sect \ref{sect:tile_detect}, can be used as a reference for how reddened the stars in each region are (on average). As shown in Fig \ref{fig:tiles_menores}, OCs tend to be found closer to the galactic plane ($-2^{\circ}$ $\lesssim b \lesssim$ $2^{\circ}$). We note a considerable lack of objects in regions affected by higher extinction, particularly within $140^{\circ}$ $\lesssim l \lesssim$ $148^{\circ}$ (higher mean colour values) and even our method was unable to recover OCs in this particular direction, suggesting that additional, highly reddened objects may remain undetected in this region.

\begin{figure}
    \includegraphics[width=0.98\linewidth]{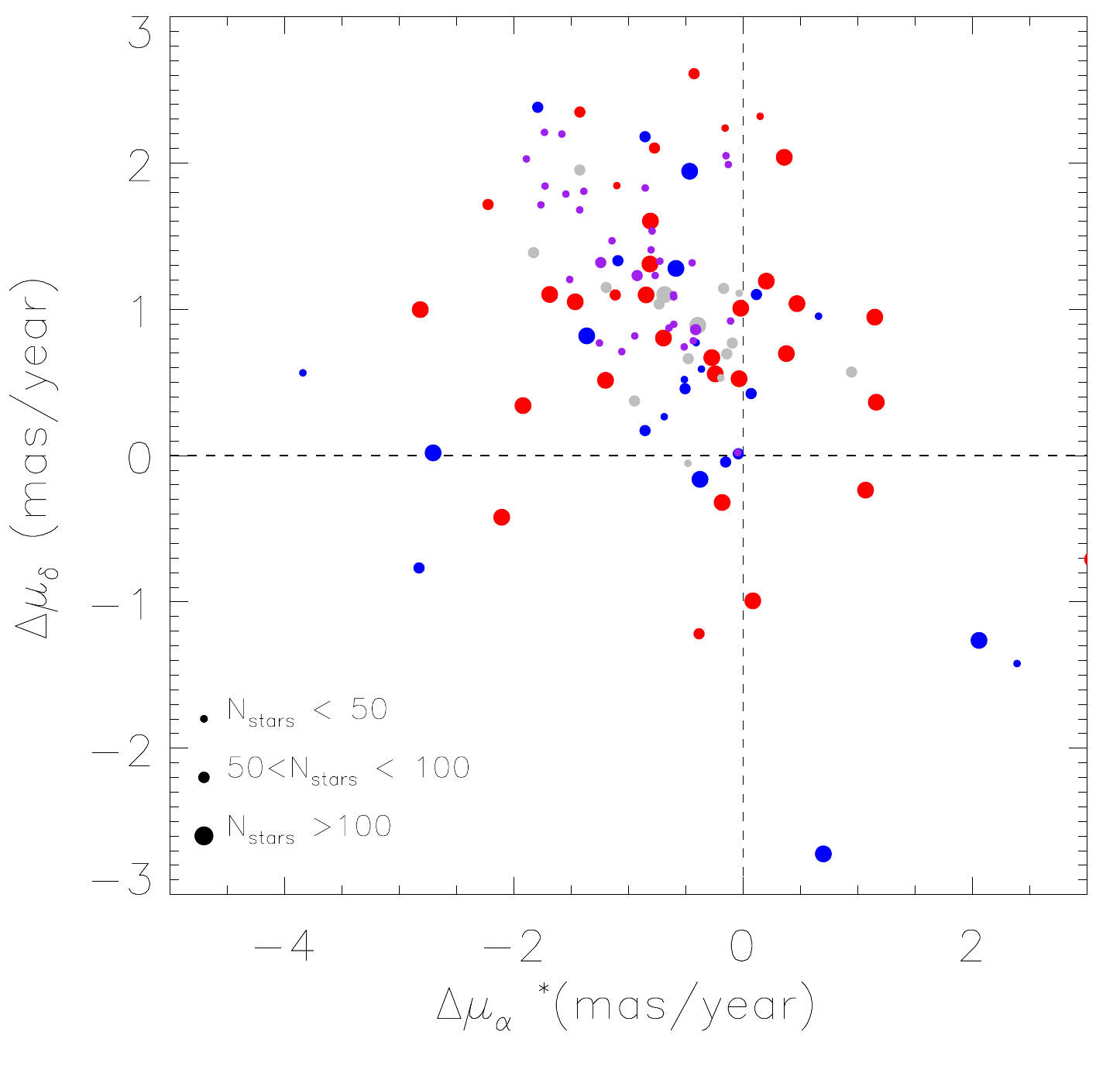}
\includegraphics[width=0.98\linewidth]{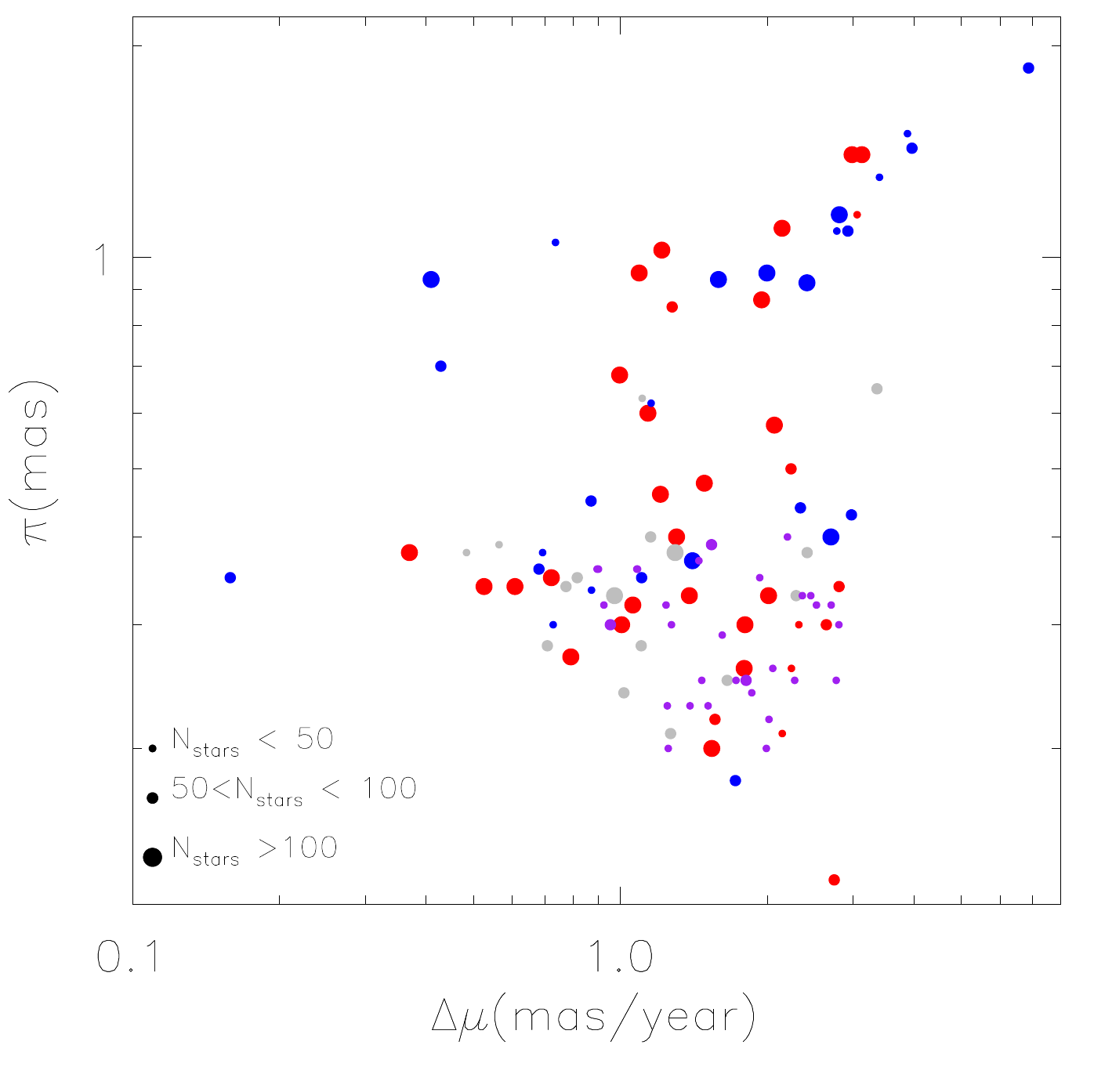}
  \caption{Top: differences in proper motion in right ascension and declination between the clusters and field population. Bottom: total difference in proper motion between the clusters and the field versus the mean OC parallax. On both panels, the OCs are represented by filled circles: known OCs before \textit{Gaia} (red), OCs discovered exclusively by \textit{Gaia} DR2 data (blue), OCs discovered exclusively by \textit{Gaia} EDR3 or DR3 data (grey) and our sample of discovered OCs (purple).}
\label{fig:detect_pm_field}
\end{figure}

During our searching procedure (see Sect. ~\ref{sect:search}), we also calculated, for the entire sample of stars of each tile, the mean values of $\mu_{\alpha}^{*}$ and $\mu_{\delta}$. These values were attributed to the mean field proper motion ($\mu_{\alpha FIELD}^{*}$ and $\mu_{\delta FIELD}$). We used them to analyse the difference on proper motion between the detected OCs and the field stars projected in their direction. For this analysis, we selected a subset of the detected OCs (see Sect. ~\ref{sect:search}) and separated them into three different samples:

\begin{enumerate}
 \item  known OCs (before \textit{Gaia} DR2 data availability), represented by all detected OCs from the following catalogues: Berkeley, Czernik, FSR, King and NGC (34 objects). They are shown as red filled circles in Fig. ~\ref{fig:detect_pm_field};
 
\item known OCs detected exclusively with \textit{Gaia} DR2 data, represented by all detected OCs from the UBC \citep{2019A&A...624A.126C,cjl19,cjl20}, UKP \citep{sla19} and Theia \citep{2019AJ....158..122K} (27 objects) catalogues. They are represented by blue filled circles in Fig. ~\ref{fig:detect_pm_field};
        
\item  known OCs detected exclusively with \textit{Gaia} EDR3 and DR3 data, from catalogues: UBC \citep{2022A&A...661A.118C} and HSC \citetalias{2023A&A...673A.114H} (grey filled circles, 37 objects) and our sample of discovered OCs (purple filled circles).

\end{enumerate}

We did not use all the detected objects to establish these groups, since the number of known OCs identified with \textit{Gaia} EDR3 and DR3 is much larger than the others. In order to obtain three groups with comparable sample sizes (a few dozens of clusters), we selected objects from the most populated catalogues only, while also ensuring that each group included objects from at least three different literature references.

For all the clusters, we calculated the differences in proper motion ($\Delta\mu_{\alpha}^{*}=\mu_{\alpha CLU}^{*}-\mu_{\alpha FIELD}^{*}$ and $\Delta\mu_{\delta}=\mu_{\delta CLU}-\mu_{\delta FIELD}$). The top panel of Fig. ~\ref{fig:detect_pm_field} shows $\Delta\mu_{\alpha}^{*}$ versus $\Delta\mu_{\delta}$. The bottom panel shows the total difference in proper motion ($\Delta\mu=\sqrt{\Delta\mu_{\alpha}^{*2}+\Delta\mu_{\delta}^{2}}$) versus the mean cluster parallax. The number of OC members is represented by the symbol size, indicating the richness of the object.

It is possible to note that, before the \textit{Gaia} era (red filled circles), the OC census in this region has mainly consisted of rich objects with a vast range of distances. The few known poor OCs also exhibit large differences in proper motion with respect to the field, helping their detectability. With \textit{Gaia} DR2 data (blue filled circles), most of the discovered OCs in this region exhibit parallaxes greater than $\sim 0.4$ mas and many of them are moderate rich or rich, although few poor and distant objects were also detected. Due the improvement in astrometric precision reached by \textit{Gaia} EDR3 and DR3 data (grey and purple filled circles), we observe a vast increase in the number of new poor and distant ($\varpi \lesssim 0.4$) clusters. 

As stated in \citetalias{hunt2025}, the proper motion of a cluster can also have a sizeable impact on its detectability, with clusters whose motion is more distinct from that of the surrounding field stars being easier to detect. In our sample, the newly discovered OCs tend to have fewer members compared to the literature OCs, on the other hand they present moderate values of $\Delta\mu_{\alpha}^{*}$ and $\Delta\mu_{\delta}$, which likely contributed to their detectability.

The lack of objects near the origin of the top panel ($\Delta\mu_{\alpha}^{*} \approx \Delta\mu_{\delta}\approx 0.0$) suggests that poor clusters with proper motions very similar to those of the Galactic disc field may still remain undetected. The few detected objects with $\mu_{CLU}\approx \mu_{FIELD}$, tend to be moderate or rich systems. As the mass plays an important role on OCs detectability \citep{hunt2025}, in this case, the number of members effectively playing the role of the visible cluster mass, providing sufficient contrast in the proper motion distribution to distinguish them from the field. The expected improvements in astrometric data precision from the future \textit{Gaia} data releases may enable the detection of these poor and distant missing OCs.

 \subsection{Comparison with confirmed OCs}
\label{sect:discu_conf_ocs}

\begin{figure}
\includegraphics[page=1,width=0.48\linewidth]{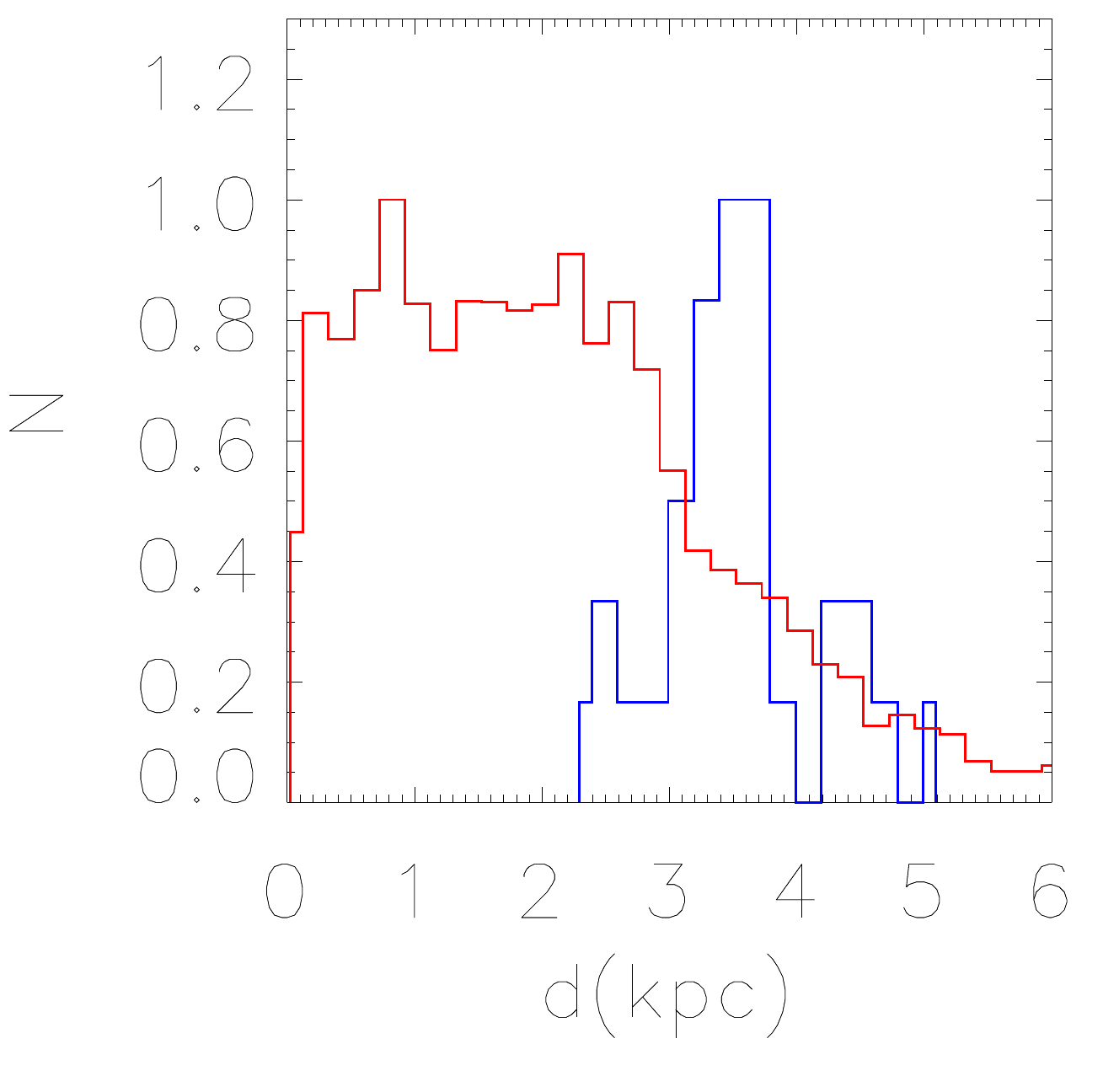}
\includegraphics[page=2,width=0.48\linewidth]{fig/discussion/astro_parameter/astro_parameters.pdf}
\includegraphics[page=3,width=0.48\linewidth]{fig/discussion/astro_parameter/astro_parameters.pdf}
\includegraphics[page=4,width=0.48\linewidth]{fig/discussion/astro_parameter/astro_parameters.pdf}

\caption{Frequency distributions of astrophysical and structural properties of the new OCs (blue) compared with those in the \citetalias{2023A&A...673A.114H} catalogue (red). Top left: heliocentric distance. Top right: the logarithmic age. Bottom left: Colour excesses. Bottom right: the concentration parameter.}
\label{fig:astro_par_hunt}
\end{figure}

Fig.~\ref{fig:astro_par_hunt} displays the normalized frequency distributions of the derived parameters for the newly discovered OCs (blue) compared with the sample of OCs reported in \citetalias{2023A&A...673A.114H} (red). The top-left panel indicates that our sample is, on average, located at larger heliocentric distances than known OCs. The top-right panel shows that the new OCs present an age distribution generally similar to those of known clusters, although our sample lacks very young ($log[t](yr)<7$) and very old objects ($log[t](yr)>9$). 

The bottom left panel shows that the new findings are more reddened than the average of known OCs. This is expected, since our search was directed towards low Galactic latitudes ($-5^{\circ}<b<5^{\circ}$) and located at larger heliocentric distances. These two factors naturally imply in longer lines of sight through the dusty Galactic disc and consequently higher extinction values. 

The bottom right panel presents the concentration parameter ($c=log[r_{t}]/log[r_{c}]$) of our sample (blue) compared with those in \cite{2022A&A...659A..59T}. We note that our OCs are, on average, less concentrated than known local OCs. These trends are consistent with that presented by the sample of OCs found in paper \citetalias{10.1093/mnras/staa1684}.

\subsection{The OCs physical reability: proper motion dispersion and sizes}
  
In this section, we analysed the physical reability of our OCs and established a comparison of their astrometric properties with literature OCs. For the literature comparison sample, we built a reliable OC sample by restricting OCs from \citetalias{2023A&A...673A.114H} with a 'cluster significance test' value $CST>5\sigma$ and a 'median CMD class' $CMDCl50 >0.5$, as recommended in that work. These criteria provide a high signal-to-noise subset of OCs, corresponding to highly likely clusters with CMDs consistent with a single stellar population. Objects with quality parameters out of these quality thresholds are considered as a low quality sample in the following analysis. For the figures of the subsequent sections, the reliable OC sample is represented by solid symbols, while the low quality objects are represented by open symbols.

\begin{figure}
  \includegraphics[width=0.98\linewidth]{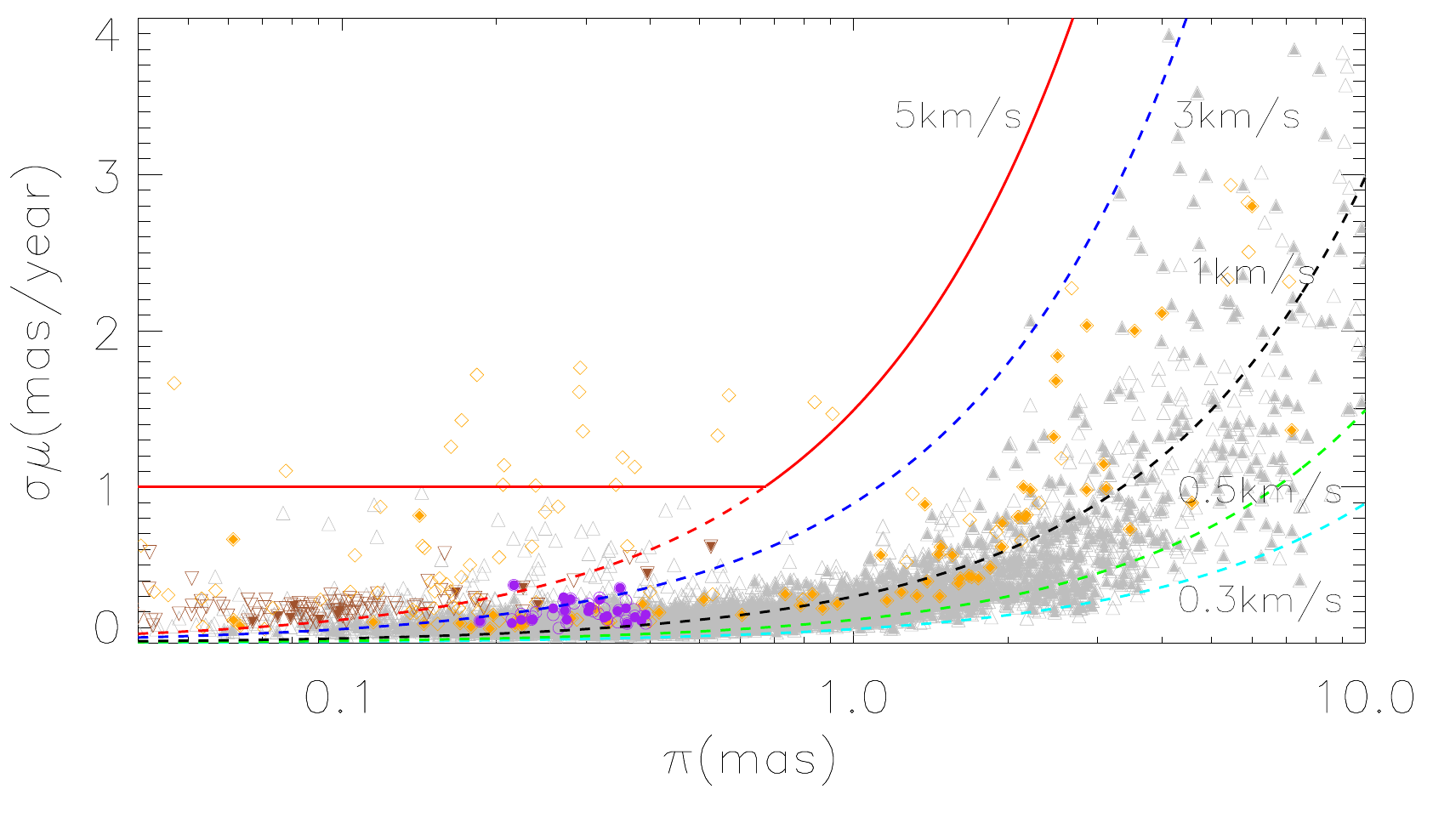}
  \caption{Dispersion in proper motion \textit{versus} parallax for confirmed OCs (grey triangles), moving groups (orange diamonds), globular clusters (brown triangles) and those reported in the present work (purple circles). The curves for real velocities dispersion are represented as follow: 0.3 km/s (cyan dotted line), 0.5 km/s (green dotted line), 1.0 km/s (black dotted line), 3.0 km/s (blue dotted line) and 5.0 km/s (red doted line for $\mu<1$ and red solid line for $\mu>1$). The horizontal red solid line delimits the proper motion dipersion at $\mu=1$.}
\label{fig:plx_versus_pm}
\end{figure}

For both our sample and the literature OCs, we calculated the total proper motion dispersion, $\Sigma \mu$, as a quadratic sum of the dispersion in $\mu_{\alpha}^{*}$ and $\mu_{\delta}$. Fig.~\ref{fig:plx_versus_pm} shows the total proper motion dispersion as a function of mean parallax for literature objects and the newly discovered OCs. The literature sample is separated according to their catalogue flags: $o$ (OC, grey triangles), $m$ (moving group, orange diamonds) and $g$ (globular cluster, brown triangles). Our sample of newly discovered OCs is represented by purple circles (OCcs are shown as opened purple circles, see Sect ~\ref{sect:density}). The expected curves for real velocity dispersion of 0.3, 0.5, 1.0, 3.0 and 5.0 km/s are also represented. The straight solid line, with $\sigma \mu = 1$ mas/yr, represents a 'forbidden zone' or an upper limit for physically coherent clusters \citep{ca20}.

    \begin{figure}
\includegraphics[width=0.98\linewidth]{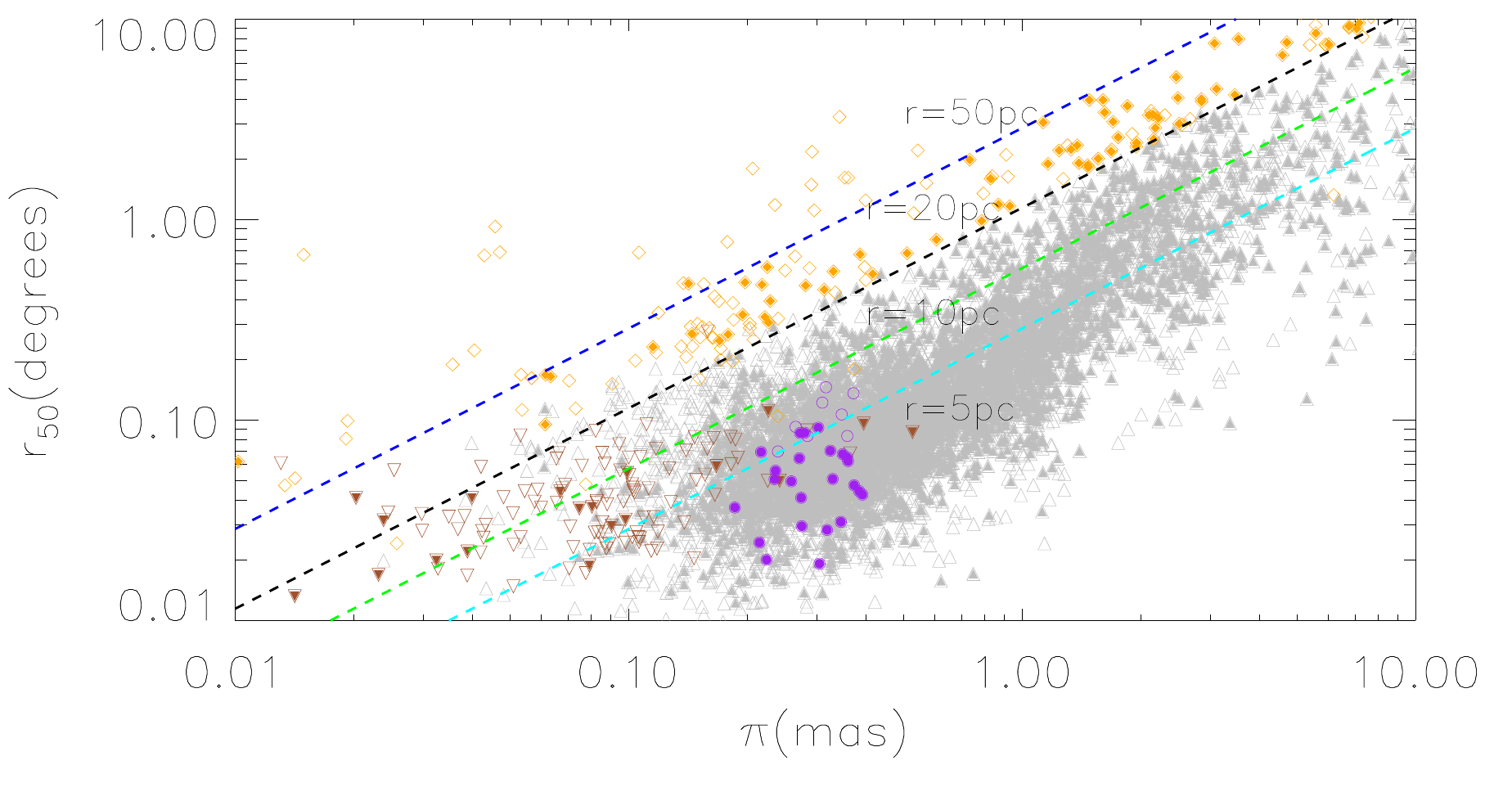}
  \caption{The cluster radius $R_{50}$ \textit{versus} parallax for confirmed OCs (grey triangles), moving groups (orange diamonds), globular clusters (brown triangles) and those reported in the present work (purple circles). theoretical sizes expected for the range of parallaxes is also represented: 5 pc (cyan dotted line), 10 pc (green dotted line), 20 pc (black dotted line) and 50 pc (blue dotted line).}
\label{fig:plx_versus_r50}
\end{figure}

For coherent stellar systems, the dispersion in proper motion is expected to correlate with parallax \citep{ca20}. However, the uncertainties in the individual stellar proper motions contribute significantly to the observed proper motion clusters dispersions, for distant objects ($d \gtrsim 1$ kpc). In such cases, the measured dispersion is largely dominated by astrometric uncertainties and the contamination by remaining field stars in the memberlists also plays a role on increasing the observed proper motion dispersion. In this way, as expected, the most distant objects in Fig.~\ref{fig:plx_versus_pm} tend to exhibit observed dispersion values higher than the theoretical ones. 

Among the sample present in \citetalias{2023A&A...673A.114H}, the only objects that exhibit dispersion in the forbidden zone, are some low quality moving groups, suggesting that these systems are not gravitationally bound. This is consistent with the analysis present in \cite{ca20}. According to their OC sample, only asterisms with memberlist available exhibit proper motion dispersions above this threshold. All the OCs identified in this work exhibit total proper motion dispersions compatible with physically coherent stellar systems, lying in the diagram region where the bulk of confirmed OCs is (grey triangles).

We also analysed the sizes of the OCs using the parameter $r_{50}$, a similar analysis present in \cite{ca20} and \citetalias{2023A&A...673A.114H}. Fig.~\ref{fig:plx_versus_r50} shows the correlation between the mean cluster parallax and $r_{50}$ (in degrees). Although known clusters span a wide range of physical sizes, we note that the bulk of confirmed OCs exihit $r_{50}(pc)\lesssim10pc$ and less OCs are located in the range $10pc\lesssim r_{50}(pc)\lesssim 20pc$.

\cite{ca20} has shown that the sparse and elongated structures identified in \cite{2019AJ....158..122K} and \cite{2020AJ....160..279K} tend to present high values of $r_{50}$. These objects have characteristic sizes of several tens of parsecs and are typically young stellar groups tracing the large-scale gas morphology of their parent molecular clouds and are not necessarily gravitationally-bound \citep{2024A&A...686A..42H}. According to \citetalias{2023A&A...673A.114H}, 28 of these objects are flagged as moving groups and the bulk of those sparse non-gravitationally bound objects tends to present $r_{50}(pc) \gtrsim 20pc$. Most objects with $r_{50}(pc) \gtrsim 50pc$ are low quality moving groups.

In this diagram, our objects (purple symbols) occupy the same region as the bulk of high-quality OCs ($r_{50}(pc)\lesssim 10 pc$). Our objects flagged as OCc (purple open circles), tend to exhibit slightly larger $r_{50}$ values than those flagged as OC (purple filled circles), indicating that, although less dense, they still appear to be physically coherent and likely gravitationally bound.

\subsection{The OCs physical reability: density}
\label{sect:density}

For the following analysis, we calculated the Galactocentric Cartesian coordinates (X, Y, Z) for both our sample of OCs and those from the literature by using their Galactic coordinates ($l$, $b$) and the heliocentric distance $d$. That is, $X = R_{\odot}-dcos(l)cos(b) , Y= dsin(l)cos(b), Z=dsin(b)$ and the Galactocentric radius $R_{GC}=\sqrt{X^{2}+Y^{2}+Z^{2}}$, adopting the Sun's Galactocentric distance at 8.34\,kpc \citep{2014ApJ...783..130R}.

 We also analysed the OCs densities. We estimated the half-mass density ($\rho = 3M/(8\pi R^{3}_{hm})$) by assuming the half-mass radius $R_{hm} \sim R_{50}$, a similar assumption adopted by \cite{2019A&A...624A...8A}. To determine the cluster mass enclosed within $R_{hm}$, we adopted the mass-luminosity relation from the same set of isochrones used to establish the clusters ages (see Sect ~\ref{sect:membership}).
 
 The PARSEC isochrones identify stellar evolutionary stages through the parameter \textit{label}. We used stars with $label=0$ and 1 (pre-main sequence and main sequence stars, respectively), to determine mass from luminosity to the clusters main sequence stars.  The absolute magnitudes of the OCs members ($M_{G}$, determined from distance and extinction) were separated in bins of 1 mag. For each bin, the stellar masses were obtained by interpolating $M_{G}$ in the mass-luminosity relation provided by the corresponding isochrone.
 
\begin{figure}
  \includegraphics[width=0.3\linewidth]{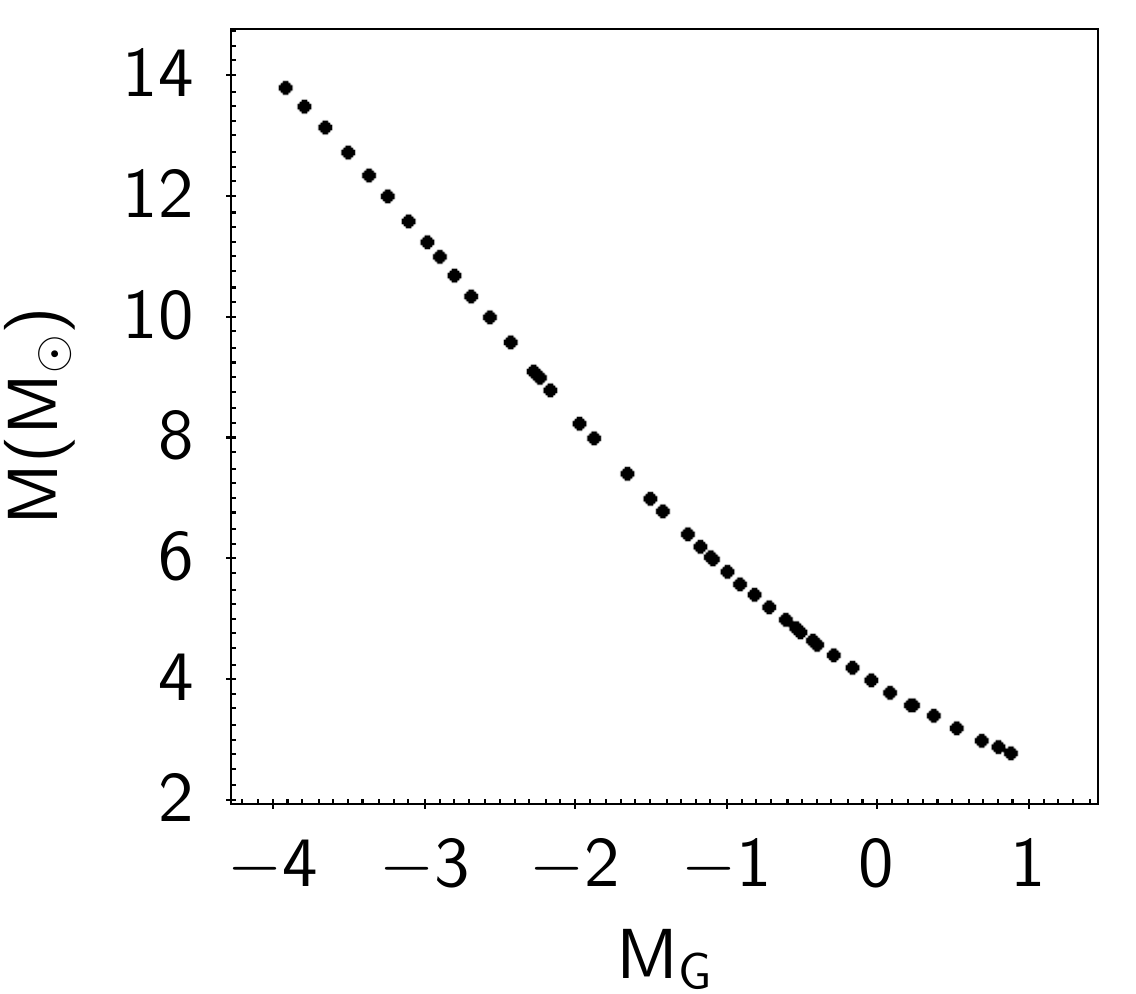}
  \includegraphics[width=0.3\linewidth]{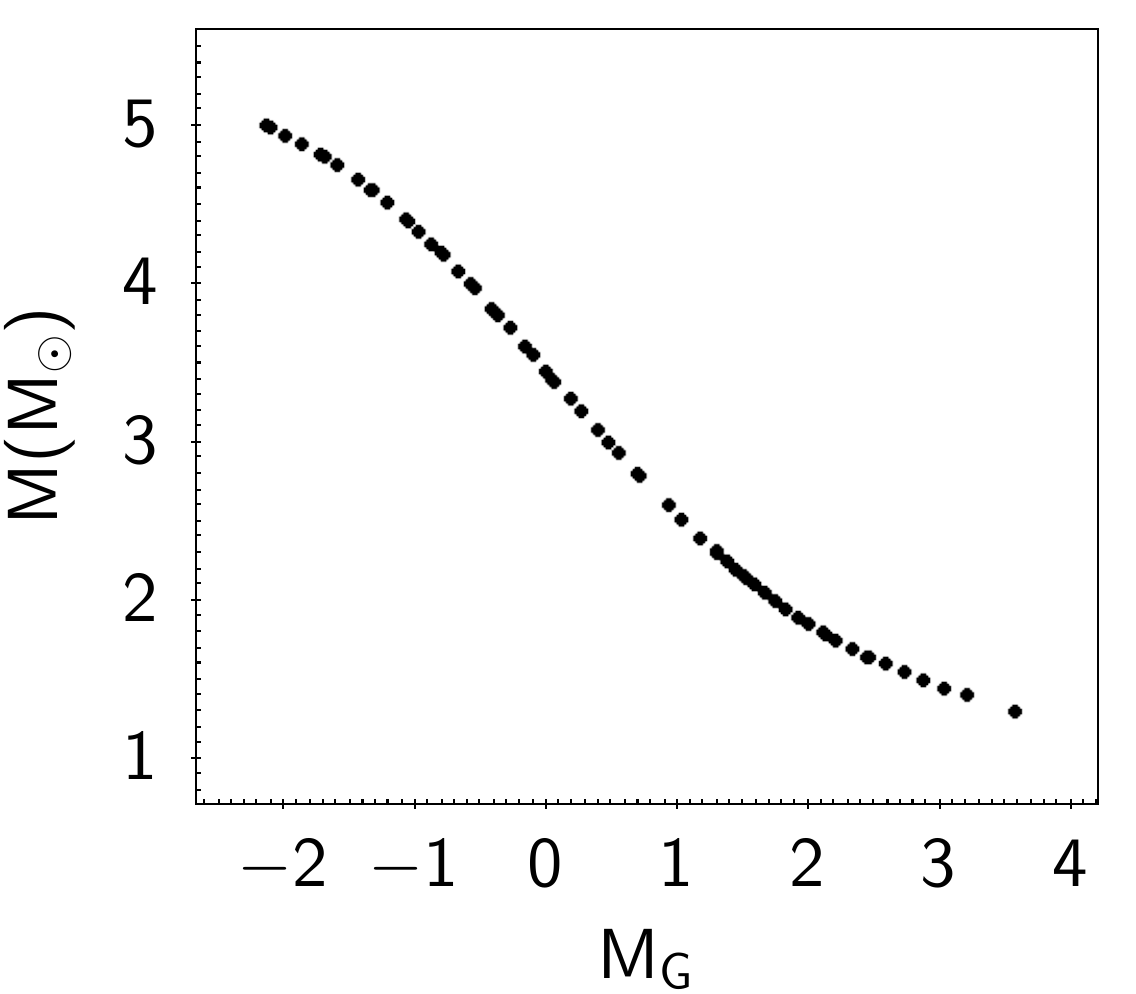}
  \includegraphics[width=0.3\linewidth]{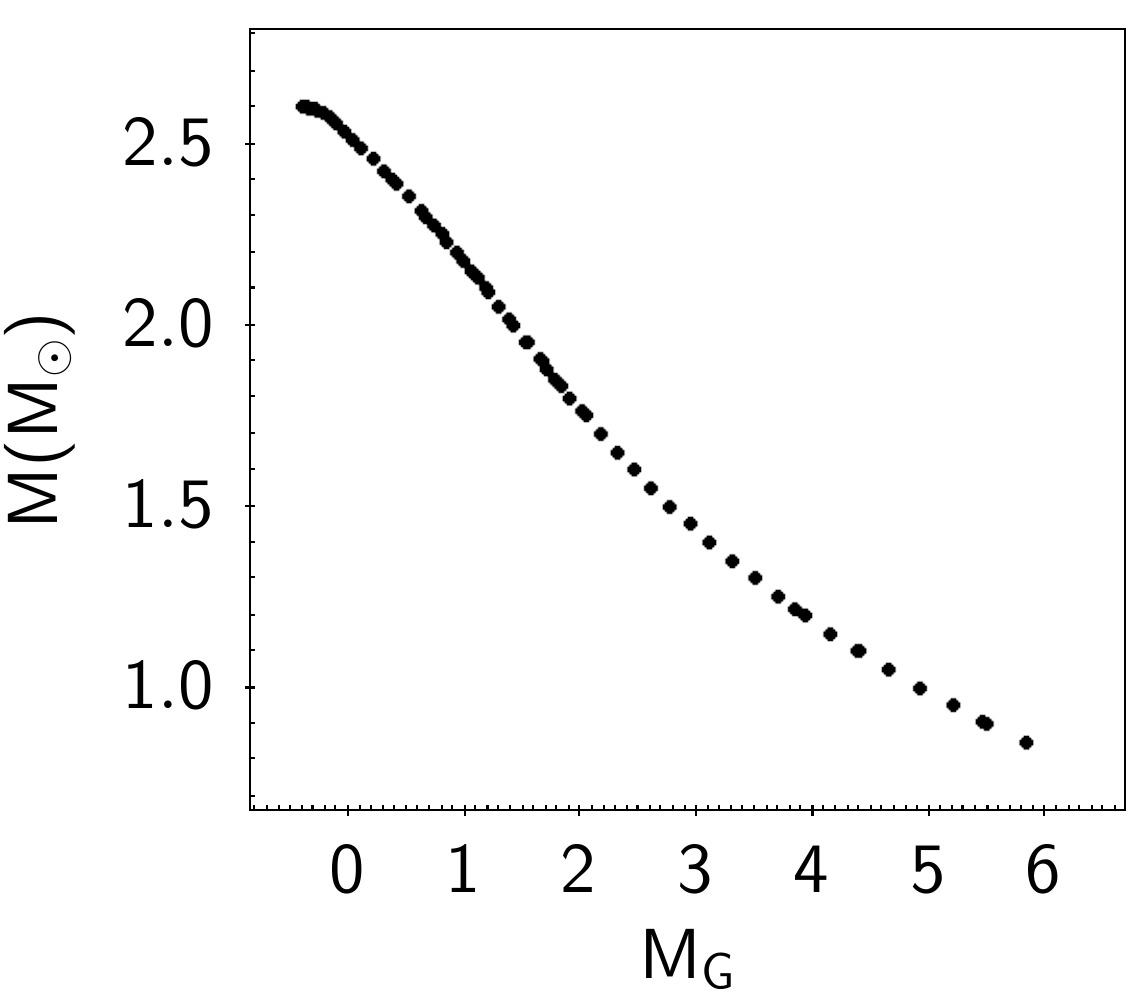}
\includegraphics[width=0.32\linewidth]{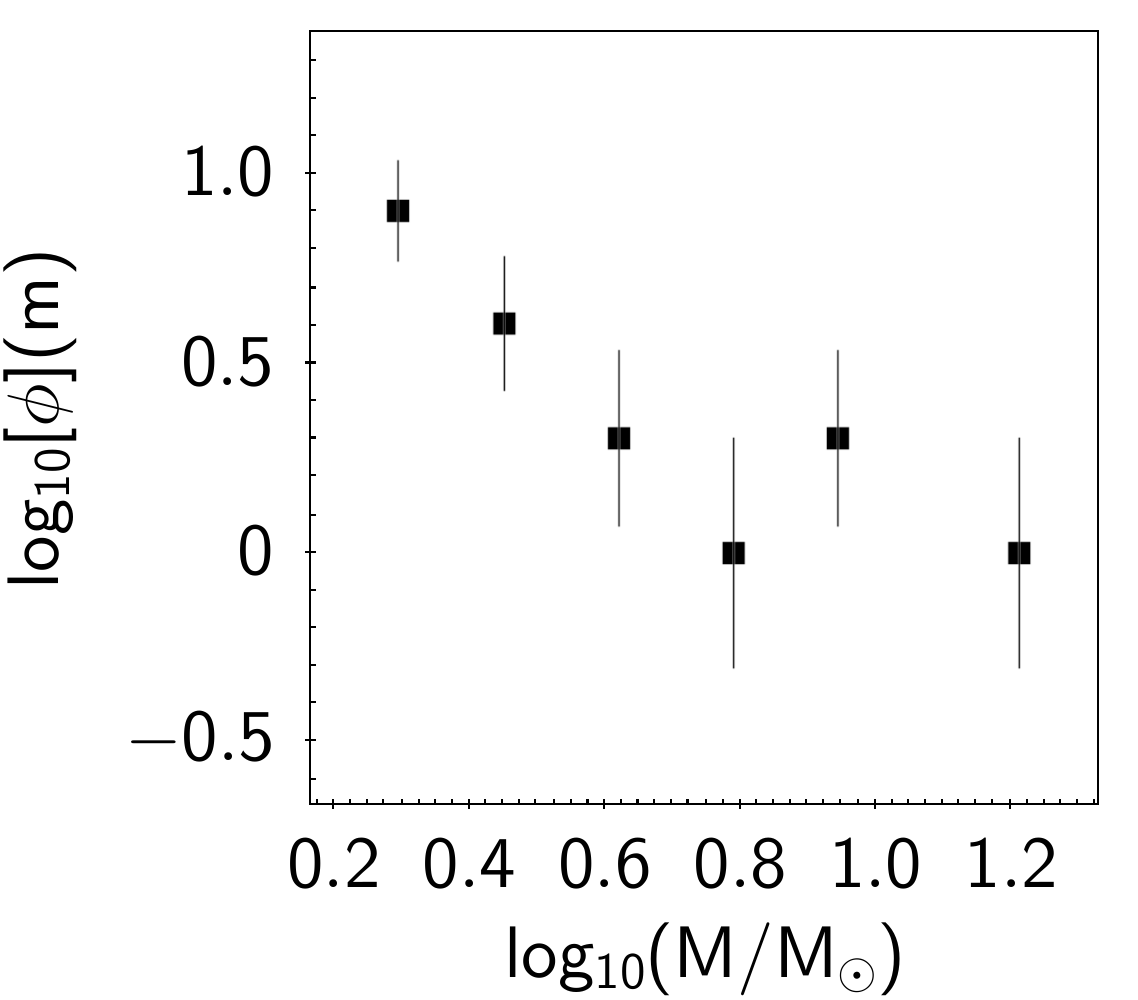}
\includegraphics[width=0.32\linewidth]{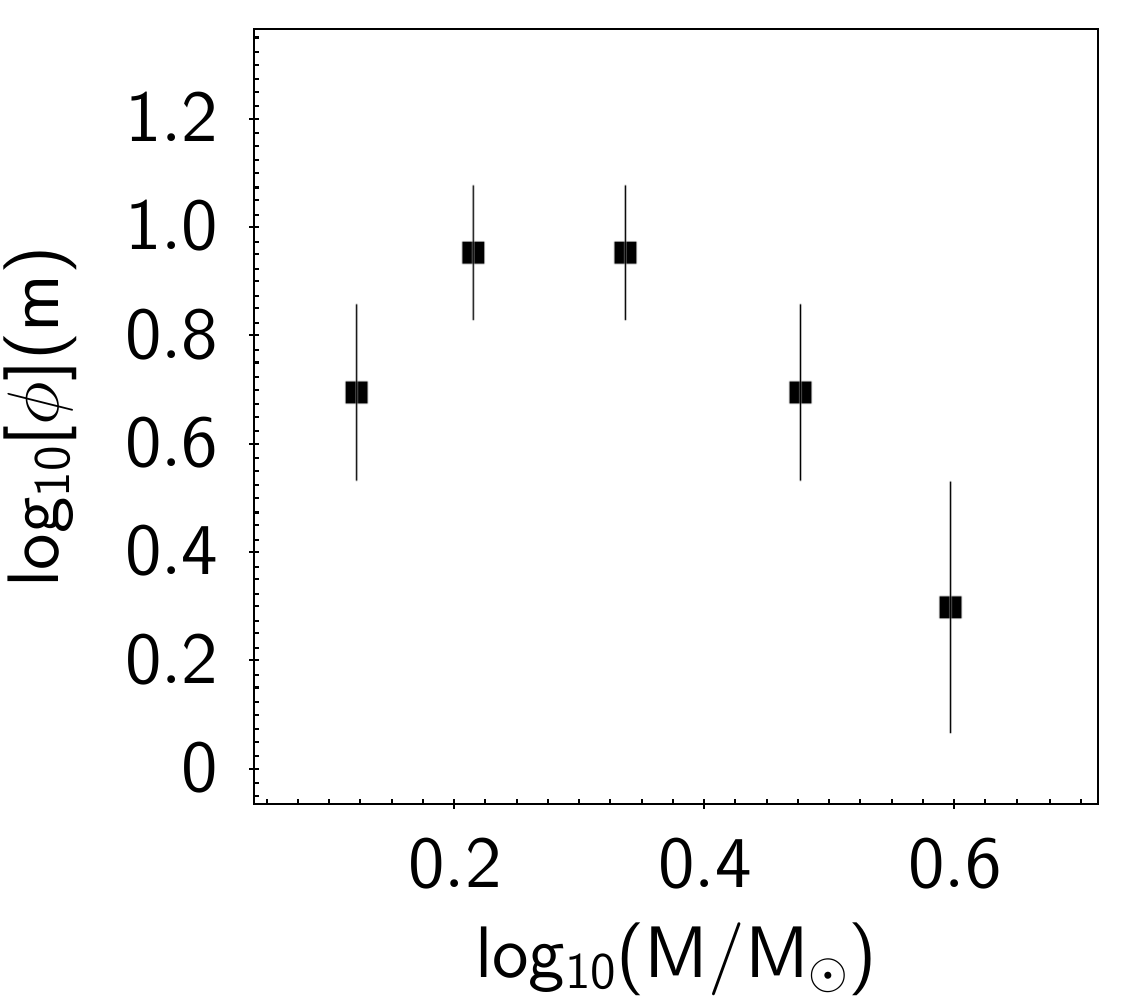}
\includegraphics[width=0.32\linewidth]{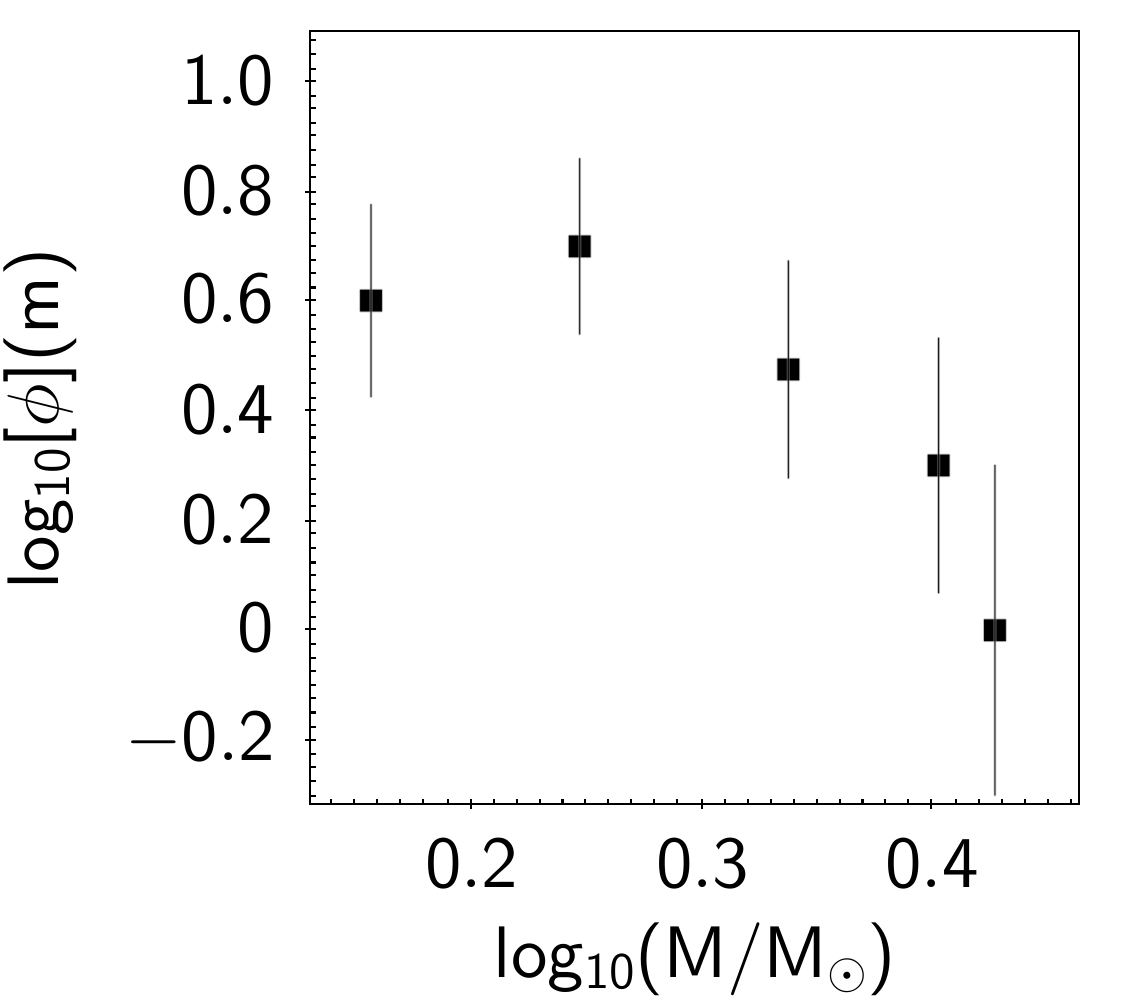}

\caption{Top panels: mass-luminosity relation for isochrones of different logarithmic ages: 7.05 dex (left), 8.0 dex (middle) and 8.75 dex (right). Bottom: examples of mass functions constructed by using OC members and these mass-luminosity relation: UFMG113 (left), UFMG122 (middle) and UFMG115 (right). }
\label{fig:mass_func}
\end{figure}

  For evolved stars ($label>1$), where the luminosity does not increase monotonically with mass, we interpolated the observed cluster magnitudes directly along the isochrone to attribute masses to the cluster stars. Fig.~\ref{fig:mass_func} shows examples of the adopted mass-luminosity relations, as well as the respective resulting mass function for different OCs. The total mass values were then obtained by integrating the mass functions. Since we did not extrapolate the mass function to account for non-observed low-mass stars, the derived mass values should be regarded as lower limits, i.e, the visible mass of the clusters.


  As the observational magnitude limit plays a crucial role in determining the visible OC masses, we also applied the same mass estimation procedure to the OCs from \citetalias{2023A&A...673A.114H}. For this comparison, we restricted their sample to objects located at distances similar to those of our sample ($2.5 kpc<d< 5kpc$), within the outer Galaxy ($80^{\circ}$ $\leq l \leq$ $260^{\circ}$) and with ages $log[t](yr)>7.15$, resulting in 1445 clusters. We did not estimate masses for the sample of globular clusters. This analysis provides a comparison between masses of previously known OCs and our new candidates.

  Gravitationally bound systems tend to present mean densities that are greater than the critical density ($\rho_{crit}$), defined as the minimum density required for survival against Galactic tidal disruption,

  \begin{equation}
     \rho_{crit} \sim \frac{M_{clu}}{(4/3)\pi R^{3}_{J}}
     \label{eqn:eq1}
  \end{equation}

  In this context, $M_{clu}$ is the cluster mass inside the Jacobi radius ($R_{J}$). That is, $\rho_{crit}$ is assumed here as the mean density for a cluster completely filling its Roche lobe. The expresssion of $R_{J}$ was taken from \cite{1957ApJ...125..451V}

 \begin{equation}
R_{J}=(\frac{M_{clu}}{3M_{G}})^{1/3} \times R_{GC}
\label{eqn:eq2}
\end{equation}

Combining Equations~\eqref{eqn:eq1} and \eqref{eqn:eq2}, the critical density may also be written as a function of the Galactocentric radius:
  \begin{equation}
  \rho_{crit}=(\frac{9}{4 \pi}) \frac{M_{G}}{R^{3}_{GC}}
  \label{eqn:eq3}
 \end{equation}

Which assumes a circular orbit around a point mass galaxy with $M_{G} \sim 1.0 \times 10^{11} M_{\odot}$ \citep{1991RMxAA..22..255A,1994A&A...288..751C,2005A&A...433..917B,2016MNRAS.461.3483T}.  We did not adopt the recent values of Jacobi mass ($M_{J}$) and $R_{J}$ from \cite{2024A&A...686A..42H}, as those results are based on a different G magnitude limit.

\begin{figure*}
  \includegraphics[width=0.28\linewidth]{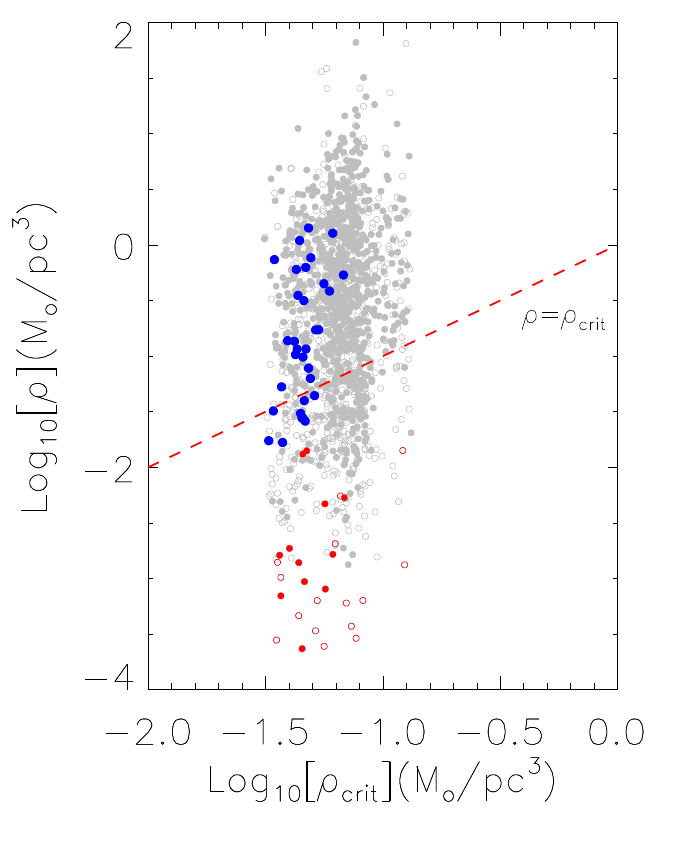}
\includegraphics[width=0.35\linewidth]{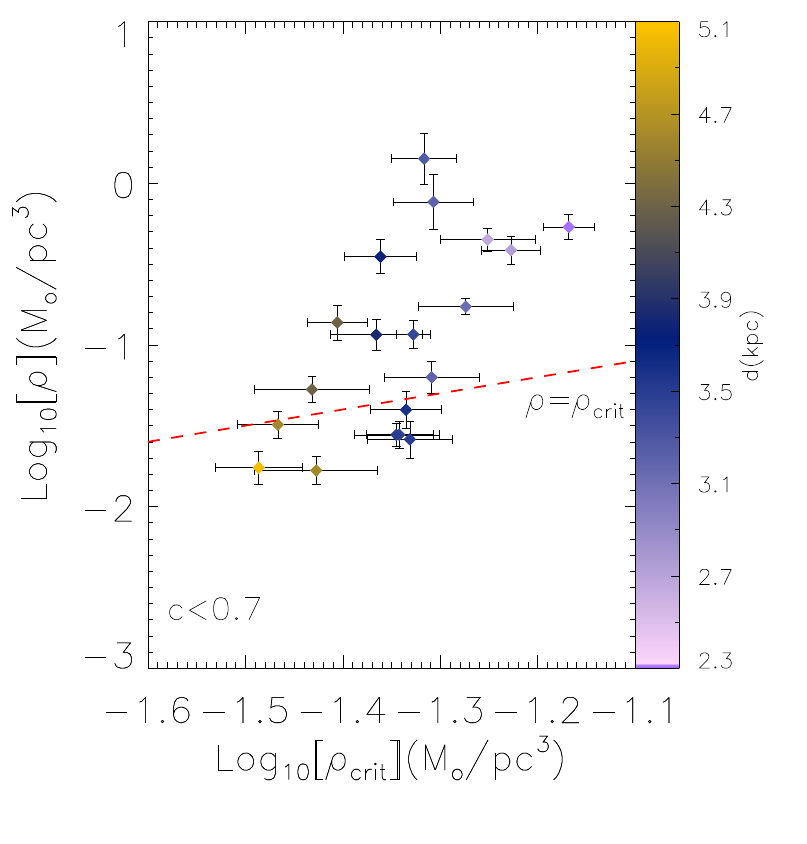}
\includegraphics[width=0.35\linewidth]{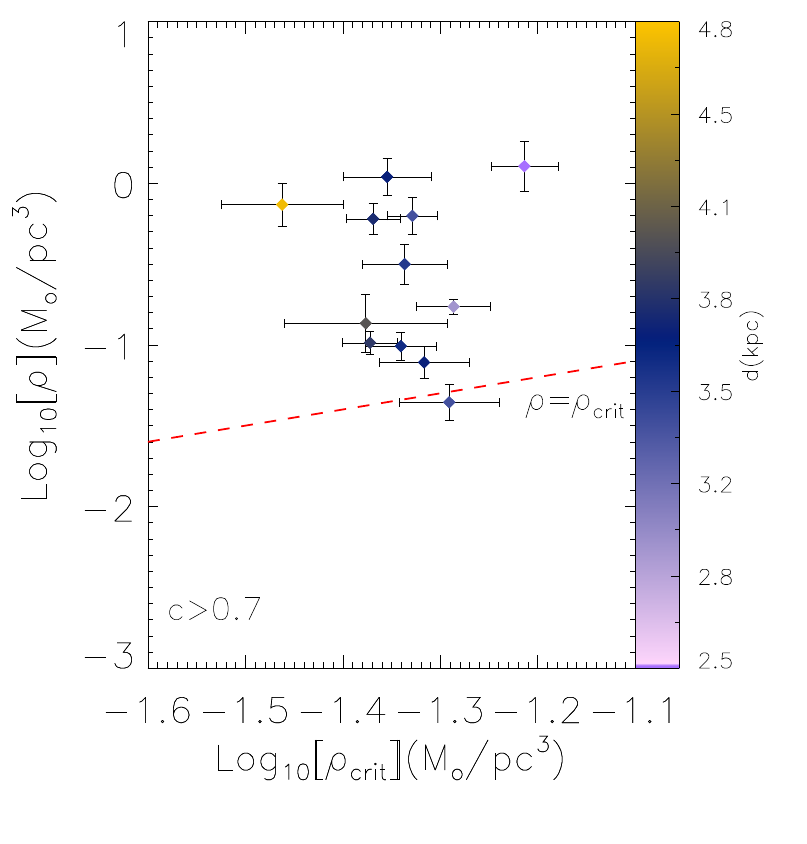}

\caption{Left: $\rho$ versus $\rho_{crit}$ for objects from \citetalias{2023A&A...673A.114H}: OCs (grey) and moving groups (red). The OCs discovered in this work are represented in blue. Middle: $\rho$ versus $\rho_{crit}$ for OCs in this work with concentration parameter $c<0.70.$. Right: $\rho$ versus $\rho_{crit}$ for OCs in this work with concentration parameter $c>0.70.$. For the middle and right panels the OC distance is represented by the colourbar and the error bars based on mass and distances uncertainties are also represented. For all the panels, the red dashed line represents $\rho=\rho_{crit}$.}
\label{fig:density_density_crit}
\end{figure*}

  According to the left panel of Fig.~\ref{fig:density_density_crit}, we note that our OCs (blue filled circles) tend to behave similarly to the literature high quality OC sample (grey filled circles): most of these objects have $\rho >\rho_{crit}$. In contrast, almost all moving groups (low and good quality samples,  open red and filled red circles, respectively) present $\rho <\rho_{crit}$. Many low quality OCs and a few good quality OCs also exhibit $\rho <\rho_{crit}$.

  The second and third panels of Figure \ref{fig:density_density_crit} show our OC sample separated into two groups based on the concentration parameter. The group with $c<0.7$ contains a larger fraction of objects with $\rho <\rho_{crit}$ than the group with $c>0.7$. We also note in both diagrams that the heliocentric distance also plays a role: all nearby objects ($d\lesssim 3kpc$, light blue sample) have $\rho >\rho_{crit}$. On the other hand, the more distant objects (dark blue and yellow samples), tend to present  $\rho <\rho_{crit}$.  
  
  Since this approach to $\rho_{\mathrm{crit}}$ depends on the Galactocentric distance and the Galactic potential (see Eq.~\ref{eqn:eq3}), $\rho/\rho_{\mathrm{crit}}$ depends solely on the intrinsic cluster properties: mass and size. As the visible mass is a lower limit to the total cluster mass, it is possible that some objects with $\rho <\rho_{crit}$ are still bound systems. A mass estimation based on deeper photometric analysis could provide a more reliable mass estimation for these cases. We flagged as open cluster candidates (OCc; see Table~\ref{Tab:clusters_prop2}) those objects that remain in the $\rho < \rho_{\rm crit}$ regime even when considering their error bars, despite other hints that they may be bound systems.

  \subsection{The OC census in the region}
  \label{sect:census_region}
  
 
 For the following analysis, we analysed the 256 objects (31 newly discovered and 225 previously known clusters) detected in this work. Their centres were cross-matched with literature catalogues and astrophysical parameters were adopted from the reference with the smallest centre separation. If the closest match did not provide astrophysical parameters, we used those from the next closest one.
 
 Objects not detected with our method, were incorporated directly from the literature. To assign the parameters of a given cluster, we gave preference to the reference in which the cluster was originally reported. Since the same cluster may appear multiple times with different names in different catalogues, we first used the cluster name as the primary identification criterion. In cases where different names referred to the same object, we compared the objects to their neighbours, following the same astrometric comparison described in Sect. \ref{sect:validation}. 
 
 In total, we identified 590 distinct clusters in this region, of which 366 have been studied using \textit{Gaia} data. The remaining objects correspond mainly to IR OCs or embedded objects, listed in \cite{dias2002}, \cite{kharchenko2013} and \cite{2019AJ....157...12B}.

\begin{figure}
\includegraphics[page=1,width=0.48\linewidth]{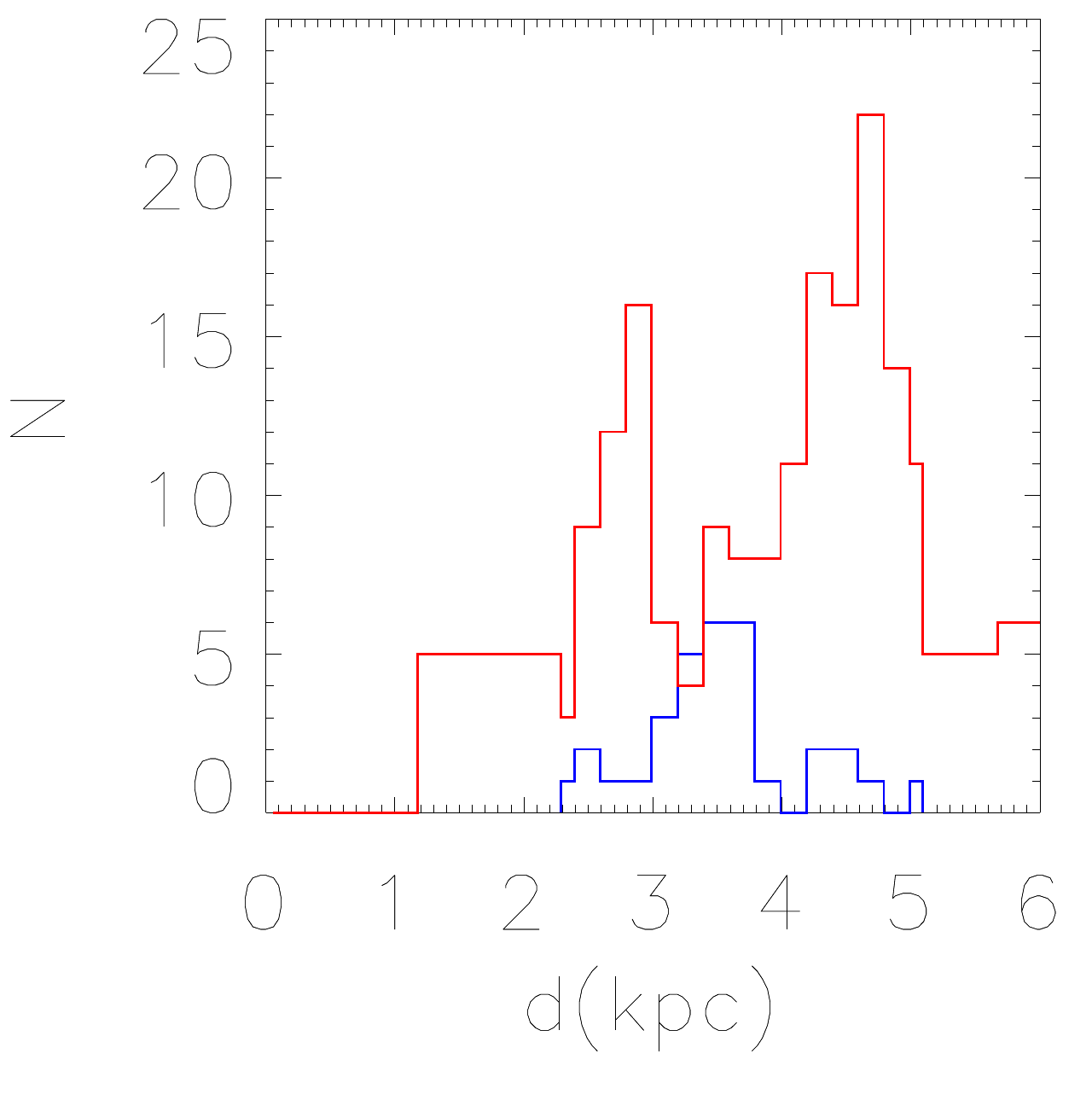}
\includegraphics[page=2,width=0.48\linewidth]{fig/discussion/perseu_arm/astro_parameters_region.pdf}
\includegraphics[page=3,width=0.48\linewidth]{fig/discussion/perseu_arm/astro_parameters_region.pdf}
\includegraphics[page=4,width=0.48\linewidth]{fig/discussion/perseu_arm/astro_parameters_region.pdf}

\caption{Histograms of astrophysical parameters of the new OCs (blue) compared with those from \citetalias{2023A&A...673A.114H} (red) towards Perseus gap region. Top left: heliocentric distance. Top right: the logarithmic age. Bottom left: Colour excesses. Bottom right: Galactocentric distance.}
\label{fig:parameters_region}
\end{figure}

Based on catalogues that provide astrometric parameters from \textit{Gaia}, our findings represent an increment of about $\sim 9\%$ in this sample and $\sim 5\%$ in the total number of objects in this region. According to the mentioned catalogues, there are 183 OCs with known distances within $d <3\,kpc$, 61 OCs within $3\,kpc <d< 4\,kpc$ and 55 beyond $d >4\,kpc$. Our new OC sample therefore represents a fractional increment of $\sim 3\%$ up to 3\,kpc, $31\%$ within $3\,kpc <d<4\,kpc$ and $13\%$ for to $d >4\,kpc$.  The increment represented by our newly found OCs is shown in the Table\ \ref{tab:estatistica}. Fig.~\ref{fig:parameters_region} shows a comparison between the new OCs and the known ones for some parameters on the region of interest. When compared with the outer Galactic disc OCs, our findings provide a significant increment of reddened ($E(B-V)>0.5$) and young-intermediate age ($7>log[t](yr)>9$) OCs with $R_{GC}\sim 11 kpc$. 

\begin{table}
\centering
\footnotesize
\caption{The impact of the newly discovered OCs in this work over the local OC population.}
\def\arraystretch{1.3}
\begin{tabular}{l c c c} \hline
Sample & fraction & increase &  \\
\hline
Total & $31/591$ & $5.2\%$ &  \\
with \textit{Gaia} data & $31/367$ & $8.4\%$ &  \\
 $d<3\ kpc$ & $5/183$  & $2.7\%$ &  \\
 $3\ kpc<d<4\ kpc$ & $19/61$ & $31\%$ &  \\
 $d>4\ kpc$ & $7/55$ & $12.7\%$ &  \\

\hline
\end{tabular}
\def\arraystretch{1.0}
\label{tab:estatistica}
\end{table}

\subsection{Do the new findings fill the gap?}

\begin{figure*}
\includegraphics[width=0.475\linewidth]{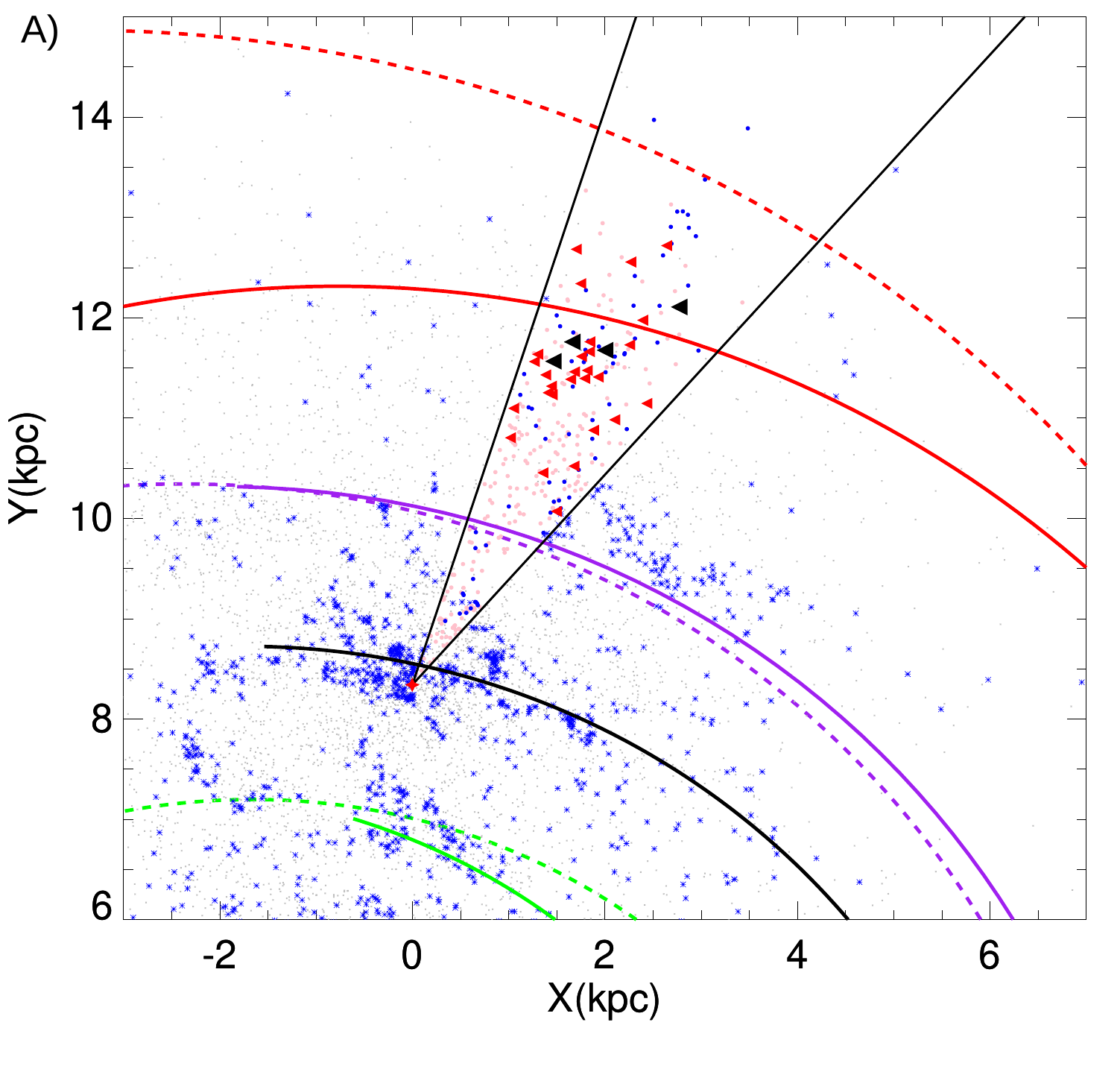}
\includegraphics[width=0.515\linewidth]{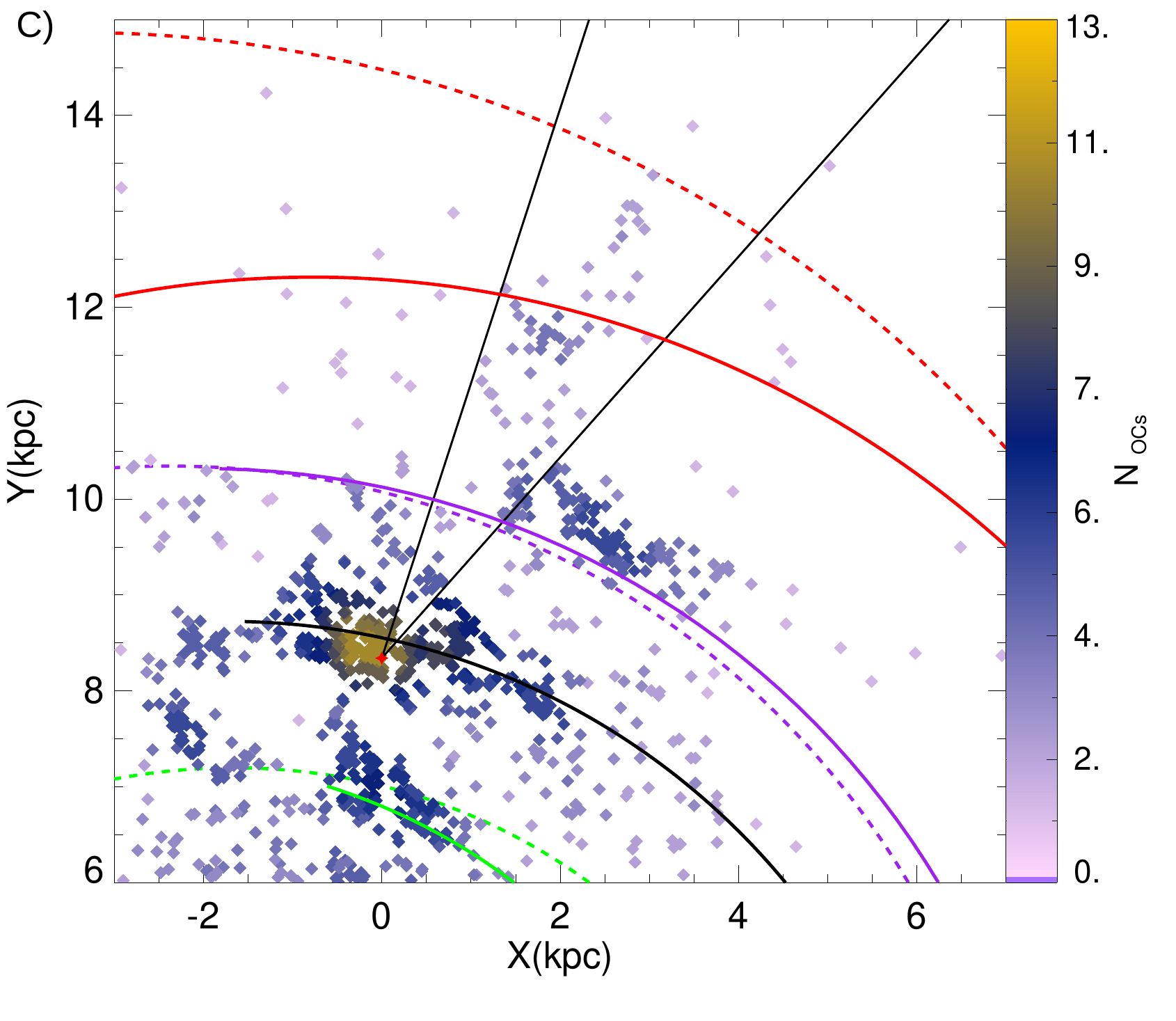}
\includegraphics[width=0.48\linewidth]{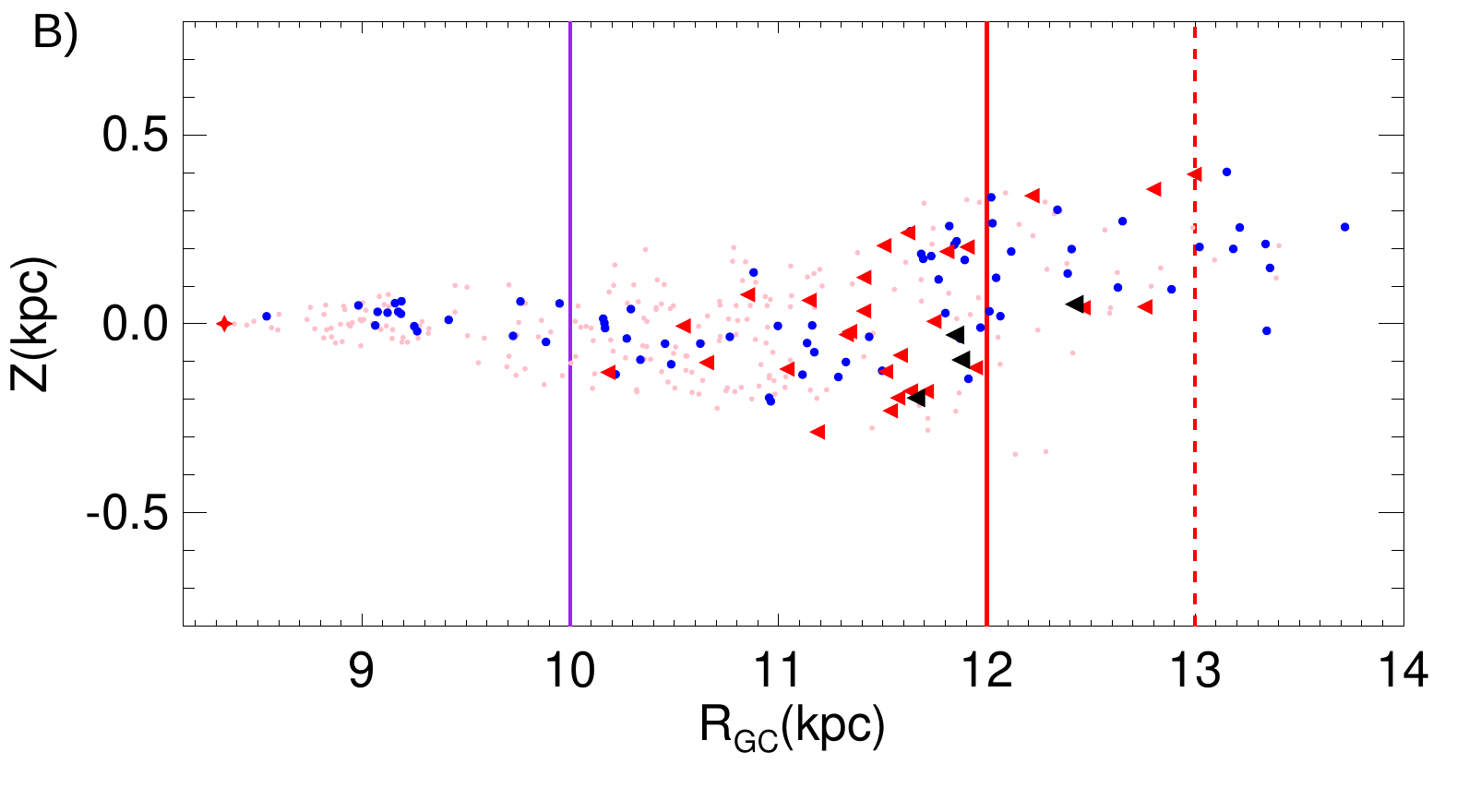}
\includegraphics[width=0.49\linewidth]{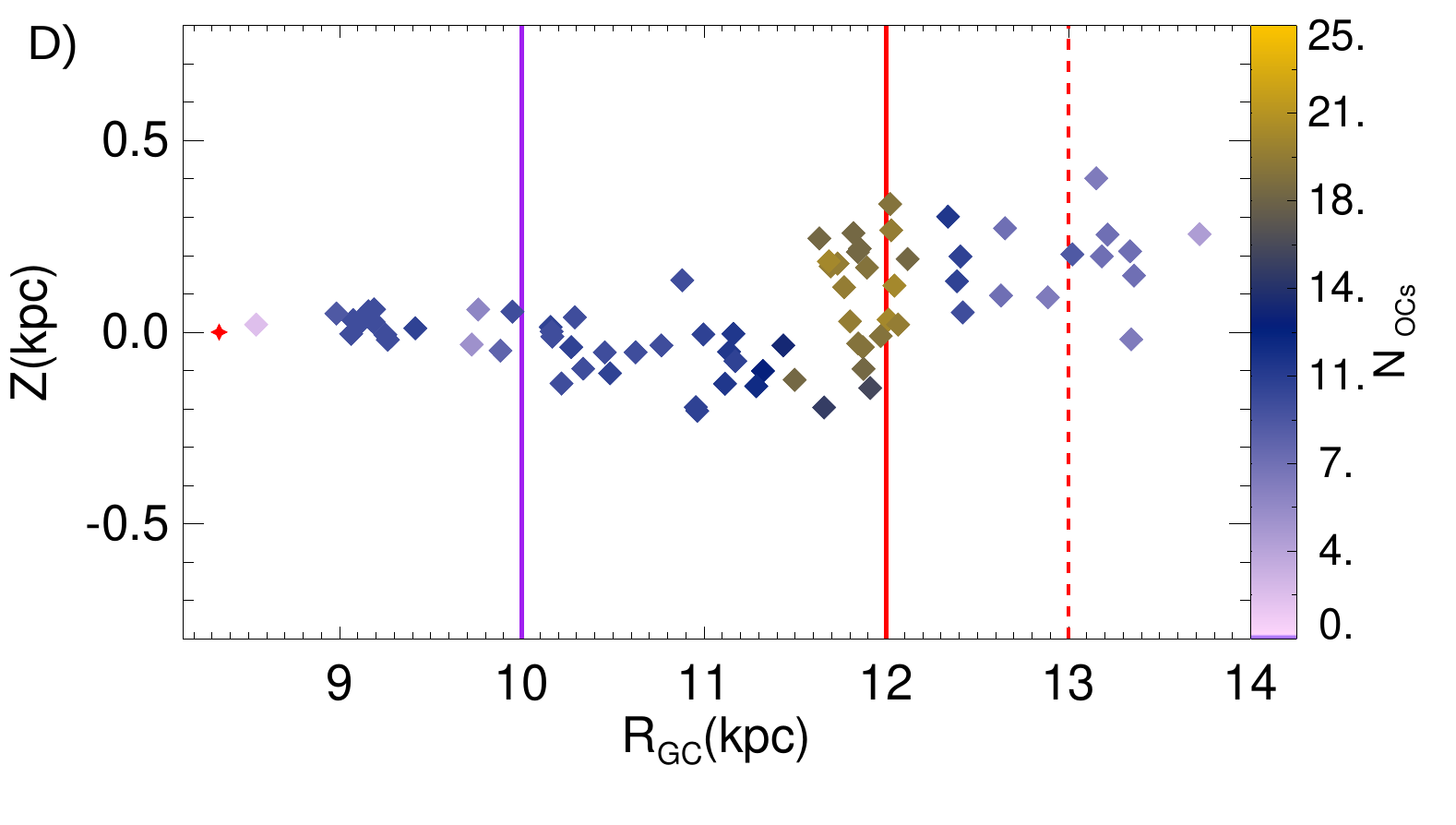}

\caption{A: schematic view of the Galactic spiral arms and the local population of known OCs according to HR2023. Blue asterisks represent their younger OCs ($log(t)<7.5$), while grey symbols indicate the older ones. Within the region of interest — delimited by two black solid lines — two OC samples are shown: those identified in the regional census (young OCs as blue filled circles, and old OCs as pink filled circles) and our sample of recently discovered OCs (young OCs as black triangles, and older OCs as red triangles). The dashed lines mark the spiral arm loci from Vallee (2020): Perseus (purple), Outer Norma (red) and Sagittarius (green). The solid lines represent the arm segments derived by Xu et al. (2023), using the same colour scheme, however the local arm is also represented by a solid black line. The Sun's position is indicated by a red star at $R_\odot=8.34$~kpc (Reid et al. 2014). B: Galactocentric distance versus distance from the Galactic plane for younger and older OCs within the region of interest. The Sun's position (red star) and the approximate locations of the Perseus (purple) and Outer Norma (red) arms are also shown. C and D: same as left, but showing only young OCs as a density map.}
\label{fig:perseu_gap_galaxy}
\end{figure*}

Fig. \ref{fig:perseu_gap_galaxy} shows how the new detections contributed to the spatial distribution of the OC population in the direction of the Perseus arm. In the panels A and B, the sample is represented as follow: OCs from \citetalias{2023A&A...673A.114H}, out of the region of interest, are represented as grey (entire sample) and blue (young OCs, $log[t](yr) < 7.5$) asterisks.  Within the region of interest we also represented the sample of OCs established in the Sect.~\ref{sect:census_region} as pink filled circles, highlighting the young ones ($log[t](yr) < 7.5$) as blue filled circles. The OCs discovered in this work are represented by red triangles (entire sample) and black triangles (young). The panel C shows only the young clusters in the region, but displayed as a density map: the number of young OCs inside squares of $0.4 \times 0.4$ kpc is calculated, where the colour bar indicates the local density. The panels B and D show the distance of the OCs from the Galactic plane as a function of the Galactocentric distance for the same sample described above.

The simple spiral model from \cite{Vallee_2020} (dashed lines) and the spiral arms segments traced by masers (\citealp{2023ApJ...947...54X}, solid lines) show good agreement in the region associated with the Perseus arm (highlighted in purple). In this way, the distribution of the new OCs suggests that they do not belong to the Perseus arm, instead, following the spiral arm segments from \cite{2023ApJ...947...54X}, locating them in the Outer Norma arm. We note that, even with the addition of the new objects in the region, there remains a noticeable lack of young OCs at Galactocentric distances $10 \lesssim R_{GC} \lesssim 11$ kpc, suggesting that either the arm is
discontinuous at this position or that young clusters in this particular direction have not yet been detected.

We also investigate the local spiral structure by verifying the presence of other standard candles in the region around $(X, Y) \approx (1, 10)$ kpc (Fig. \ref{fig:perseu_gap_galaxy}), exploring catalogues of H\,II regions and classical Cepheids data. We adopted the sample of 459 H\,II regions with parallax-based distances from \cite{2025A&A...696A..67S} and the sample of 2857 dynamically young classical Cepheids studied by \cite{2025ApJS..278...57S} and \cite{2025A&A...698A.230D}. For both databases, we calculated the Galactic cartesian coordinates and Galactocentric distances based on the informed distances and Galactic coordinates, adopting the same Solar Galactocentric radius used for the OC samples ($R_{Sun}=8.34$~kpc). 
 
\begin{figure*}
\includegraphics[width=0.48\linewidth]{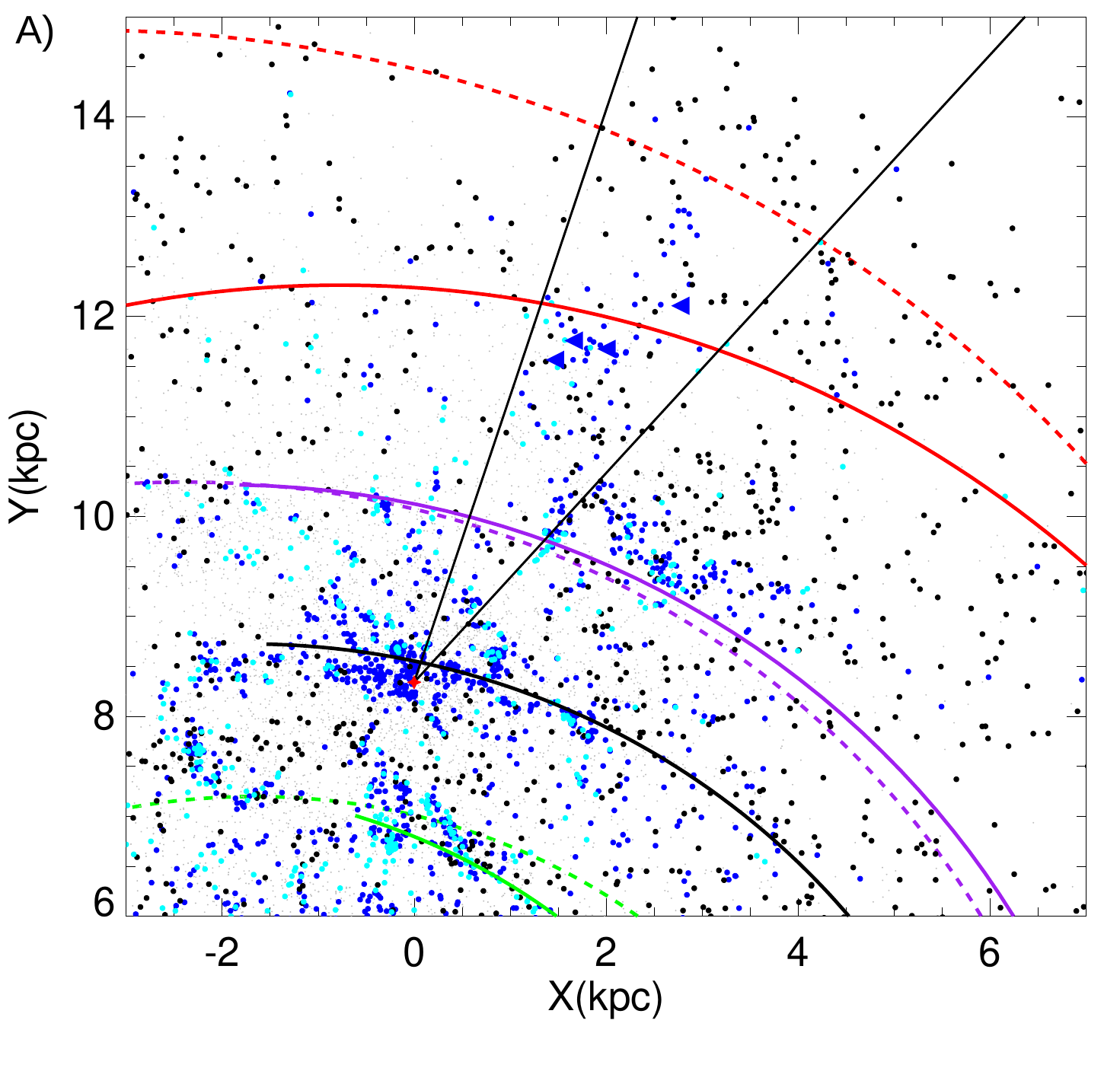}
\includegraphics[width=0.49\linewidth]{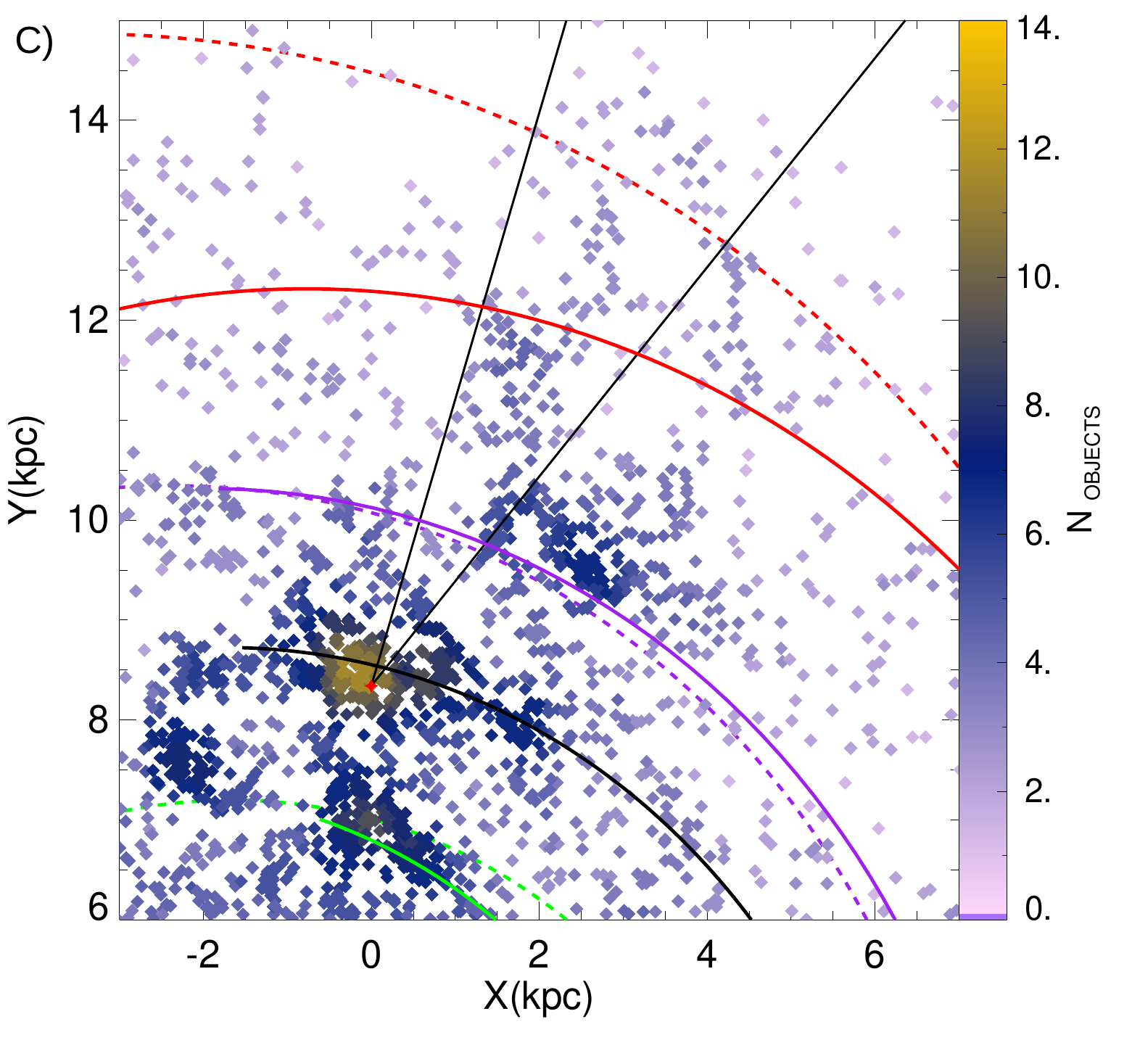}
\includegraphics[width=0.48\linewidth]{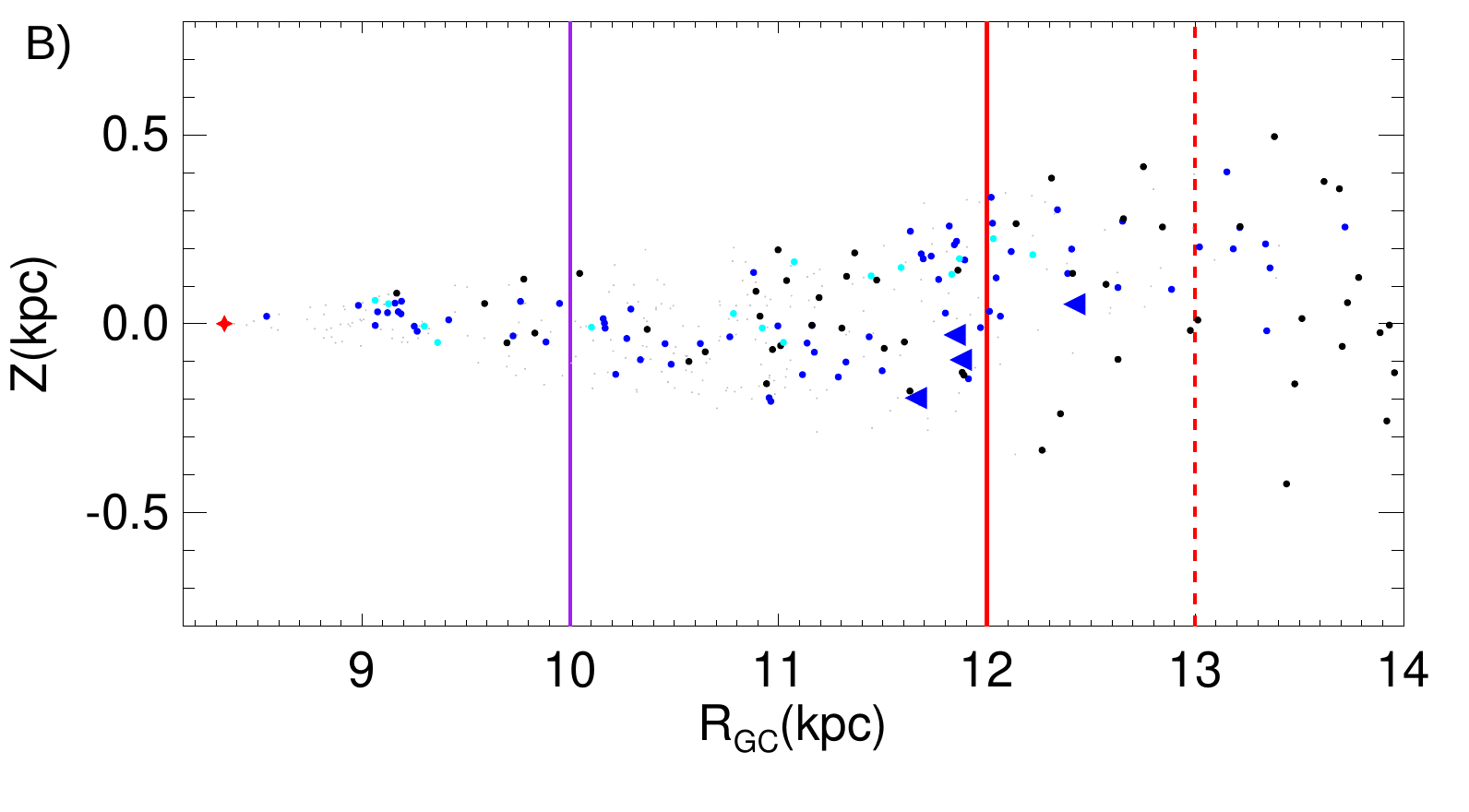}
\includegraphics[width=0.49\linewidth]{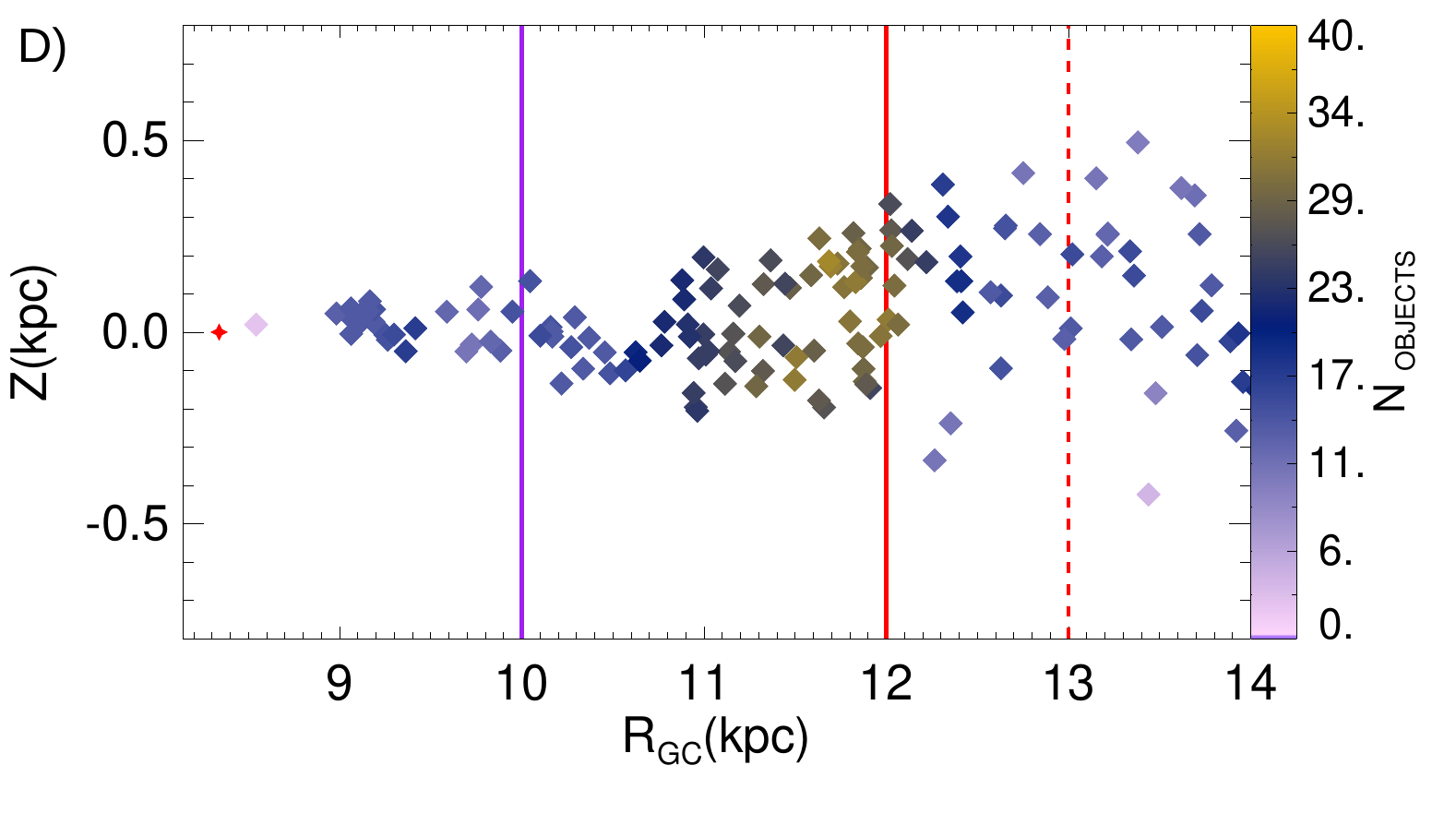}

\caption{Same as Fig \ref{fig:perseu_gap_galaxy}, however, the panels also represent the sample of classical Cepheids and H\,II regions. All young OCs are represented by blue: those from literature are represented by blue dots, while our sample of new young OCs by blue triangles. Older OCs from these samples are represented by grey. Classical Cepheids are represented as black and H\,II regions as cyan. Right panels: same as left, but showing all young objects together as a density map.}
\label{fig:perseu_gap_galaxy_cef_hII}
\end{figure*}

Figure \ref{fig:perseu_gap_galaxy_cef_hII} shows the addition of the sample of classical Cepheids and H\,II regions to the same region. All young OCs: those from \citetalias{2023A&A...673A.114H}, the clusters included in the the census of the region (see Sect.~\ref{sect:census_region}) and our sample of new OCs are represented as blue symbols. Classical Cepheids are represented as black and H\,II as cyan. Density maps analogous to those shown in the panels C and D of Figure \ref{fig:perseu_gap_galaxy} are also shown. The aim of this comparison is not to draw the spiral arms, as such different spiral tracers, do not always trace exactly the same spiral arm segments, but rather to verify whether there is a significant presence of young OCs and standard candles in this specific region. Even after including these populations, the apparent absence of objects remains, reinforcing the indication that the Perseus arm is interrupted in this direction.

Arm interruptions are a common feature in multi-arm, patchy and flocculent spiral galaxies (such as NGC 2841 or NGC 7793). The four-arm picture of the Milky Way estimated by many authors \citep[e.g.][]{1976A&A....49...57G,2003A&A...397..133R,2017PASP..129i4102K,10.1093/mnras/stad3350} approachs better the view that our Galaxy  resembles a flocculent spiral rather than a grand design spiral, with multiple dynamic arms induced by local gravitational instabilities \citep{2014PASA...31...35D}.

\section{CONCLUDING REMARKS}
\label{sect:concl}

In this work, we discovered 31 OCs projected towards the Galactic anticentre with \textit{Gaia} DR3 data. They are located in the second Galactic quadrant, at low latitudes and immersed in dense stellar fields. Our method was capable of disentangling genuine cluster members from the high-density Galactic disc foreground/background. From the clusters’ stellar spatial distributions, centres, sizes and structural parameters were obtained. By performing solar metallicity isochrone fittings over their decontaminated CMDs, distances, ages and colour excesses were derived. 

Our analysis shows that the discovered objects are distant ($3\,kpc - 5 \,kpc$) and reddened ($0.5<E(B-V)<1.5$.) systems and lie in the same loci of the bulk of confirmed OCs in terms of ages (from 20\,Myr to 1\,Gyr) and exhibit less concentrated structures than the average. Eight of the newly identified clusters show low stellar densities ($\rho < \rho_{\rm crit}$), which could be an artefact of observational limits or may indicate that they exhibit characteristics of open cluster remnants (that is, dissolving clusters). Nevertheless, their proper motion dispersions and sizes are consistent with the high-quality OCs reported by \citetalias{2023A&A...673A.114H}, and they present a clear contrast with the sparse moving groups, exhibiting much more concentrated structures.

Our search methodology has proven effective do detect fiducial OCs: nearly $100\%$ of the OCs reported in papers \citetalias{2019MNRAS.483.5508F}, \citetalias{10.1093/mnras/staa1684} and \citetalias{2021MNRAS.502L..90F} using \textit{Gaia} DR2 data were later recovered by \citetalias{2023A&A...673A.114H} with \textit{Gaia} DR3 data. In this work, our approach, combined with the high precision of \textit{Gaia} DR3 astrometric and photometric data, , enabled the identification of OCs directly in the astrometric parameter space. The method also proved sensitive to less concentrated systems, showing that visual inspection remains a useful tool for detecting less massive systems in crowded fields, although further development of specialized automated searches is still required. Since proper motion plays a significant role in the cluster detectability, we note that the newly detected clusters, despite being poorer than many literature OCs, exhibit moderate values of $\Delta\mu_{\alpha}^{*}$ and $\Delta\mu_{\delta}$, which likely contributed to their detectability.

The discovered OCs represent a significant increment in the census of Galactic clusters, particularly for objects with $3kpc<d<4kpc$ ($31\%$) and objects with  $d>4kpc$ ($12\%$). This suggests that a non-negligible fraction of OCs may have been missed by automatic detection algorithms. These clusters do not belong to the Perseus arm and may be distributed along the Outer Norma arm, where the few detected young objects are valuable tracers for future studies of the outer Galactic structure. The Gulf of Camelopardalis region seems to be an interruption of the Perseus arms, showing an almost complete absence of young OCs and standard candles (classical Cepheids and H\,II regions). Such interruptions on the arms structure could suggest different scenarios on those regions: low star formation rate, the presence of dust clouds or even that the Milky Way is not a grand design spiral galaxy with prominent and well-defined continuous spiral arms, but a patchy, multi-arm or a flocculent spiral galaxy.

\section*{Acknowledgements}
We thank the referee for constructive comments that helped us improve the clarity and quality of this manuscript. The authors wish to thank the Brazilian financial agencies FAPEMIG, CNPq and CAPES (finance code 001). W.Corradi acknowledges the support from CNPq - BRICS 440142/2022-9, FAPEMIG APQ 02493-22  and FNDCT/FINEP/REF 0180/22. 
F.F.S.M. acknowledges financial support from Conselho Nacional de Desenvolvimento Cient\'ifico e Tecnol\'ogico -- CNPq (proc.\ 404482/2021-0) and from FAPERJ (proc.\ E-26/201.386/2022 and E-26/211.475/2021).
This research has made use of the VizieR catalogue access tool, CDS, Strasbourg, France. This work has made use of data from the European Space Agency (ESA) mission \textit{Gaia} (\url{https://www.cosmos.esa.int/gaia}), processed by the \textit{Gaia} Data Processing and Analysis Consortium (DPAC, \url{https://www.cosmos.esa.int/web/gaia/dpac/consortium}). Funding for the DPAC has been provided by national institutions, in particular the institutions participating in the \textit{Gaia} Multilateral Agreement. This  research  has  made  use  of  TOPCAT \citep{Taylor:2005}.

\section*{Data availability}
The data underlying this article is publicly available (\textit{Gaia DR3}) or is available in the article.



\bibliographystyle{mnras}
\input{main_paper.bbl}  

\bsp	
\label{lastpage}
\end{document}